\documentclass[final,3p,times]{elsarticle}

\usepackage[T1]{fontenc}
\usepackage[utf8]{inputenc}
\usepackage{amsmath}
\usepackage{amssymb}
\usepackage{bm}
\usepackage{mathtools}
\usepackage{physics}
\usepackage{cleveref}
\crefformat{equation}{#2Eq. (#1)#3}
\crefrangeformat{equation}{#3Eqs. (#1)#4 to #5(#2)#6}
\crefmultiformat{equation}{#2Eqs. (#1)#3}{ and #2(#1)#3}{, #2(#1)#3}{ and #2(#1)#3}
\crefrangemultiformat{equation}{#3Eqs. ((#1))#4 to #5((#2))#6}{ and #3(#1)#4 to #5(#2)#6}{, #3(#1)#4 to #5(#2)#6}{ and #3(#1)#4 to #5(#2)#6}
\usepackage{caption}
\Crefformat{figure}{#2Fig.~#1#3}
\usepackage{float}
\usepackage{subfig}
\usepackage{booktabs}
\usepackage{threeparttable}
\usepackage{tikz}
	\usetikzlibrary{positioning,arrows}
	\definecolor{gray1}{RGB}{215, 215, 215}
\usepackage{xspace}
\newcommand*{\ie}{i.e.\@\xspace}
\newcommand*{\viz}{viz.\@\xspace}
\usepackage{amsthm}
\newtheorem*{remark}{Remark}

\newcommand{\rom}[1]{\romannumeral #1}

\usepackage[titletoc]{appendix}
\allowdisplaybreaks

\makeatletter
\def\ps@pprintTitle{%
	\let\@oddhead\@empty
	\let\@evenhead\@empty
	\let\@oddfoot\@empty
	\let\@evenfoot\@oddfoot
}
\makeatother

\begin{document}

\begin{frontmatter}



\title{A theoretical framework for multi-physics modeling of poro-visco-hyperelasticity-induced time-dependent fracture of blood clots} 


\author[1]{Dongxu Liu\corref{cor1}}
\ead{Dongxu.Liu@bsd.uchicago.edu}

\author[1]{Nhung Nguyen\corref{cor1}}
\ead{nhungng@bsd.uchicago.edu}

\author[2,3]{Tinh Quoc Bui}

\author[1]{Luka Pocivavsek}

\cortext[cor1]{Corresponding author}

\affiliation[1]{organization={The University of Chicago},
	addressline={5841 S. Maryland Avenue}, 
	city={Chicago},
	postcode={60637}, 
	state={IL},
	country={United States}}

\affiliation[2]{organization={Duy Tan Research Institute for Computational Engineering (DTRICE), Duy Tan University},
	addressline={6 Tran Nhat Duat, Dist. 1}, 
	city={Ho Chi Minh City},
	postcode={700000}, 
	country={Vietnam}}

\affiliation[3]{organization={Faculty of Civil Engineering, Duy Tan University},
	city={Da Nang City},
	postcode={550000}, 
	country={Vietnam}}

\begin{abstract}
Fracture resistance of blood clots plays a crucial role in physiological hemostasis and pathological thromboembolism. Although recent experimental and computational studies uncovered the poro-viscoelastic property of blood clots and its connection to the time-dependent deformation behavior, the effect of these time-dependent processes on clot fracture and the underlying time-dependent fracture mechanisms are not well understood. This work aims to formulate a thermodynamically consistent, multi-physics theoretical framework for describing the time-dependent deformation and fracture of blood clots. This theory concurrently couples fluid transport through porous fibrin networks, non-linear visco-hyperelastic deformation of the solid skeleton, solid/fluid interactions, mechanical degradation of tissues, gradient enhancement of energy, and protein unfolding of fibrin molecules. The constitutive relations of tissue constituents and the governing equation of fluid transport are derived within the framework of porous media theory by extending non-linear continuum thermodynamics at large strains. A physics-based, compressible network model is developed for the fibrin network of blood clots to describe its mechanical response. The kinetic equations of the internal variables, introduced for describing the non-linear viscoelastic deformation, non-local damage driving force and protein unfolding, are formulated according to the thermodynamics principles by incorporating a non-equilibrium energy of fibrin networks, a gradient-enhanced energy, and a stretch-induced internal energy of fibrin molecules, respectively, into the total free energy density function. An energy-based damage model is developed to predict the damage and fracture of blood clots, and an evolving regularization parameter is proposed to limit the damage zone bandwidth. The proposed model is implemented into finite element code by writing subroutines and is experimentally validated using single-edge cracked clot specimens with different constituents. The fracture of blood clots subject to different loading conditions is simulated, and the mechanisms of clot fracture are systematically analyzed. Computational results show that this model can accurately capture the experimentally measured deformation and fracture. The viscoelasticity and fluid transport play essential roles in the fracture of blood clots under physiological loading.
\end{abstract}



\begin{keyword}
Blood clots \sep Multi-physics modeling \sep Solid/fluid biphase \sep Fracture \sep Evolving regularization parameter \sep Protein unfolding


\end{keyword}

\end{frontmatter}


\section{Introduction}
\label{sec:intro}
\begin{figure}[H]
	\centering
	\subfloat[]{
		\begin{minipage}[b]{0.235\linewidth}
			\includegraphics[width=1\linewidth]{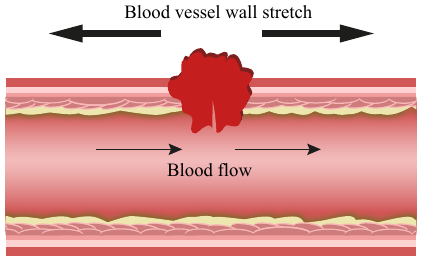} 
		\end{minipage}
		\label{}
	}
	\subfloat[]{
		\begin{minipage}[b]{0.235\linewidth}
			\includegraphics[width=1\linewidth]{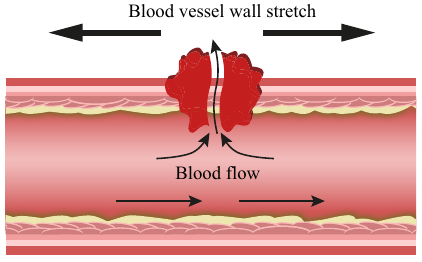} 
		\end{minipage}
		\label{}
	}
	\subfloat[]{
		\begin{minipage}[b]{0.235\linewidth}
			\includegraphics[width=1\linewidth]{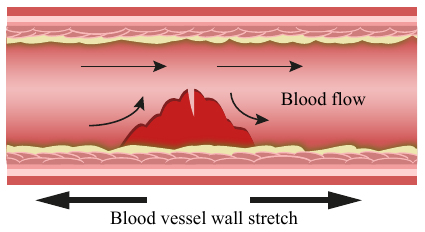} 
		\end{minipage}
		\label{}
	}
	\subfloat[]{
		\begin{minipage}[b]{0.235\linewidth}
			\includegraphics[width=1\linewidth]{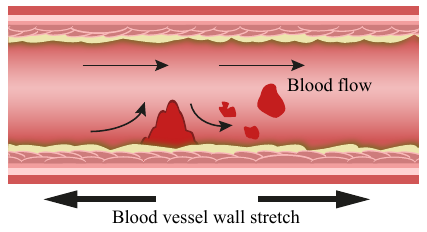} 
		\end{minipage}
		\label{}
	}
	\caption{Schematic of location, loading conditions, and incipient rupture of blood clots: (a) physiological hemostasis; (b) rupture of a blood clot at the wounded tissue; (c) pathological thrombosis and (d) rupture of an intravascular blood clot. Diastole and systole kinematics of blood vessels and blood flow result in large stretches and shear strain in blood clots. Under the physiological loadings, the fracture resistance of blood clots plays a crucial role in health maintenance.}
	\label{fig:1}
\end{figure}

As an essential biological tissue, a blood clot is a three-dimensional polymeric network composed of a fibrin scaffold, water and blood cells. Blood clots forming at the wounded tissues can stop bleeding, \ie, physiological hemostasis (\Cref{fig:1}a), while intravascular clots can narrow blood vessels and disorder the circulatory system, \ie, pathological thrombosis (\Cref{fig:1}c). The tissue can be digested by lytic enzymes via fibrinolysis, which results in primary initial defects \citep{collet2005elasticity}. Subject to mechanical loading from hydrodynamic forces of blood flow, blood vessel wall fluctuations, extravascular muscle contraction and other forces, the defects can evolve into significant cracks or even rupture \citep{tutwiler2020rupture}. In addition, tensile stresses by retrieving thrombi in mechanical thrombectomy procedures can also stretch and break clots \citep{fereidoonnezhad2021blood}. On the one hand, rupture of blood clots at injured sites is related to hemostasis issues, which could result in dangerous blood loss and slow wound healing (\Cref{fig:1}b). On the other hand, fractured intravascular blood clots may fragment into smaller particles, which migrate in the bloodstream, obstruct blood flow in vessels and consequently cause thromboembolism (\Cref{fig:1}d). Embolization may cause ischemic stroke, heart attack and pulmonary emboli. Blood clot fracture is closely correlated with the high mortality rate of cerebrovascular and cardiovascular diseases. Despite its vital role in maintaining health, the mechanisms of blood clot fracture remain inconclusive. Understanding the mechanisms is crucial for developing new strategies to prevent, diagnose and treat thrombotic diseases. It can also advance the development of novel wound-dressing biomaterials \citep{jimenez2023multiscale,nour2019review}.
\par
The physiological and pathological importance motivates experimental and computational studies on mechanical and fracture behaviors of blood clots. \citet{tutwiler2020rupture} experimentally investigated the mechanical and structural rupture of human fibrin clots with different initial crack lengths, where the toughness of physiologically relevant fibrin clots is calculated. Moreover, composition-dependent fracture tests are conducted \citep{fereidoonnezhad2021blood}, and novel experimental approaches, such as lab shear tests and double cantilever beam tests, are developed to quantify the fracture behavior of blood clots \citep{liu2021fracture}. Experimental and histological investigations uncovered the complex mechanical and structural characteristics of the tissue \citep{brown2009multiscale}. Therein, the time-dependent deformation has been observed in multiple experimental tests \citep{sugerman2020nonlinear}, which is commonly explained by the intrinsic viscoelasticity of an individual fibrin fiber and other conformational changes of fibrin, such as the rearrangement, sliding and disentanglement of fibrin fibers \citep{ghezelbash2022blood,chaudhuri2020effects,collet2005elasticity,gersh2009fibrin,feller2022fibrin,schmitt2011characterization}. Furthermore, the most recent studies demonstrate that fluid transport and its interaction with the solid skeleton also play essential roles in time-dependent deformation \citep{Liu2024,ghezelbash2022blood,he2022viscoporoelasticity,varner2023elasticity}. These time-dependent processes are closely correlated with the damage and fracture of blood clots \citep{Liu2024,rausch2021hyper}.
\par
Designing and conducting well-controlled experiments to study clot fracture under physiological loading conditions remain challenging due to its fragile, wet and low-strength characteristics \citep{sugerman2021whole,liu2021fracture,van2016constitutive}. These limits motivate the development of computational modeling and numerical simulation as a promising approach to gain insight into time-dependent deformation and fracture of blood clots. Viscoelastic models have been developed for describing complex deformation of thrombi. \citet{van2008non} presented a non-linear viscoelastic constitutive model and used it to study the variation of mechanical properties throughout the thickness of abdominal aortic aneurysm thrombi caused by structural changes. Combining dynamic ultrasound elastography and typical rheological models, \citet{schmitt2011characterization} proposed a method to characterize the viscoelasticity of whole blood clots by quantitating the viscoelastic parameters. In whole blood clots, the main solid components, \ie, fibrin fibers, red blood cells and platelets, dominate the viscoelasticity \citep{ghezelbash2022blood,van2016constitutive}. To understand the contributions from these components, \citet{van2016constitutive} and \citet{tashiro2021finite} developed constitutive models based on a generalized Maxwell model for accurately capturing the non-linear viscoelasticity of blood clots with different compositions. While viscoelastic models have been well established, the time-dependent non-linear deformation of blood clots cannot be accurately captured by the simple combination of viscoelastic and hyperelastic models \citep{he2022viscoporoelasticity,tashiro2021finite}. Additional dissipation mechanisms, such as fluid flow, plasticity and damage, have received increased attention recently in order to obtain a more accurate prediction \citep{ghezelbash2022blood,he2022viscoporoelasticity,tashiro2022numerical}.
\par
A blood clot is a fluid-saturated porous medium \citep{du2021computational,selvadurai2017inflation}. Subject to physiological loading, inhomogeneous deformation leads to nonuniform distributions of pore pressure of interstitial fluid, which drives the migration of fluid molecules. Fluid transport interacts with the deformation of polymerized fibrin networks and blood cells and contributes to the time-dependent deformation. Due to the biphasic material properties and high fluid content, the contribution from the fluid transport to the time-dependent deformation cannot be ignored \citep{ghezelbash2022blood,he2022viscoporoelasticity}. Despite its importance, only a few models were developed that considered the solid/fluid coupling to
describe the time-dependent deformation behavior of blood clots. \citet{noailly2008poroviscoelastic} analytically and numerically compared the large strain poroelastic and poroviscoelastic constitutive models for describing time-dependent mechanical behaviors of fibrin gels, concluding that fibrin viscoelasticity and fluid flow are both necessary for an accurate prediction. \citet{selvadurai2017inflation} formulated a poro-hyperelastic model by extending the work of \citep{suvorov2016poro,selvadurai2016coupled} for studying the classical Lam\'e-type problem. This model considered the porous skeleton-pore fluid coupling and was applied to describe the evolving behavior of intra-luminal thrombus caused by fluid transport. \citet{du2021computational} developed a two-phase continuum model to study thrombus stability under flow. This model considered thrombi as porous viscoelastic material, and the viscoelasticity is determined by modeling breaking and formation of interplatelet molecular bonds. \citet{he2022viscoporoelasticity} and \citet{ghezelbash2022blood} systematically conducted experimental and computational studies to reveal the mechanisms of time-dependent deformation of blood clots. They hypothesized the importance of coupling poroelasticity and viscoelasticity in modeling the deformation behavior of blood clots.
\par
To correlate the complex non-linear deformations with the damage and fracture behavior of blood clots, different damage modeling methods were proposed. A viscoelastic damage model was established to study the influence of non-linear viscoelastic behavior on the damage of blood clots under large deformation \citep{rausch2021hyper}, where a gradient-enhanced damage modeling approach was combined to a hyper-viscoelastic framework. This model accurately predicted the deformation and damage of whole blood clots under uniaxial tension, cyclic tension and stress relaxation loadings. \citet{fereidoonnezhad2021blood} used a cohesive zone model to simulate crack initiation and propagation of clot specimens with different red blood cell/fibrin ratios, where blood clots were modeled as an anisotropic hyperelastic fibrous soft tissue using the constitutive model proposed by \citet{fereidoonnezhad2020new}. In order to accurately capture the loading/unloading hysteresis and the permanent deformation of blood clots, \citet{tashiro2022numerical} introduced an elastoplastic damage model by applying the damage and plastic models proposed by \citet{simo1987fully} and \citet{miehe1995discontinuous}.
\par
Furthermore, blood clots are highly extensible supramolecular protein polymers due to the hierarchical microstructures of fibrin fibers \citep{spiewak2022biomechanical,feller2022fibrin,brown2009multiscale}. The primary and essential component of a fibrin fiber is the fibrin monomer
or the fibrinogen molecule that is composed of $\alpha$A, $\beta$B and $\gamma$ polypeptide chains \cite{Stamboroski2021,yesudasan2020multiscale}. At large stretches, the configuration change of an individual fiber drives the extension of fibrin monomers, leading to the unfolding of the coiled-coils and the domains of a $\gamma$-nodule \citep{brown2009multiscale,zhmurov2011mechanism,Vos2020}. The unfolding behavior of fibrin monomers dissipates energy and contributes to the high stretchability of blood clots. While the influence of protein unfolding on the macroscopic deformation of clots has been experimentally and theoretically studied \cite{brown2009multiscale,purohit2011protein,yesudasan2020multiscale}, its role in blood clot fracture is still unknown.
\par
Owing to the complicated multi-physics coupling in blood clots, the mechanism of their time-dependent mechanical behaviors is still not well understood. Modeling the complete coupling of those physical processes during the deformation and crack propagation in solid/fluid biphasic soft biological tissues remains challenging. In this regard, a multi-physics computational model concomitantly coupling non-linear viscoelasticity, fluid transport, material damage and protein unfolding is imperative for understanding the time-dependent fracture behavior of blood clots under complex physiological loading conditions. To this end, a thermodynamically consistent, physics-based poro-visco-hyperelastic damage model is developed in this work using the theory of porous media at large strains for describing the multi-mechanism time-dependent deformation and fracture of blood clots. A compressible network model is formulated for the equilibrium and non-equilibrium responses of fibrin networks by extending a classical physics-based hyperelastic model. The unfolding of the $\gamma$ chain in fibrin is described by introducing an internal variable to the total stretch of the fibrin networks, and an unfolding energy is incorporated into the Helmholtz free energy density of fibrin networks to derive the evolution equation of the internal variable. A novel energy-based network alteration model is derived according to the maximum dissipation principle to describe the damage evolution of fibrin networks. Moreover, an evolving gradient-enhanced constitutive framework is used to diminish the damage localization and mesh sensitivity in numerical simulations. For fluid transport, the governing equation for fluid flow through fibrin networks is derived within the thermodynamic framework. This model is implemented into an implicit finite element procedure on different geometries subject to tension and shear/tension mixed-mode loading conditions to show the predictive capability of the model and study the mechanisms of time-dependent fracture of blood clots.
\par
This paper is organized as follows. In \Cref{sec:Modelling_TDF}, a generalized thermodynamic framework is formulated. The kinetic equations for non-equilibrium deformation, protein unfolding stretch, damage variable, non-local thermodynamic forces, and fluid transport are derived. In \Cref{sec:specification_energy_function}, the constitutive relations are specified for fibrin networks and red blood cells. In \Cref{sec:computational_example}, mesh objectivity, model validation, time dependence of fracture, and mix-mode fracture are demonstrated. \Cref{sec:conclusion} summarizes the highlights, findings, and conclusions of this work.
%
\section{Thermodynamic formulation of the constitutive equations}
\label{sec:Modelling_TDF}
\begin{figure}[H]
	\centering
	\includegraphics[width=0.9\linewidth]{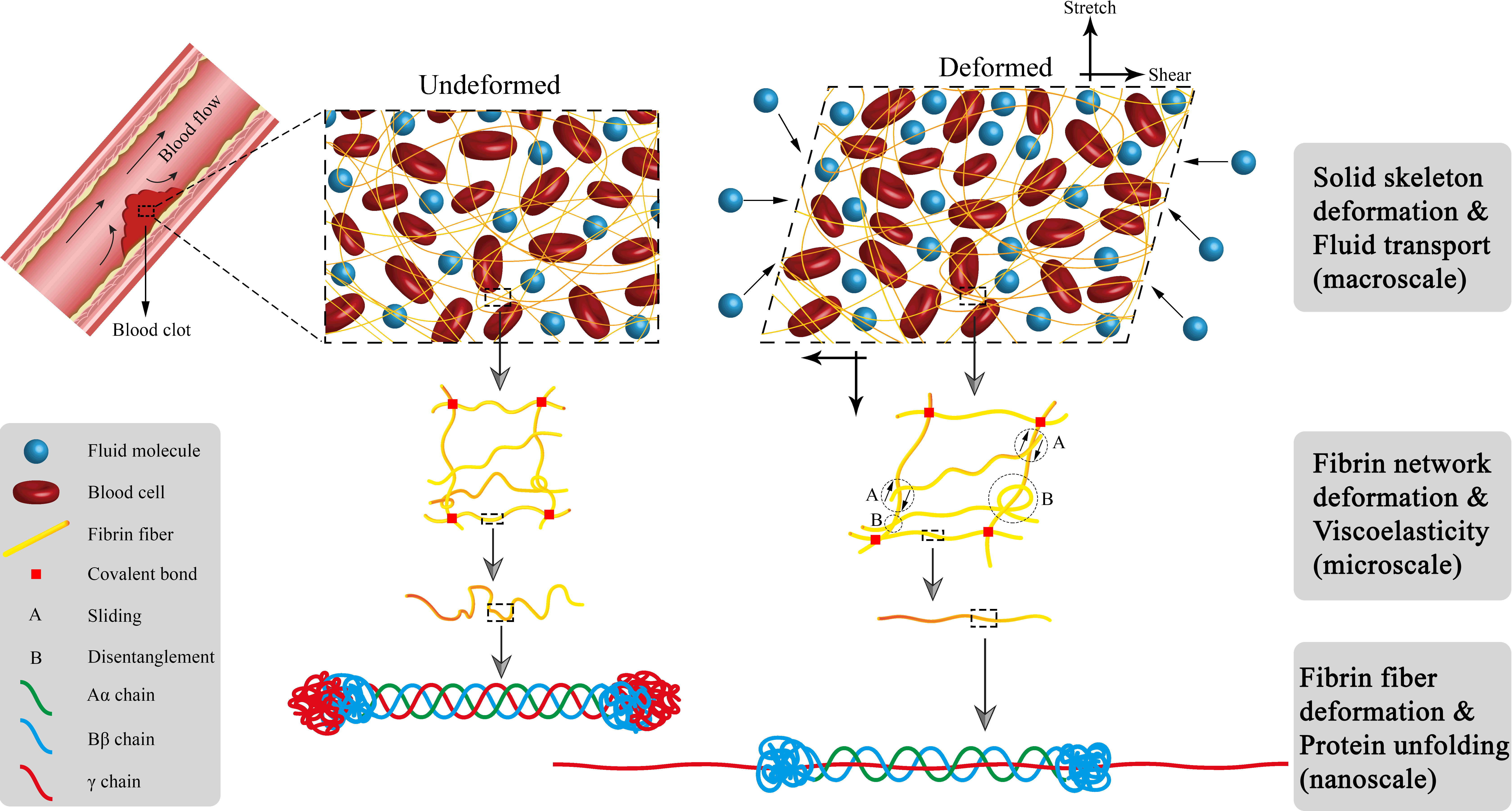}
	\caption{A schematic representative of the multi-physics problem in the time-dependent mechanical responses of blood clots. On the macroscale, physiological loading drives the time-dependent solid  skeleton deformation and fluid transport. On the microscale, the deformation of a fibrin network demonstrates viscoelasticity due to the viscoelastic fiber stretch, relative sliding between fibers, and fiber disentanglement. On the nanoscale, the force-driven hierarchical structure variation causes the protein unfolding of fibrin monomers.}
	\label{fig:2}
\end{figure}

A blood clot is a biopolymer with a natural hydrogel microstructure. Whole blood clots are mainly composed of red blood cells and fibrin networks filled with water (see \Cref{fig:2}). Subject to complex physiological loadings, such as shear force and stretch, clots demonstrate time-dependent deformation. First, The fibrin network demonstrates time-dependent deformation because of the viscoelasticity of a single fibrin fiber, relative sliding between fibers, and their disentanglement behavior. Furthermore, the inhomogeneous solid deformation results in fluid transport through the fibrin networks. This procedure is also time-dependent. Solid skeleton deformation interacts with fluid transport, concurrently contributing to the mechanical behavior of the bulk tissue. Due to the biphasic nature of clots, the theory of porous media is applicable to describe the solid/fluid coupling in the tissue. Moreover, the fibrin monomer in a single fibrin fiber can unfold due to stretch, as shown in \Cref{fig:2}. All these macroscale, microscale and nanoscale mechanical behaviors play essential roles in the damage and fracture of blood clots. Based on porous media theory, a generalized biphasic continuum modeling framework of the constitutive behavior of blood clots is presented in this section. 
\subsection{Mixture theory and the concept of volume fraction}
\label{sec:Mixture_volume_fraction}
The theory of porous media is the combination of the mixture theory and the concept of volume fraction. According to the mixture theory, a blood clot is treated as a fluid-saturated two-phase porous material which is a macroscopically homogenized mixture $\varphi$ of immiscible and interacted constituents $\varphi^{\alpha}$ ($\alpha=S$: solid or $F$: fluid), \ie, $\varphi = \varphi^{S} \cup \varphi^{F}$ \citep{de2012theory}. The mixture theory is restricted by the concept of volume fraction which is introduced to describe the local composition. The volume fraction of $\varphi^{\alpha}$ at any time $t$ is defined as $n^{\alpha} \coloneqq \dd{v}^{\alpha}/\dd{v}$, where $\dd{v}^{\alpha}$ and $\dd{v}$ are the bulk volume elements of $\varphi$ and the partial volume element of $\varphi^{\alpha}$, respectively. Since blood clots are assumed to be fluid-saturated, the volume fractions are constrained by the saturation condition
\begin{equation}
	\label{equ:saturation_condition}
	\sum_{\alpha}^{S,F} n^{\alpha} = 1 \,.
\end{equation}
\par
With these definitions at hand, the realistic density and apparent density of $\varphi^{\alpha}$ can be defined as $\rho^{\alpha R} \coloneqq \dd{m}^{\alpha}/\dd{v}^{\alpha}$ and $\rho^{\alpha} \coloneqq \dd{m}^{\alpha}/\dd{v}$, respectively, where $\dd{m}^{\alpha}$ denotes the mass element of $\varphi^{\alpha}$. The relationship between $\rho^{\alpha}$ and $\rho^{\alpha R}$ is expressed as $\rho^{\alpha} = n^{\alpha} \rho^{\alpha R}$. The mixture density $\rho$ of $\varphi$ can be given as $\rho \coloneqq \sum_{\alpha}^{S,F} \rho^{\alpha}$. In this work, the solid and fluid constituents are assumed to be materially incompressible due to the significantly larger change of intervoid space, induced by fluid flux, than the materially volumetric compressibility. Therefore, the realistic density remain constant, \ie, $\rho^{\alpha R}$= const. It should be noted that material volume expansions or contractions due to fluid influx or efflux from the solid skeleton can lead to the change of the partial density and the mixture density, although both constituents are assumed to be intrinsically incompressible.
\subsection{Kinematics}
\label{sec:Kinematics}
To start, a brief summary of the kinematics of large strain visco-hyperelasticity, fluid flow and notations adopted is provided. The position vector $\vb{x}$ of an arbitrary material point in the current configuration is concomitantly occupied by fluid and solid, which originate from their position $\vb{X}_{\alpha}$ in the initial configuration following their own motion $\vb{x} = \vb*{\chi}_{\alpha} (\vb{X}_{\alpha},t)$ and velocity $\acute{\vb{x}}_{\alpha}$ \citep{markert2010comparison}. With these definitions at hand, the solid displacement vector $\vb{u}_{S} \coloneqq \vb{x}-\vb{X}_{S}$ and the seepage velocity vector $\vb{w}_{FR} \coloneqq \acute{\vb{x}}_{F}-\acute{\vb{x}}_{S}$ are introduced for describing the solid kinematics and fluid motion, respectively. The fluid motion within the porous solid correlates with solid deformation via the seepage velocity.
\par
The solid deformation gradient tensor is defined as $\vb{F}_{S} \coloneqq \pdv*{\vb{x}}{\vb{X}_{S}}$ within the large deformation framework, with its determinant $J_{S} \coloneqq \det \vb{F}_{S} > 0$. The right and left Cauchy--Green deformation tensors of the solid constituent are introduced as $\vb{C}_{S} \coloneqq \vb{F}_{S}^\intercal \vb{F}_{S}$ and $\vb{B}_{S} \coloneqq \vb{F}_{S} \vb{F}_{S}^\intercal$, respectively, where $(\bullet)^\intercal$ denotes the transposition of a tensor. The first invariant of $\vb{C}_{S}$ and $\vb{B}_{S}$ is defined as $I_{S1} \coloneqq \tr(\vb{C}_{S}) =\tr(\vb{B}_{S})$, where $\tr(\bullet)$ represents the trace of a tensor. The spatial velocity gradient tensor is denoted by $\vb{L}_{\alpha} \coloneqq \grad_{\vb{x}}{\acute{\vb{x}}_{\alpha}}$, where $\grad_{\vb{x}} (\bullet)$ denotes the spatial gradient operator with respect to the spatial position vector $\vb{x}$. The spatial divergence operator is given as $\grad_{\vb{x}} \vdot(\bullet)$.
\par
To describe the non-equilibrium mechanical response of the solid skeleton of blood clots, the solid deformation gradient $\vb{F}_{S}$ is multiplicatively decomposed into elastic and viscous parts, \ie, $\vb{F}_{S}=\vb{F}_{S}^{e,i} \vb{F}_{S}^{v,i}$, with their determinants $J_{S}^{e,i} \coloneqq \det \vb{F}_{S}^{e,i} > 0$ and $J_{S}^{v,i} \coloneqq \det \vb{F}_{S}^{v,i} > 0$, where $i=m$ or $fib$ represent the main blood cells or fibrin, respectively. The decomposition $\vb{F}_{S}=\vb{F}_{S}^{e,i} \vb{F}_{S}^{v,i}$ leads to the introduction of an intermediate configuration $\widehat{\mathcal{B}}^{S,i}$. The viscous part $\vb{F}_{S}^{v,i}$ links the deformation from the initial configuration $\mathcal{B}_{0}^{S,i}$ to the intermediate configuration $\widehat{\mathcal{B}}^{S,i}$, while the elastic part $\vb{F}_{S}^{e,i}$ relates the deformation from the intermediate configuration $\widehat{\mathcal{B}}^{S,i}$ to the current configuration $\mathcal{B}^{S,i}$. In this work, $(\widehat{\bullet})$ represents a quantity $(\bullet)$ defined in the intermediate equilibrium configuration. The elastic right Cauchy--Green deformation tensor $\widehat{\vb{C}}_{S}^{e,i}$ and the elastic left Cauchy--Green deformation tensor $\vb{B}_{S}^{e,i}$ are defined as $\widehat{\vb{C}}_{S}^{e,i} \coloneqq \pqty{\vb{F}_{S}^{e,i}}^\intercal \vb{F}_{S}^{e,i}$ and $\vb{B}_{S}^{e,i} \coloneqq \vb{F}_{S}^{e,i} \pqty{\vb{F}_{S}^{e,i}}^\intercal$, respectively. The first invariant of $\widehat{\vb{C}}_{S}^{e,i}$ and $\vb{B}_{S}^{e,i}$ is denoted by $I_{S1}^{e,i} \coloneqq \tr(\widehat{\vb{C}}_{S}^{e,i})=\tr(\vb{B}_{S}^{e,i})$.
\par
The total strain tensor $\widehat{\vb{\Gamma}}_{S}^{i}$ of the solid constituent $i$ in the intermediate configuration is additively decomposed into an elastic part $\widehat{\vb{\Gamma}}_{S}^{e,i} \coloneqq \pqty{\widehat{\vb{C}}_{S}^{e,i}-\vb{I}}/2$ and an inelastic part $\widehat{\vb{\Gamma}}_{S}^{v,i} \coloneqq \bqty{\vb{I}-\pqty{\widehat{\vb{B}}_{S}^{v,i}}^{-1}}/2$, \ie, $\widehat{\vb{\Gamma}}_{S}^{i} =\widehat{\vb{\Gamma}}_{S}^{e,i}+\widehat{\vb{\Gamma}}_{S}^{v,i}$, where $\widehat{\vb{B}}_{S}^{v,i} \coloneqq \vb{F}_{S}^{v,i} \pqty{\vb{F}_{S}^{v,i}}^\intercal$ is the inelastic left Cauchy--Green deformation tensor, and $\vb{I}$ is the second-order identity tensor.
\subsection{Principle of virtual power and balance equations}
\label{sec:virtual_power}
The power of internal forces in a domain $\Omega$ in the current configuration is given as
\begin{equation}
	\label{equ:power_int}
	P_{\mathrm{int}} = \int_{\Omega} \pqty{\vb{T}^{S} \vdot \vb{L}_{S} + \vb{T}^{F} \vdot \vb{L}_{F}} \dd{v} + \int_{\Omega} \pqty{ -\tilde{\vb{p}}^{F} \vdot \acute{\vb{x}}_{F} -\tilde{\vb{p}}^{S} \vdot \acute{\vb{x}}_{S}} \dd{v} + \int_{\Omega} \bqty{b \pqty{\overline{Y}_{eqv}^{S}}'_{S} + \vb{y} \vdot \grad_{\vb{x}} \pqty{\overline{Y}_{eqv}^{S}}'_{S} } \dd{v} + \int_{\Omega} f_{fm} \pqty{\lambda_{fm}}'_{S} \dd{v} \,,
\end{equation}
where $\vb{T}^\alpha$ denotes the Cauchy stress, the vector $\tilde{\vb{p}}^{\alpha}$ is introduced to describe the volume-specific local interaction force between the involved constituents with the constraint $\tilde{\vb{p}}^{F} + \tilde{\vb{p}}^{S} = \vb{0}$, $b$ is the internal microforce, $\vb{y}$ is the microtraction flux vector, $\overline{Y}_{eqv}^{S}$ is an internal variable representing the non-local equivalent thermodynamic force, $f_{fm}$ is the microforce associated with the fibrin monomer stretch $\lambda_{fm}$. The power of the external forces is introduced as
\begin{equation}
	\label{equ:power_ext}
	P_{\mathrm{ext}} = \int_{\Omega} \pqty{ \rho^{S} \acute{\vb{x}}_{S} + \rho^{F} \acute{\vb{x}}_{F}} \vdot \vb{b} \dd{v} + \int_{\partial \Omega} \pqty{\vb{t}^{S} \vdot \acute{\vb{x}}_{S} + \vb{t}^{F} \vdot \acute{\vb{x}}_{F}} \dd{a} + \int_{\Omega} a \pqty{\overline{Y}_{eqv}^{S}}'_{S} \dd{v} + \int_{\partial \Omega} k \pqty{\overline{Y}_{eqv}^{S}}'_{S} \dd{a} \,,
\end{equation}
where $\vb{b}$ denotes the mass-specific body force, $\vb{t}^{\alpha}$ is the surface traction, $a$ and $k$ represent the external volumetric and surface microforces, respectively. According to the principle of virtual power, a balance equation regarding the power of internal and external forces in \cref{equ:power_int,equ:power_ext} is given as,
\begin{equation}
	\label{equ:balance_equation_powers}
	\delta P_{\mathrm{ext}} - \delta P_{\mathrm{int}} = 0.
\end{equation}
Integrating \cref{equ:balance_equation_powers} by parts and applying the divergence theorem, one derives the following balance equations with the corresponding boundary conditions:
\begin{itemize}
	\item the balance equations of the linear momentum for the solid skeleton and fluid:
	\begin{equation}
		\label{equ:balance_momentum}
		\grad_{\vb{x}} \vdot \vb{T}^{\alpha} + \rho^{\alpha} \vb{b} + \tilde{\vb{p}}^{\alpha} = \vb{0} \quad \mathrm{in} \, \Omega \,, \quad \vb{T}^{\alpha} \vb{n} = \vb{t}^{\alpha} \quad \mathrm{on} \, \partial \Omega \,;
	\end{equation}
	\item the balance equation of the internal microforce $b$:
	\begin{equation}
		\label{equ:balance_b}
		\grad_{\vb{x}} \vdot \vb{y} - b + a = 0 \quad \mathrm{in} \, \Omega \,, \quad \vb{y} \vdot \vb{n} = k \quad \mathrm{on} \, \partial \Omega \,;
	\end{equation}
	\item the balance equation of the internal microforce $f_{fm}$:
	\begin{equation}
		\label{equ:balance_f_fm}
		f_{fm}=0 \,.
	\end{equation}
\end{itemize}
\subsection{Micromechanical modeling of blood clots}
\label{sec:Balance_equs_entropy_inequa}
In the Theory of Porous Media, the constitutive models of the constituents $\varphi^{\alpha}$ are constrained by their balance equation of mass and momentum. The local momentum balance equations are given in \cref{equ:balance_momentum}. The local mass balance equations of the constituents are written as \cite{liu2022modelling}
\begin{equation}
	\label{equ:balance_equations}
	\pqty{\rho^{\alpha}}'_{\alpha} + \rho^{\alpha} \grad_{\vb{x}} \vdot \acute{\vb{x}}_{\alpha}=0 \,,
\end{equation}
where the material time derivative of a scalar function $(\bullet)'_{\alpha}$ following the motion of $\varphi^{\alpha}$ is defined as $(\bullet)'_\alpha = \pdv*{(\bullet)}{t} + \grad_{\vb{x}} (\bullet) \vdot \acute{\vb{x}}_\alpha$. Using the relation $\rho^{\alpha} = n^{\alpha} \rho^{\alpha R}$, the mass balance equation \cref{equ:balance_equations} can be reformulated as the volume balance equation, \ie,
\begin{equation}
	\label{equ:volume_balance}
	(n^\alpha)'_\alpha + n^\alpha \grad_{\vb{x}} \vdot \acute{\vb{x}}_\alpha=0 \,,
\end{equation}
where the assumption of material incompressibility is used, \ie, $\rho^{\alpha R}$= const.
\par
Different types of blood clots with distinct compositions can form under specific mechanobiological environment. The solid skeleton in a whole blood clot mainly contains fibrin network and blood cells. The mechanical stability of a blood clot is sensitive to its composition \citep{garyfallogiannis2023fracture}. To derive the constitutive relation of a blood clot, we postulate the existence of the Helmholtz free energy $\psi^{\alpha}$ per unit mass of the constituent $\varphi^{\alpha}$. For a generalized whole blood clot consisting of fibrin networks ($i=fib$) and main blood cells ($i=m$), the Helmholtz free energy densities of the solid skeleton $\psi^{S}$ and the fluid $\psi^{F}$ are given as
\begin{equation}
	\label{equ:free_energy_solid_fluid}
	\psi^{S} = \hat{\psi}^{S}\pqty{\vb{C}_{S},\widehat{\vb{C}}_{S}^{e,i}, d, Y_{eqv}^{S}, \, \overline{Y}_{eqv}^{S}, \, \grad_{\vb{x}}{\overline{Y}_{eqv}^{S}}, \lambda_{fm}} \qq{and} \psi^{F}=\hat{ \psi}^{F}(-) \,,
\end{equation}
respectively. In \cref{equ:free_energy_solid_fluid}, the solid free energy density $\psi^{S}$, under the isothermal condition, is assumed to be a function of the right Cauchy--Green deformation tensor $\vb{C}_{S}$, the elastic right Cauchy--Green deformation tensor $\widehat{\vb{C}}_{S}^{e,i}$, the damage variable $d$, the local equivalent thermodynamic force $Y_{eqv}^{S}$, the non-local equivalent thermodynamic force $\overline{Y}_{eqv}^{S}$ and its gradient $\grad_{\vb{x}}{\overline{Y}_{eqv}^{S}}$, and the fibrin monomer stretch $\lambda_{fm}$ of fibrin associated with protein unfolding. By assuming the material incompressibility and negligible viscosity of water in blood clots, the fluid free energy density $\psi^{F}$ in \cref{equ:free_energy_solid_fluid} is independent of any process variable \cite{ehlers2009extended,ehlers2017phase,Liu2019}.
\par
The total Helmholtz free energy density $\psi^{S}$ of the solid skeleton in blood clots is decomposed into four parts to clarify their contributions, \ie,
\begin{equation}
	\label{equ:free_energy_solid_total}
	\psi^{S} = \psi_{eq}^{S} + \psi_{neq}^{S} + \psi_{fm}^{S} +\psi_{nl}^{S} \,,
\end{equation}
with
\begin{equation}
	\label{equ:free_energy_solid_parts}
	\begin{aligned}
		\psi_{eq}^{S} &= \hat{\psi}_{eq}^{S}\pqty{d,\vb{C}_{S},\lambda_{fm}} = \sum_{i}^{m,fib} \hat{\psi}_{eq}^{i} \pqty{d,\vb{C}_{S},\lambda_{fm}} = \sum_{i}^{m,fib} n^{i} \hat{f}^{i}(d) \hat{\psi}_{eq0}^{i} \pqty{\vb{C}_{S},\lambda_{fm}}\,, \\
		\psi_{neq}^{S} &= \hat{\psi}_{neq}^{S}\pqty{d,\widehat{\vb{C}}_{S}^{e,i},\lambda_{fm}} = \sum_{i}^{m,fib} \hat{\psi}_{neq}^{i} \pqty{d,\widehat{\vb{C}}_{S}^{e,i},\lambda_{fm}} = \sum_{i}^{m,fib} n^{i} \hat{f}^{i}(d) \hat{\psi}_{neq0}^{i} \pqty{\widehat{\vb{C}}_{S}^{e,i},\lambda_{fm}} \,,\\
		\psi_{fm}^{S} &= \hat{\psi}_{fm}^{S} \pqty{d, \lambda_{fm}} = n^{fib} \hat{f}^{fib}(d) \hat{\psi}_{fm0}^{S}\pqty{\lambda_{fm}} \qq{and} \\
		\psi_{nl}^{S} &= \hat{\psi}_{nl}^{S}\pqty{Y_{eqv}^{S}, \, \overline{Y}_{eqv}^{S}, \, \grad_{\vb{x}}{\overline{Y}_{eqv}^{S}}}\,.
	\end{aligned}
\end{equation}
The solid free energy density $\psi^{S}$ is assumed to be an isotropic function. The equilibrium part $\psi_{eq}^{S}$ in \cref{equ:free_energy_solid_parts}$_1$ includes the free energy in fibrin networks $\psi_{eq}^{fib} = \hat{\psi}_{eq}^{fib} \pqty{d,\vb{C}_{S}}$ and the free energy in main blood cells $\psi_{eq}^{m} = \hat{\psi}_{eq}^{m} \pqty{d,\vb{C}_{S}}$. The non-equilibrium contribution $\psi_{neq}^{S}$ in \cref{equ:free_energy_solid_parts}$_2$ describes the viscoelasticity of the solid skeleton. A single fibrin fiber shows viscoelasticity according to previous micromechanical tests \citep{litvinov2017fibrin}. The dissociation of crosslinks, sliding between protofibrils and configurational rearrangement of fibrin chains in fibrin networks also lead to viscoelasticity of blood clots \citep{ryan1999structural}. In addition, red blood cells demonstrate viscoelasticity \citep{puig2007viscoelasticity}. Therefore, $\psi_{neq}^{S}$ is expressed as the sum of the free energy in fibrin networks $\psi_{neq}^{fib} = \hat{\psi}_{neq}^{fib} \pqty{d,\widehat{\vb{C}}_{S}^{e,fib}}$ and the free energy in main blood cells $\psi_{neq}^{m} = \hat{\psi}_{neq}^{m} \pqty{d,\widehat{\vb{C}}_{S}^{e,m}}$. Moreover, $n^{i} \coloneqq \dd{v}^{i}/\dd{v}^{S}$ denotes the volume fraction of different solid components, with the restriction $n^{fib}+n^{m}=1$. It is assumed that both the equilibrium part and the non-equilibrium part are subject to damage. In \cref{equ:free_energy_solid_parts}$_1$ and \cref{equ:free_energy_solid_parts}$_2$, $f^{i} = \hat{f}^{i}(d)$ represents an appropriate damage function coupled to the damage-free Helmholtz free energy functions $\psi_{eq0}^{i} = \hat{\psi}_{eq0}^{i} \pqty{\widehat{\vb{C}}_{S}^{e,i}}$ and $\psi_{neq0}^{i} = \hat{\psi}_{neq0}^{i} \pqty{\widehat{\vb{C}}_{S}^{e,i}}$ by $\psi_{eq}^{i} = f^{i} \psi_{eq0}^{i}$ and $\psi_{neq}^{i} = f^{i} \psi_{neq0}^{i}$, respectively. When blood clots are stretched and individual protofibril fibers approach their stretch limit, the fibrin monomers that make up the fibers are forced to extended, \ie, protein unfolding \cite{brown2009multiscale}. To describe the protein unfolding effect, an energy density function $\psi_{fm}^{S}$ is introduced in \cref{equ:free_energy_solid_parts}$_3$, which is a function of the monomer stretch $\lambda_{fm}$ and the damage variable $d$. The damage-free counterpart is denoted by $\psi_{fm0}^{S} = \hat{\psi}_{fm0}^{S}\pqty{\lambda_{fm}}$. Finally, a gradient-enhanced non-local free energy function $\psi_{nl}^{S}$ is introduced in \cref{equ:free_energy_solid_parts}$_4$ to describe the non-local damage of the solid skeleton in blood clots.
\par
To ensure the thermodynamic consistence, the constitutive model should satisfy the entropic inequality. Proceeding from the isothermal condition, the Clausius--Planck inequality reads
\begin{equation}
	\label{equ:entropy_inequ_generalized}
	\dv{_{S}}{t}\int_{\Omega} \rho^{S} \psi^{S} \dd{v} + \dv{_{F}}{t}\int_{\Omega} \rho^{F} \psi^{F} \dd{v} \leqslant P_{\mathrm{ext}} = P_{\mathrm{int}} \,.
\end{equation}
Recalling the power of internal forces \cref{equ:power_int}, inserting the expression of $P_{\mathrm{int}}$ into \cref{equ:entropy_inequ_generalized} and taking into account the saturation condition \cref{equ:saturation_condition}, one obtains the entropy inequality,
\begin{equation}
	\label{equ:entropy_inequ_with_Saturatiion_1}
	\mathfrak{D}_{int}=\vb{T}^{S} \vdot \vb{D}_{S} + \vb{T}^{F} \vdot \vb{D}_{F} +f_{fm} \pqty{\lambda_{fm}}'_{S} + b \pqty{\overline{Y}_{eqv}^{S}}'_{S} + \vb{y} \vdot \grad_{\vb{x}} \pqty{\overline{Y}_{eqv}^{S}}'_{S} - \rho^{S}\pqty{\psi^{S}}'_{S}-\rho^{F}\pqty{\psi^{F}}'_{F} - \tilde{\vb{p}}^{F} \vdot \vb{w}_{FR} - p\pqty{n^{S} + n^{F}}'_{S} \geqslant 0 \,,
\end{equation}  
where $\vb{D}_{\alpha} \coloneqq \pqty{\vb{L}_{\alpha} + \vb{L}_{\alpha}^\intercal}/2$ is the symmetric part of the spatial velocity gradient tensor $\vb{L}_{\alpha} \coloneqq \grad_{\vb{x}}{\acute{\vb{x}}_{\alpha}}$, $f_{fm}$ and $b$ are the internal microforce, $\vb{y}$ is the flux vector, and $n^{\alpha}$ is the volume fraction of $\varphi^{\alpha}$ which is defined in \Cref{sec:Mixture_volume_fraction}. The last term in \cref{equ:entropy_inequ_with_Saturatiion_1} ensures that the saturation constraint in \cref{equ:saturation_condition} is fulfilled based on the assumptions of biphasic saturation and materially incompressibility, where $p$ acts as a Lagrangean multiplier that can be interpreted as the pore pressure governing the incompressibility constraint.
\par
The material time derivatives of $\psi^{S}$, $\psi^{F}$ and $\pqty{n^{S} + n^{F}}$ in \cref{equ:entropy_inequ_with_Saturatiion_1} are calculated and presented in \ref{app:time_derivative}. Inserting \cref{equ:mat_time_diritive_Saturation_cond,equ:time_derivative_energy} into \cref{equ:entropy_inequ_with_Saturatiion_1}, and making use of \cref{equ:time_derivative_energy_eq_neq,equ:time_derivative_energy_neq_4}, the Clausius--Planck inequality is rewritten as
\begin{equation}
	\label{equ:dissipation}
	\begin{aligned}
		\mathfrak{D}_{int} &= \bqty{\vb{T}^{S} + n^{S} p \vb{I} - \sum_{i}^{m,fib} 2 \rho^{S} \vb{F}_{S} \pdv{\psi_{eq}^{i}}{\vb{C}_{S}} \vb{F}_{S}^\intercal - \sum_{i}^{m,fib} 2 \rho^{S} \vb{F}_{S}^{e,i} \pdv{\psi_{neq}^{i}}{\widehat{\vb{C}}_{S}^{e,i}} \pqty{\vb{F}_{S}^{e,i}}^\intercal - 2 \rho^{S} \vb{F}_{S} \pdv{\psi_{nl}^{S}}{Y_{eqv}^{S}} \pdv{Y_{eqv}^{S}}{\vb{C}_{S}} \vb{F}_{S}^\intercal} \vdot \vb{D}_S + \pqty{\vb{T}^{F} + n^{F} p\vb{I}} \vdot \vb{D}_{F} \\
		&\quad + \sum_{i}^{m,fib} 2 \rho^{S} \pqty{\widehat{\vb{C}}_{S}^{e,i} \pdv{\psi_{neq}^{i}}{\widehat{\vb{C}}_{S}^{e,i}}} \vdot \widehat{\vb{L}}_{S}^{v,i} - \sum_{i}^{m,fib} \rho^{S} \pdv{\psi_{eq}^{i}}{d} \pqty{d}'_{S} - \sum_{i}^{m,fib} \rho^{S} \pdv{\psi_{neq}^{i}}{d} \pqty{d}'_{S} - \rho^{S} \pdv{\psi_{fm}^{S}}{d} \pqty{d}'_{S} \\
		&\quad + \pqty{b-\rho^{S} \pdv{\psi_{nl}^{S}}{\overline{Y}_{eqv}^{S}}} \pqty{\overline{Y}_{eqv}^{S}}'_{S} + \pqty{\vb{y}- \rho^{S} \pdv{\psi_{nl}^{S}}{\grad_{\vb{x}}{\overline{Y}_{eqv}^{S}}}} \vdot \grad_{\vb{x}} \pqty{\overline{Y}_{eqv}^{S}}'_{S} - \pqty{\tilde{\vb{p}}^{F} -p \grad_{\vb{x}}{n^{F}}} \vdot \vb{w}_{FR} \\ &\quad + \pqty{f_{fm} - \sum_{i}^{m,fib} \rho^{S} \pdv{\psi_{eq}^{i}}{\lambda_{fm}} - \sum_{i}^{m,fib} \rho^{S} \pdv{\psi_{neq}^{i}}{\lambda_{fm}} - \rho^{S} \pdv{\psi_{fm}^{S}}{\lambda_{fm}}} \pqty{\lambda_{fm}}'_{S} \geqslant 0 \,.
	\end{aligned}
\end{equation}
Exploiting the standard Coleman--Noll procedure \citep{collman1963thermodynamics} yields the solid Cauchy stress $\vb{T}^{S}$ and the fluid Cauchy stress $\vb{T}^{F}$, \ie,
\begin{equation}
	\label{equ:cauchy_Stress_derive}
	\vb{T}^{S} = -n^{S} p \vb{I} + \vb{T}_{E}^{S} \qq{and} \vb{T}^{F} = -n^{F} p \vb{I} \,,
\end{equation}
respectively, where $p$ is identified as the pore pressure \citep{markert2008biphasic} and $\vb{T}_{E}^{S}$ is introduced as the effective solid stress. According to \cref{equ:dissipation,equ:cauchy_Stress_derive}, the effective solid stress tensor $\vb{T}_{E}^{S}$ is composed of the equilibrium stresses and non-equilibrium stresses of fibrin networks and blood cells, and the stress related to the non-local free energy density function, which is expressed as
\begin{equation}
	\label{equ:effective_solid_stress}
	\vb{T}_{E}^{S}= \sum_{i}^{m,fib} \vb{T}_{E,eq}^{i} + \sum_{i}^{m,fib} \vb{T}_{E,neq}^{i} + \vb{T}_{E,nl}^{S} \,,
\end{equation}
with
\begin{equation}
	\label{equ:effective_solid_stress_1}
	\vb{T}_{E,eq}^{i} \coloneqq 2 \rho^{S} \vb{F}_{S} \pdv{\psi_{eq}^{i}}{\vb{C}_{S}} \vb{F}_{S}^\intercal \,, \quad \vb{T}_{E,neq}^{i} \coloneqq 2 \rho^{S} \vb{F}_{S}^{e,i} \pdv{\psi_{neq}^{i}}{\widehat{\vb{C}}_{S}^{e,i}} \pqty{\vb{F}_{S}^{e,i}}^\intercal \qq{and} \vb{T}_{E,nl}^{S} \coloneqq 2 \rho^{S} \vb{F}_{S} \pdv{\psi_{nl}^{S}}{Y_{eqv}^{S}} \pdv{Y_{eqv}^{S}}{\vb{C}_{S}} \vb{F}_{S}^\intercal \,.
\end{equation}
Likewise, by means of the Coleman--Noll procedure, the flux vector $\vb{y}$ of internal work of $\overline{Y}_{eqv}^{S}$ is expressed as
\begin{equation}
	\label{equ:nonlocal_stress}
	\vb{y} = \rho^{S} \pdv{\psi_{nl}^{S}}{\grad_{\vb{x}}{\overline{Y}_{eqv}^{S}}} \,.
\end{equation}
\par
Substituting \cref{equ:cauchy_Stress_derive,equ:nonlocal_stress} back into \cref{equ:dissipation} and making use of \cref{equ:effective_solid_stress,equ:effective_solid_stress_1}, the dissipation inequality can be reduced and expressed for arbitrary admissible processes as,
\begin{equation}
	\label{equ:entropy_inequ_last_term}
	\begin{aligned}
		\sum_{i}^{m,fib} 2 \rho^{S} \pqty{\widehat{\vb{C}}_{S}^{e,i} \pdv{\psi_{neq}^{i}}{\widehat{\vb{C}}_{S}^{e,i}}} \vdot \widehat{\vb{L}}_{S}^{v,i} &\geqslant 0 \,, \\
		- \sum_{i}^{m,fib} \rho^{S} \pdv{\psi_{eq}^{i}}{d} \pqty{d}'_{S} - \sum_{i}^{m,fib} \rho^{S} \pdv{\psi_{neq}^{i}}{d} \pqty{d}'_{S} - \rho^{S} \pdv{\psi_{fm}^{S}}{d} \pqty{d}'_{S} &\geqslant 0 \,, \\
		\pqty{b - \rho^{S} \pdv{\psi_{nl}^{S}}{\overline{Y}_{eqv}^{S}}} \pqty{\overline{Y}_{eqv}^{S}}'_{S} &\geqslant 0 \,, \\
		-\pqty{\tilde{\vb{p}}^{F} - p \grad_{\vb{x}}{n^{F}}} \vdot \vb{w}_{FR} &\geqslant 0 \,, \\
		\pqty{f_{fm} - \sum_{i}^{m,fib} \rho^{S} \pdv{\psi_{eq}^{i}}{\lambda_{fm}} - \sum_{i}^{m,fib} \rho^{S} \pdv{\psi_{neq}^{i}}{\lambda_{fm}} - \rho^{S} \pdv{\psi_{fm}^{S}}{\lambda_{fm}}} \pqty{\lambda_{fm}}'_{S} &\geqslant 0 \,.
	\end{aligned}
\end{equation}
Since the term $\widehat{\vb{C}}_{S}^{e,i} \pdv*{\psi_{neq}^{i}}{\widehat{\vb{C}}_{S}^{e,i}}$ is symmetric, \cref{equ:entropy_inequ_last_term}$_1$ can be rewritten as
\begin{equation}
	\label{equ:entropy_inequ_viscosity}
	\sum_{i}^{m,fib} 2 \rho^{S} \pqty{\widehat{\vb{C}}_{S}^{e,i} \pdv{\psi_{neq}^{i}}{\widehat{\vb{C}}_{S}^{e,i}}} \vdot \widehat{\vb{L}}_{S}^{v,i} = \sum_{i}^{m,fib} 2 \rho^{S} \pqty{\widehat{\vb{C}}_{S}^{e,i} \pdv{\psi_{neq}^{i}}{\widehat{\vb{C}}_{S}^{e,i}}} \vdot \widehat{\vb{D}}_{S}^{v,i} \geqslant 0 \,,
\end{equation}
where the inelastic strain rate tensor $\widehat{\vb{D}}_{S}^{v,i}$ is defined in \cref{equ:Oldroyd_rate_of_Gamma_5}. The residual inequalities in \Cref{equ:entropy_inequ_last_term} are used to derive the evolution laws of the involved internal variables to ensure their thermodynamic consistence, see \Cref{sec:evolution_equ}.
\subsection{Evolution equations of internal variables}
\label{sec:evolution_equ}
\subsubsection{Visco-hyperelastic deformations of solid constituent}
\label{sec:evolution_equ_visco}
Viscoelasticity is a time-dependent local event in the solid constituent. Both fibrin fibers and blood cells show viscoelastic deformation according to experimental data \citep{litvinov2017fibrin,puig2007viscoelasticity}. To describe the viscoelasticity, the solid deformation gradient tensor $\vb{F}_{S}$ is decomposed into two parts, \ie, $\vb{F}_{S}=\vb{F}_{S}^{e,i} \vb{F}_{S}^{v,i}$. Proceeding from the multiplicative decomposition, one obtains the dissipation inequality \cref{equ:entropy_inequ_viscosity}, which can be used to derive the inelastic solid deformation. To simplify the dissipation inequality \cref{equ:entropy_inequ_viscosity}, an overstress tensor $\hat{\bm{\tau}}_{E,neq}^{i}$, which is defined in the intermediate configuration, is introduced as
\begin{equation}
	\label{equ:overstress}
	\hat{\bm{\tau}}_{E,neq}^{i} \coloneqq 2 \rho_{0S}^{S} \pdv{\psi_{neq}^{i}}{\widehat{\vb{C}}_{S}^{e,i}} \,,
\end{equation}
where $\rho_{0S}^{S} \coloneqq n_{0S}^{S} \rho^{SR}$ is the partial density of the solid skeleton in the solid reference configuration, with $n_{0S}^{S}$ the initial solid volume concentration in the solid reference configuration. As a consequence, the inequality \cref{equ:entropy_inequ_viscosity} can be written as
\begin{equation}
	\label{equ:entropy_inequ_last_term_2}
	\frac{1}{J_{S}} \sum_{i}^{m,fib} \pqty{\widehat{\vb{C}}_{S}^{e,i} \hat{\bm{\tau}}_{E,neq}^{i}} \vdot \widehat{\vb{D}}_{S}^{v,i} \geqslant 0 \,,
\end{equation}
where the relation $\rho^{S}=\rho_{0S}^{S}/J_{S}$ is used, with $J_{S}>0$, and $\widehat{\vb{C}}_{S}^{e,i} \hat{\bm{\tau}}_{E,neq}^{i}$ is treated as the Mandel stress tensor. A simple sufficient condition is utilized for fulfilling the inequality \cref{equ:entropy_inequ_last_term_2}, which is given as
\begin{equation}
	\label{equ:entropy_inequ_last_term_a}
	\widehat{\vb{D}}_{S}^{v,i} = \frac{1}{\phi^{i}} \widehat{\vb{C}}_{S}^{e,i} \hat{\bm{\tau}}_{E,neq}^{i} \,,
\end{equation}
where $\phi^{i}>0$ represents the viscosity of the solid constituent $\varphi^{i}$. It is noted that the inelastic strain rate tensor $\widehat{\vb{D}}_{S}^{v,i}$ is defined in the intermediate configuration. By applying the push-forward operation to \cref{equ:entropy_inequ_last_term_a} with $\vb{F}_{S}^{e,i}$, one obtains its expression in the current configuration, \ie,
\begin{equation}
	\label{equ:entropy_inequ_last_term_a_Eulerian}
	\vb{F}_{S}^{e,i} \widehat{\vb{D}}_{S}^{v,i} \pqty{\vb{F}_{S}^{e,i}}^\intercal = \frac{1}{\phi^{i}} \vb{F}_{S}^{e,i} \widehat{\vb{C}}_{S}^{e,i} \hat{\bm{\tau}}_{E,neq}^{i} \pqty{\vb{F}_{S}^{e,i}}^\intercal = \frac{1}{\phi^{i}} \vb{B}_{S}^{e,i} \vb*{\tau}_{E,neq}^{i} \qq{with} \vb*{\tau}_{E,neq}^{i} \coloneqq \vb{F}_{S}^{e,i} \hat{\bm{\tau}}_{E,neq}^{i} \pqty{\vb{F}_{S}^{e,i}}^\intercal \,,
\end{equation}
where $\vb*{\tau}_{E,neq}^{i}$ denotes the non-equilibrium Kirchhoff stress of the solid constituent $\varphi^{i}$. Substituting $\widehat{\vb{D}}_{S}^{v,i}$ in \cref{equ:entropy_inequ_last_term_a_Eulerian} by the relation $\widehat{\vb{D}}_{S}^{v,i} = - \vb{F}_{S}^{v,i} \bqty{\pqty{\vb{C}_{S}^{v,i}}^{-1}}'_{S} \pqty{\vb{F}_{S}^{v,i}}^\intercal /2$, and after some mathematical manipulations, \cref{equ:entropy_inequ_last_term_a_Eulerian} is rewritten as
\begin{equation}
	\label{equ:entropy_inequ_last_term_a_Eulerian_1}
	\vb{F}_{S} \bqty{\pqty{\vb{C}_{S}^{v,i}}^{-1}}'_{S} \vb{F}_{S}^\intercal = -\frac{2}{\phi^{i}} \vb{B}_{S}^{e,i} \vb*{\tau}_{E,neq}^{i} \,,
\end{equation}
with $ \vb{C}_{S}^{v,i} \coloneqq \pqty{\vb{F}_{S}^{v,i}}^\intercal \vb{F}_{S}^{v,i}$. It is noted that the left-hand side of \cref{equ:entropy_inequ_last_term_a_Eulerian_1} is the Oldroyd rate of $\vb{B}_{S}^{e,i}$ \citep{reese1998theory}, which is defined as
\begin{equation}
	\label{equ:oldroyd_Be}
	\pqty{\vb{B}_{S}^{e,i}}_{S}^{\triangledown} \coloneqq \pqty{\vb{B}_{S}^{e,i}}'_{S} - \vb{B}_{S}^{e,i} \vb{L}_{S}^\intercal - \vb{L}_{S} \vb{B}_{S}^{e,i} \,,
\end{equation}
and thereby obtaining the following relation,
\begin{equation}
	\label{equ:oldroyd_Be_1}
	\vb{F}_{S} \bqty{\pqty{\vb{C}_{S}^{v,i}}^{-1}}'_{S} \vb{F}_{S}^\intercal = \pqty{\vb{B}_{S}^{e,i}}_{S}^{\triangledown} = \pqty{\vb{B}_{S}^{e,i}}'_{S} - \vb{B}_{S}^{e,i} \vb{L}_{S}^\intercal - \vb{L}_{S} \vb{B}_{S}^{e,i} \,.
\end{equation}
Combining \cref{equ:entropy_inequ_last_term_a_Eulerian_1,equ:oldroyd_Be_1}, the evolution equation for the elastic part of the left Cauchy--Green deformation tensor can be expressed as
\begin{equation}
	\label{equ:evolution_equ_Be}
	\pqty{\vb{B}_{S}^{e,i}}'_{S} = \vb{B}_{S}^{e,i} \vb{L}_{S}^\intercal + \vb{L}_{S} \vb{B}_{S}^{e,i} - \frac{2}{\phi^{i}} \vb{B}_{S}^{e,i} \vb*{\tau}_{E,neq}^{i} \,.
\end{equation}
\subsubsection{Damage initiation and evolution laws}
\label{sec:damage_derive}
\par
For a whole blood clot, the fracture of the bulk material includes the rupture of fibrin networks and failure of blood cells. On the one hand, subject to physiological loading, fibrin fibers are stretched until the fibrin monomer reaches a critical stretch value, after which fibrin fibers break and fibrin networks rupture \citep{tutwiler2020rupture,brown2009multiscale}. On the other hand, physiological loading can also lead to damage of blood cells due to the interaction of blood cells with fibrin networks, cell-cell interaction, high shear rates on blood clot surfaces, etc \citep{leverett1972red,weisel2019red}. In order to describe the continuous mechanical damage of fibrin networks and blood cells of whole blood clots, an energy-driven damage evolution equation is proposed. The local thermodynamic force $Y^{S}$, which is conjugated to the damage variable $d$, can be derived from \cref{equ:entropy_inequ_last_term}$_2$ as
\begin{equation}
	\label{equ:dam_driv_Force}
	Y^{S} \coloneqq - \sum_{i}^{m,fib} \rho^{S} \pdv{\psi_{eq}^{i}}{d} - \sum_{i}^{m,fib} \rho^{S} \pdv{\psi_{neq}^{i}}{d} - \rho^{S} \pdv{\psi_{fm}^{S}}{d}
	= - \sum_{i}^{m,fib} \rho^{S} n^{i} \psi_{eq0}^{i} \dv{f^{i}}{d} - \sum_{i}^{m,fib} \rho^{S} n^{i} \psi_{neq0}^{i} \dv{f^{i}}{d} - \rho^{S} n^{fib} \psi_{fm0}^{S} \pdv{f^{fib}}{d} \,.
\end{equation}
The scalar damage function $f^{i}$ is coupled to the damage-free Helmholtz free energy density $\psi_{eq0}^{i}$, $\psi_{neq0}^{i}$ and $\psi_{fm0}^{S}$, see \cref{equ:free_energy_solid_parts}, for describing the irreversible damage of the solid skeleton in blood clots. $f^{i}$ ranges between 1 and 0, where in undamaged material $f^{i}=1$, and material is fully damaged upon $f^{i}=0$. The damage function is defined as $f^{i} \coloneqq 1 - H_{TC}^{i} \tilde{d}^{\alpha^{i}}$, with $\tilde{d} = d/d_{cri}$, where $d_{cri}$ denotes a critical damage value. The step function $H_{TC}^{i}$ regulates damage-induced tension-compression asymmetry due to the different damage behaviors under tension and compression. The specific form of $H_{TC}^{i}$ is given in \Cref{sec:specification_energy_function}. The derivative of the damage function with respect to the damage variable is calculated as $\dv*{f^{i}}{d} = - H_{TC}^{i} \alpha^{i} d^{\alpha^{i}-1}/d_{cri}^{\alpha^{i}}$. For clarity, $Y^{S}$ can be alternatively expressed as the sum of the contribution $Y^{fib}$ from fibrin networks and the contribution $Y^{m}$ from blood cells, \ie,
\begin{equation}
	\label{equ:damage_driving_contributions}
	Y^{S} = Y^{fib} + Y^{m}, \qq{with}	
	Y^{fib} \coloneqq - \rho^{S} n^{fib} \dv{f^{fib}}{d} \pqty{\psi_{eq0}^{fib} + \psi_{neq0}^{fib} + \psi_{fm0}^{S}} \qq{and} Y^{m} \coloneqq - \rho^{S} n^{m} \dv{f^{m}}{d} \pqty{\psi_{eq0}^{m} + \psi_{neq0}^{m}} \,.
\end{equation} 
\par
The fibrin fibers and blood cells demonstrate different resistance against damage due to their different mechanical properties and damage modes subject to physiological loading \cite{fereidoonnezhad2021blood,chernysh2020structure}. To consider the distinct damage resistance behavior of fibrin fibers and blood cells, a local equivalent damage driving force $Y_{eqv}^{S}$ is introduced as
\begin{equation}
	\label{equ:equivalent_driving_force}
	Y_{eqv}^{S} \coloneqq \frac{Y^{fib}}{\pqty{n^{fib}}^{\hbar} Y_0^{fib}} + \varpi \frac{Y^{m}}{\pqty{n^{m}}^{\hbar} Y_0^{m}}\,,
\end{equation}
where $Y_0^{fib}$ and $Y_0^{m}$ are the initial resistance against damage of fibrin fibers and blood cells, respectively. The model parameter $\hbar$ is introduced to modulate the impact of constituent volume fractions on the damage initiation. The contribution of these two solid constituents to the clot damage is distinct \cite{fereidoonnezhad2021blood}. A weighting parameter $\varpi$ is included into \cref{equ:equivalent_driving_force} to characterizes the ratio of contribution of blood cells to that of fibrin fibers. A damage potential function $F_{\mathrm{dam}}^{S} = \hat{F}_{\mathrm{dam}}^{S} \pqty{Y_{eqv}^{S},d}$ is defined to derive the damage evolution law and evaluate the initiation damage criterion, which is given as
\begin{equation}
	\label{equ:Fdam}
	F_{\mathrm{dam}}^{S} \coloneqq Y_{eqv}^{S}-R^{S} \,,
\end{equation}
where $R^{S} = \hat{R}^{S}(d)$ is a damage resistance function. Using the maximum dissipation principle, the damage evolution equation is expressed as
\begin{equation}
	\label{equ:dam_evo_equ}
	(d)'_{S} = (\varkappa)'_{S} \pdv{F_{\mathrm{dam}}^{S}}{Y_{eqv}^{S}}= (\varkappa)'_{S} \,,
\end{equation}
where $(\varkappa)'_{S}$ is a damage multiplier. It is noted that the damage rate $(d)'_{S}$ is equal to the damage multiplier $(\varkappa)'_{S}$ according to the damage potential $F_{\mathrm{dam}}^{S}$ introduced in \cref{equ:Fdam}. The damage multiplier $(\varkappa)'_{S}$ is non-negative under loading/unloading conditions according to the Karush--Kuhn--Tucker (KKT) conditions
\begin{equation}
	\label{equ:dam_multiplier}
	(\varkappa)'_{S} \geqslant 0, \quad F_{\mathrm{dam}}^{S} \leqslant 0, \quad (\varkappa)'_{S} F_{\mathrm{dam}}^{S}=0 \,.
\end{equation}
On the one hand, the KKT conditions result in an irreversible evolution of damage, \ie, $(d)'_{S} \geqslant 0$. On the other hand, the damage driving force is always non-negative. These restrictions ensure that the Clausius--Planck dissipation inequality is satisfied. In the case of damage growth, the damage multiplier $(\varkappa)'_{S}$ is determined by the damage consistency condition, $\pqty{F_{\mathrm{dam}}^{S}}'_{S}=0$.
\par
The damage model depends on the specification of the damage resistance function $R^{S}$. In this work, a logarithmic-type function is proposed as
\begin{equation}
	\label{equ:dam_resistance}
	R^{S} = 1 +\frac{1}{b^{S}} \ln \frac{d_{cri}}{d_{cri}-d} \,,
\end{equation}
where $b^{S}$ is a model parameter controlling the evolution rate of damage resistance and $d_{cri}$ denotes the critical value of damage. The damage potential function can be rewritten as
\begin{equation}
	\label{equ:dam_poten_equ}
	F_{\mathrm{dam}}^{S} = Y_{eqv}^{S}-R^{S} = Y_{eqv}^{S} - \pqty{1 +\frac{1}{b^{S}} \ln \frac{d_{cri}}{d_{cri}-d}} \,.
\end{equation}
\par
In the case of damage growth, the damage rate $(d)'_{S}$ is determined by the damage consistency condition
\begin{equation}
	\label{equ:dam_poten_equ_diff}
	\pqty{F_{\mathrm{dam}}^{S}}'_{S} = \pdv{F_{\mathrm{dam}}^{S}}{Y_{eqv}^{S}} \pqty{Y_{eqv}^{S}}'_{S} + \pdv{F_{\mathrm{dam}}^{S}}{R^{S}} \pdv{R^{S}}{d} \pqty{d}'_{S}=0 \,.
\end{equation}
Combining \cref{equ:dam_poten_equ,equ:dam_poten_equ_diff}, we obtain the damage evolution equation
\begin{equation}
	\label{equ:dam_evo_equ_fibre_Final}
	(d)'_{S}= d_{cri} b^{S} \exp[-b^{S} \ev{Y_{eqv}^{S}-1}] \pqty{Y_{eqv}^{S}}'_{S} \,,
\end{equation}
where $\ev{\bullet}$ denotes the Macaulay bracket, \ie, $\pqty{\vqty{\bullet}+\bullet}/2$. The damage evolution law is characterized by the initial damage resistance $Y_0^{m}$ and $Y_0^{fib}$, the damage development parameter $b^{S}$ and the critical value of damage $d_{cri}$. Subject to a continuous loading, when the equivalent damage driving force increases to the damage initial resistance, damage starts to accumulate from the intact state ($d = 0$) to a critical damage state ($d = d_{cri}$). By integrating the damage evolution equation \cref{equ:dam_evo_equ_fibre_Final}, one obtains the damage model, \ie,
\begin{equation}
	\label{equ:dam_model}
	d= d_{cri} \bqty{ 1 - \exp(-b^{S} \ev{Y_{eqv}^{S}-1})} \,.
\end{equation}
\par
\begin{remark}
	\label{remark:damage_physical_measing}
	Fibrin networks in blood clots are composed of fibrin fibers, which are covalently crosslinked with the transglutaminase, \ie, the blood clotting factor XIIIa \cite{tutwiler2020rupture,brown2009multiscale}. Subject to stretch, the force in fibrin fibers results in fiber breakage. The number of connected, unconnected and total fibrin fibers per unit volume are denoted by $N_{c}^{fib}$,$N_{u}^{fib}$ and $N_{t}^{fib}$, respectively, with the constraint $N_{t}^{fib} = N_{c}^{fib} + N_{u}^{fib}$. The physical meaning of the damage function in fibrin networks can be interpreted as the ratio of the number density of connected fibrin fibers $N_{c}^{fib}$ to the number density of total fibrin fibers $N_{t}^{fib}$, \ie, $f^{fib} = N_{c}^{fib}/N_{t}^{fib}$. In this regard, the proposed damage model \cref{equ:dam_model} coincides with the classical network alteration theory \cite{marckmann2002theory,zhao2012theory}, which describes the damage of polymeric materials.
\end{remark}
\subsubsection{Evolution equation for non-local damage}
\label{sec:evolution_equ_nonlocal}
To alleviate damage localization and mesh dependence, a gradient-enhanced regularization method is adopted in this work \cite{peerlings1996gradient}. To this end, a gradient-enhanced non-local free energy density function $\psi_{nl}^{S}$, which is a function of the non-local internal variable $\overline{Y}_{eqv}^{S}$, is incorporated into the total solid free energy density function $\psi^{S}$, see \cref{equ:free_energy_solid_total}. The derivation of the evolution equation for $\overline{Y}_{eqv}^{S}$ relies on the specification of $\psi_{nl}^{S}$, which is composed of a gradient term and a penalty term. In the gradient term, the gradient of $\overline{Y}_{eqv}^{S}$ denotes the first-order term of a Taylor series of $\overline{Y}_{eqv}^{S}$ \cite{de2016gradient}. The penalty term correlates $\overline{Y}_{eqv}^{S}$ to its local counterpart $Y_{eqv}^{S}$. The non-local free energy $\psi_{nl}^{S}$ is given as
\begin{equation}
	\label{equ:free_energy_nonlocal}
	\psi_{nl}^{S} = \frac{1}{\rho_{0S}^{S}} \bqty{\frac{c_{d} h}{2} \pqty{\grad_{\vb{x}}{\overline{Y}_{eqv}^{S}}}^2 + \frac{\beta_{d} h}{2} \pqty{\overline{Y}_{eqv}^{S}-{Y}_{eqv}^{S}}^2 } \,,
\end{equation}
where the regularization parameter $c_{d}$ is related to the internal length of materials, which modulates the width of the damage zone; the penalty parameter $\beta_{d}$ penalizes the distinction between ${Y}_{eqv}^{S}$ and $\overline{Y}_{eqv}^{S}$, and enforces the local quantity to coincide with the non-local counterpart; the coupling modulus $h$ signifies the interaction between the local and non-local variables.
\par
\begin{figure}[H]
	\centering
	\subfloat[]{
		\begin{minipage}[b]{0.45\linewidth}
			\includegraphics[width=1\linewidth]{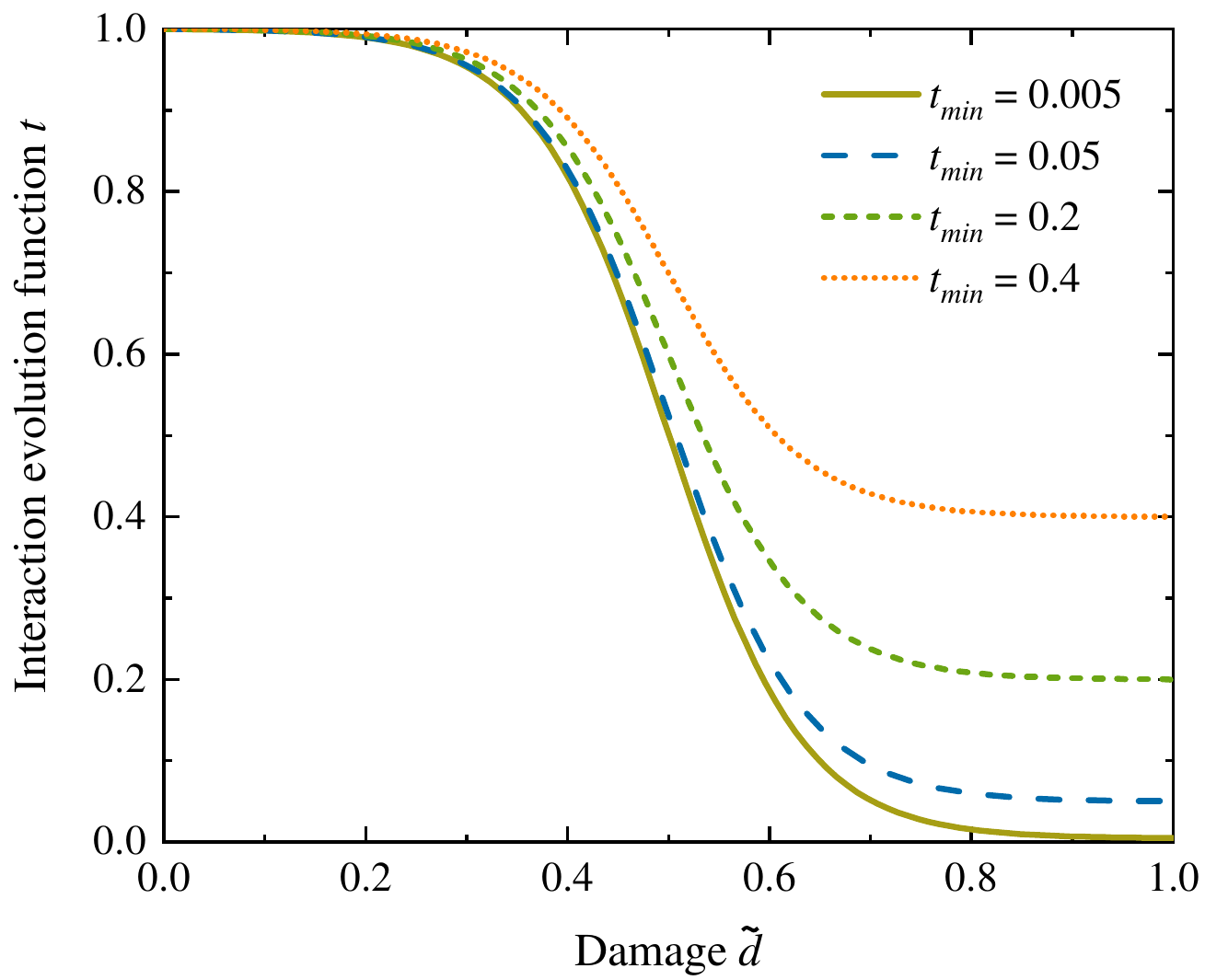} 
		\end{minipage}
		\label{}
	}
	\quad
	\subfloat[]{
		\begin{minipage}[b]{0.45\linewidth}
			\includegraphics[width=1\linewidth]{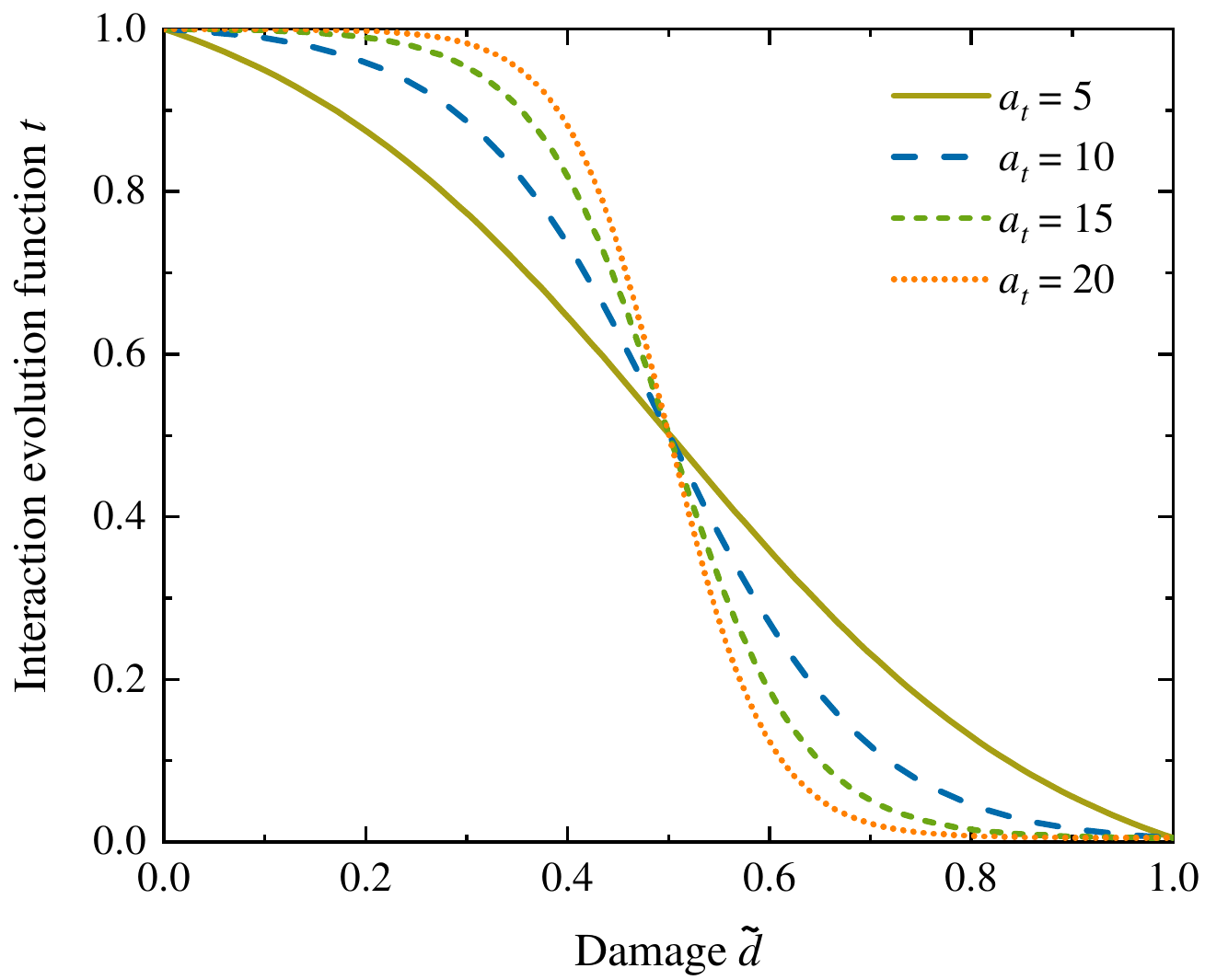}
		\end{minipage}
		\label{}
	}
	\caption{Effect of model parameters on the shape of evolution function $t$: (a) $t_{min}$ and (b) $a_{t}$. The values $t_{min}=0.005$ and $a_{t}=15$ are adopted throughout this paper.}
	\label{fig:3}
\end{figure}

It is assumed that the large deformation results in multiple and diffuse initial microdamages in the solid skeleton at an early stage of material degradation. With further loading, the fibrin network aligns in the direction of the exploited load to accommodate deformation \cite{brown2009multiscale,purohit2011protein}. The microdamage propagation of fibrin fibers aligned in this direction is more pronounced, thereby confining the microdamages to a narrower bandwidth. As loading proceeds to a critical level, the microdamages within this band significantly coalesce and interact, leading to the formation of macrocracks. The microdamages and macrocracks are associated with the local driving force $Y_{eqv}^{S}$ and the non-local counterpart $\overline{Y}_{eqv}^{S}$, respectively. The interaction domain size of the microdamages can be controlled by the regularization parameter $c_{d}$. Since the bandwidth, where fibrin fibers align in the loading direction and are highly stretched, continuously evolves as deformation proceeds, the damage process zone bandwidth of macrocracks correspondingly narrows. Therefore, the regularization parameter $c_{d}$ should evolve with material degradation rather than being a constant that causes non-physical spurious damage growth. To this end, an interaction evolution function $t = \hat{t}\pqty{\tilde{d}}$ is introduced and coupled to $c_{d}$ to characterize the evolution of the regularization parameter, \ie,
\begin{equation}
	\label{equ:evolving_regularization_para}
	c_{d} = c_{d0} t \qq{with} t = \hat{t}\pqty{\tilde{d}} \,,
\end{equation}
where $c_{d0}$ is a constant model parameter denoting the initial regularization parameter before the damage evolves, \ie, $\tilde{d}=0$. The interaction evolution function $t$ has the following property,
\begin{equation}
	\label{equ:transient_nonlocal_function}
	t = \hat{t}\pqty{\tilde{d}} = 
	\begin{cases}
		1, & \mbox{if } {\tilde{d}=0},\\
		t_{min}, & \mbox{if } {\tilde{d}=1} \,.
	\end{cases} 
\end{equation}
The differential function $t$ decreases with damage accumulation, and is constrained by \cref{equ:transient_nonlocal_function}. By introducing the function $t$, the regularization parameter approaches to zero when damage accumulates to a critical value, and the traditional gradient-enhanced damage model is recovered. Moreover, the introduction of $t$ provides a solution for the damage bond widening issue \cite{geers1998strain}. To fulfill this constraint condition in \cref{equ:transient_nonlocal_function}, an evolution function $t$ is proposed as,
\begin{equation}
	\label{equ:modified_Sigmoid_1_sim_sym}
	t = \pqty{1-t_{min}} \frac{1-\exp[-a_{t}\pqty{0.5-\tilde{d}}]}{2\Bqty{1+\exp[-a_{t}\pqty{0.5-\tilde{d}}]}} \frac{1+\exp(-0.5a_{t})}{1-\exp(-0.5a_{t})} + \frac{1+t_{min}}{2} \,,
\end{equation}
where the parameter $t_{min}$ controls the minimum value, and $a_{t}$ modulates the decreasing slope. The influence of the parameters $t_{min}$ and $a_{t}$ on the interaction evolution function $t$ is shown in \Cref{fig:3}.
\par
Recalling the flux vector $\vb{y}$ in \cref{equ:nonlocal_stress} and the inequality \cref{equ:entropy_inequ_last_term}$_3$ related to the internal microforce $b$ of the non-local internal variable $\overline{Y}_{eqv}^{S}$, and substituting the non-local free energy $\psi_{nl}^{S}$ by \cref{equ:free_energy_nonlocal}, one obtains the expressions of $\vb{y}$ and $b$, \ie,
\begin{equation}
	\label{equ:y}
	\vb{y} = \rho^{S} \pdv{\psi_{nl}^{S}}{\grad_{\vb{x}}{\overline{Y}_{eqv}^{S}}} = \frac{h}{J_{S}} c_{d} \grad_{\vb{x}}{\overline{Y}_{eqv}^{S}} \,,
\end{equation}
\begin{equation}
	\label{equ:b}
	b = \rho^{S} \pdv{\psi_{nl}^{S}}{\overline{Y}_{eqv}^{S}} = \frac{h}{J_{S}} \beta_{d} \pqty{\overline{Y}_{eqv}^{S}-{Y}_{eqv}^{S}} \,.
\end{equation}
Replacing the flux vector $\vb{y}$ and the internal microforce $b$ in the force balance equation \cref{equ:balance_b} with \cref{equ:y,equ:b}, respectively, and ignoring the external microforce $a$ yields a second order differential equation that governs the evolution of the non-local internal variable $\overline{Y}_{eqv}^{S}$, \ie,
\begin{equation}
	\label{equ:evolution_nonlocal_Y}
	- c_{d} \grad_{\vb{x}}^2{\overline{Y}_{eqv}^{S}} = \beta_{d} \pqty{Y_{eqv}^{S}-\overline{Y}_{eqv}^{S}} \,.
\end{equation}
\par
Incorporating \cref{equ:free_energy_nonlocal} into \cref{equ:effective_solid_stress_1}$_3$, one obtains the stress $\vb{T}_{E,nl}^{S}$,
\begin{equation}
	\label{equ:nonlocal_stress_specific}
	\vb{T}_{E,nl}^{S} = \frac{h}{J_{S}} \beta_{d} \pqty{Y_{eqv}^{S}-\overline{Y}_{eqv}^{S}} 2 \vb{F}_{S} \pdv{Y_{eqv}^{S}}{\vb{C}_{S}} \vb{F}_{S}^\intercal \,,
\end{equation}
which is associated with the non-local free energy density.
\begin{remark}
	The introduction of the non-local free energy in \cref{equ:free_energy_nonlocal} leads to the stress response in \cref{equ:nonlocal_stress_specific}. With the increase of damage $d$, the residual stress is high since the stress in \cref{equ:nonlocal_stress_specific} is not coupled with $d$. Therefore, a small value of the coupling modulus $h$ should be adopted to keep the residual stress minimum with the damage evolution \cite{peerlings2004thermodynamically}.
\end{remark}
\subsubsection{Fluid transport}
\label{sec:evolution_equ_fluid}
A blood clot is a natural porous material including a porous solid skeleton and interstitial fluid. Complex physiological loading can lead to inhomogeneous deformation of blood clots and thereby driving the transport of fluid through the porous fibrin networks. The fluid transport interacts with the non-linear deformation of fibrin networks and blood cells. The fluid transport is restricted by the dissipation inequality \cref{equ:entropy_inequ_last_term}$_4$, which can be fulfilled by proposing the following form for $\tilde{\vb{p}}^{F}$,
\begin{equation}
	\label{equ:interaction_Force_Fluid}
	\tilde{\vb{p}}^{F} = p \grad_{\vb{x}}{n^{F}} - \vb{H}^{F}\vb{w}_{FR} \,,
\end{equation}
where $\vb{H}^{F}$ denotes a second-order positive definite material parameter tensor. Recalling the fluid momentum balance equation in \cref{equ:balance_momentum} and substituting the quantities $\vb{T}^{F}$ and $\tilde{\vb{p}}^{F}$ with their expressions in \cref{equ:cauchy_Stress_derive}$_2$ and \cref{equ:interaction_Force_Fluid}, respectively, one obtains
\begin{equation}
	\label{equ:momentum_Fluid_rewritten}
	\grad_{\vb{x}} \vdot(-n^{F} p \vb{I}) + n^{F}\rho^{FR}\vb{b} + p \grad_{\vb{x}}{n^{F}} - \vb{H}^{F} \vb{w}_{FR}=\vb{0} \,.
\end{equation}
The first term on the left of \cref{equ:momentum_Fluid_rewritten} is reformulated as
\begin{equation}
	\label{equ:momentum_Fluid_rewritten_first_term}
	\grad_{\vb{x}} \vdot(-n^{F} p \vb{I}) = -n^{F}\grad_{\vb{x}}{p} - p \grad_{\vb{x}}{n^{F}} \,,
\end{equation}
where $\grad_{\vb{x}} \vdot(z \vb{A}) = z \grad_{\vb{x}} \vdot(\vb{A}) + \vb{A}\grad_{\vb{x}}{z}$, with $z$ an arbitrary scalar and $\vb{A}$ an arbitrary second-order tensor, is used. Replacing the first term in \cref{equ:momentum_Fluid_rewritten} with \cref{equ:momentum_Fluid_rewritten_first_term}, and substituting the fluid seepage velocity $\vb{w}_{FR}$ in \cref{equ:momentum_Fluid_rewritten} with the fluid filter velocity $\vb{w}_{F}$ according to their relations $\vb{w}_{FR} = \pqty{n^{F}}^{-1} \vb{w}_{F}$, one obtains, after some algebraic manipulations, the Darcy's law, \ie,
\begin{equation}
	\label{equ:momentum_Fluid_rewritten_1}
	\vb{w}_{F} = -\pqty{n^{F}}^2 \pqty{\vb{H}^{F}}^{-1} \pqty{\grad_{\vb{x}}{p} - \rho^{FR}\vb{b}} \,,
\end{equation}
which governs the fluid transport in blood clots. The inverse relationship between $\vb{w}_{F}$ and $\grad_{\vb{x}}{p}$ indicates the fluid flows direction in blood clots, \ie, the negative gradient of pore pressure. The fluid flow behavior can be modulated by specifying an appropriate tensor $\vb{H}^{F}$. By assuming isotropic flow, the positive definite material parameter tensor $\vb{H}^{F}$ is defined as $\vb{H}^{F} \coloneqq \pqty{n^{F}}^2 k^{-1}\vb{I}$, where $k$ is the hydraulic permeability. Experimental study shows that the permeability of blood clots is influenced by the fluid content \cite{wufsus2013hydraulic,garyfallogiannis2023fracture}. To capture the constituent-dependent permeability, $k$ is assumed to be a function of fluid volume fraction, \ie, $k=\hat{k}(n^{F})$. To this end, a permeability function is proposed by extending the Davies's equation \cite{Xu2010,wufsus2013hydraulic}, which is given as
\begin{equation}
	\label{equ:permeability}
	k \coloneqq \frac{D_{eqv}^{2}}{64\pqty{1-n^{F}}^{1.5} \bqty{1.0+56\pqty{1-n^{F}}^{3}} \eta^{F}} \qq{with} D_{eqv} \coloneqq 2\Bqty{\frac{1}{\bqty{n^{fib} \pqty{1-n^{F}}}^{-1/3} D_{fib}} + \frac{1}{\bqty{n^{m} \pqty{1-n^{F}}}^{-1/3} D_{m}}}^{-1} \,,
\end{equation}
where $D_{eqv}$ is the equivalent diameter of a fibrin fiber and a blood cell, $\eta^{F}$ is the dynamic viscosity of fluid phase, and $D_{fib}$ and $D_{m}$ denote the diameter of a fibrin fiber and a blood cell, respectively. In this work, the values of $\eta^{F}=8.9\times 10^{-10}$ MPa \cite{Moldoveanu2018}, $D_{fib}=500$ nm \cite{Diamond1993} and $D_{m}=8$ $\mu$m \cite{DiezSilva2010} are adopted.
\subsubsection{Evolution equation for fibrin monomer stretch}
\label{sec:evolution_equ_monomer_stretch}
\begin{figure}[H]
	\centering
	\includegraphics[width=1\linewidth]{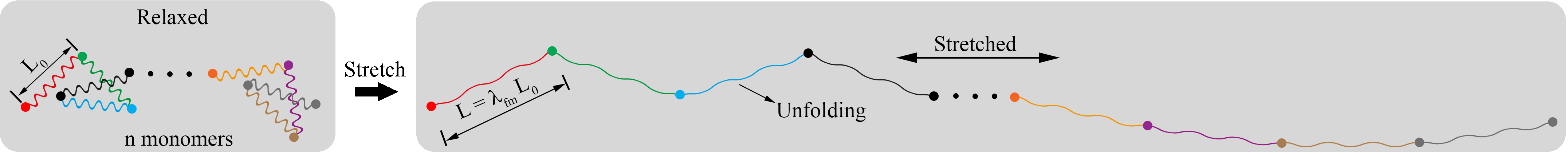}
	\caption{A schematic representation of fibrin fiber deformation and fibrin monomer unfolding.}
	\label{fig:4}
\end{figure}

The approach to modeling the protein unfolding is inspired by the idea of an overstretched form of DNA molecules proposed by \citet{Smith1996}, where the freely joined chain model was extended by assuming stretchable Kuhn segments. The mechanism of force-driven protein unfolding of fibrin chains is illustrated in \Cref{fig:4}. In the initial relaxed state, the length of a monomer is denoted by $L_{0}$. The monomers rearrange as the fibrin fiber is stretched, and the protein structure unfolds simultaneously, leading to the extension of the monomer length to $L=\lambda_{fm}L_{0}$. The evolution of the monomer stretch $\lambda_{fm}$ is constrained by the dissipation inequality \cref{equ:entropy_inequ_last_term}$_5$.
The inequality \cref{equ:entropy_inequ_last_term}$_5$ can be fulfilled by assuming a simple sufficient expression, \viz,
\begin{equation}
	\label{equ:evolution_equ_monomer_stretch}
	\frac{H}{H_{fm}} \pqty{J_{S} f_{fm} - \sum_{i}^{m,fib} \rho_{0S}^{S} \pdv{\psi_{eq}^{i}}{\lambda_{fm}} - \sum_{i}^{m,fib} \rho_{0S}^{S} \pdv{\psi_{neq}^{i}}{\lambda_{fm}} - \rho_{0S}^{S} \pdv{\psi_{fm}^{S}}{\lambda_{fm}}} = \pqty{\lambda_{fm}}'_{S}\,,
\end{equation}
where $H_{fm} > 0$ is a viscosity-like parameter that controls the dynamics of the evolution of $\lambda_{fm}$. The unfolding of fibrin monomers can only be activated when the fibrin fiber is subject to tension \cite{brown2009multiscale,liu2010mechanical}, while a fibrin monomer stretch $\lambda_{fm}$ does not evolve when subject to compression, \ie $\pqty{\lambda_{fm}}'_{S}=0$. To this end, a Heaviside function $H(\Lambda_{0}^{fib}-1)$ (\ie, $H = 1$ if $\Lambda_{0}^{fib} \geqslant 1$; $H = 0$ if $\Lambda_{0}^{fib} < 1$) is coupled to the evolution equation. Herein, $\Lambda_{0}^{fib}$ is the stretch of fibrin fibers, excluding unfolding-induced fibrin monomer stretch. The definition of $\Lambda_{0}^{fib}$ will be given in \Cref{sec:constitituve_model_equilibrium}. Recalling the balance equation \cref{equ:balance_f_fm} for microforce $f_{fm}$, \cref{equ:evolution_equ_monomer_stretch} reduces to
\begin{equation}
	\label{equ:evolution_equ_monomer_stretch_reduced}
	\frac{H}{H_{fm}} \pqty{- \sum_{i}^{m,fib} \rho_{0S}^{S} \pdv{\psi_{eq}^{i}}{\lambda_{fm}} - \sum_{i}^{m,fib} \rho_{0S}^{S} \pdv{\psi_{neq}^{i}}{\lambda_{fm}} - \rho_{0S}^{S} \pdv{\psi_{fm}^{S}}{\lambda_{fm}}} = \pqty{\lambda_{fm}}'_{S}\,,
\end{equation}
which is a thermodynamically consistent evolution equation for the fibrin monomer stretch $\lambda_{fm}$. A specific form of the evolution equation can be derived depending on the specification of the free energy density functions, which will be discussed in detail in \Cref{sec:evolution_equ_monomer_stretch_specified}.
\section{Specification of constitutive relation for blood clots}
\label{sec:specification_energy_function}
The solid constituent of a whole blood clot mainly contains fibrin networks and blood cells. To derive the constitutive relation for whole blood clots within the theoretical framework proposed in \Cref{sec:Modelling_TDF}, the free energy density functions for fibrin networks and blood cells are specified in this section.
\subsection{Constitutive relation for fibrin networks}
\label{sec:specification_energy_function_fibrin}
A fibrin network is the scaffold of a blood clot, which maintains the mechanical and structural integrity of the bulk tissue. A fibrin network is composed of fibrin fibers, which are covalently crosslinked through intra- and intermolecular interactions, \ie, covalent clotting factor XIIIa \cite{brown2009multiscale,purohit2011protein}. The covalent crosslinking is essential for the mechanical strength and structural stability of a blood clot. From the microstructural point of view, the fibrin network in a blood clot exhibits similarities to a polymer network in polymeric materials, such as elastomers and hydrogels \cite{he2022viscoporoelasticity,ghezelbash2022blood,noailly2008poroviscoelastic}. Physics-based models, which are initially developed for polymeric materials, have been used to describe the deformation of blood clots due to their direct connection to the molecular microstructure \cite{purohit2011protein,yesudasan2020multiscale,spiewak2022biomechanical}. To model the constitutive relation of the fibrin network, a physics-based constitutive model is extended in this section by incorporating non-linear viscoelasticity, compressiblility of porous media, internal energy of fibrin monomers, entropy changes due to fluid-flux-induced swelling/de-swelling, and the compaction point concept due to fluid-exflux-induced consolidation.
\par
The physics-based model is derived using statistical mechanics based on microstructural parameters of molecules. We start with introducing the damage-free free energy density function of a fibrin network measured per unit mass, involving fibrin monomer stretch-induced internal energy, \viz,
\begin{equation}
	\label{equ:free_energy_entropy}
	\psi_{0}^{fib} \coloneqq \psi_{fm0}^{S} - T^{S} \mathcal{H}^{fib} \,,
\end{equation}
with
\begin{equation}
	\label{equ:free_energy_entropy_comp}
	\psi_{fm0}^{S} \coloneqq \frac{n_{0S}^{S}}{\rho_{0S}^{S}} \sum_{1}^{N_{t0}^{fib}} \sum_{1}^{n} \epsilon_{fm} \qq{and} \mathcal{H}^{fib} = \mathcal{H}_{eq}^{fib} + \mathcal{H}_{neq}^{fib} \coloneqq \frac{n_{0S}^{S}}{\rho_{0S}^{S}} \sum_{1}^{N_{t0}^{fib}} \eta_{eq}^{fib} + \frac{n_{0S}^{S}}{\rho_{0S}^{S}} \sum_{1}^{N_{t0}^{fib}} \eta_{neq}^{fib} \,,
\end{equation}
where the free energy density of the fibrin network $\psi_{0}^{fib}$ is composed of three contributions, \ie, $\psi_{0}^{fib} = \psi_{eq0}^{fib} + \psi_{neq0}^{fib} + \psi_{fm0}^{S}$; the absolute temperature is denoted by $T^{S}$; $\mathcal{H}^{fib}$ is the change in the total configurational entropy for the fibrin network, $\mathcal{H}_{eq}^{fib} \coloneqq \pqty{n_{0S}^{S} \sum_{1}^{N_{t0}^{fib}} \eta_{eq}^{fib}}/ \rho_{0S}^{S} $ and $\mathcal{H}_{neq}^{fib} \coloneqq \pqty{n_{0S}^{S} \sum_{1}^{N_{t0}^{fib}} \eta_{neq}^{fib}}/ \rho_{0S}^{S}$ are the contributions in the equilibrium and non-equilibrium states; the quantity $N_{t0}^{fib}$ represents the number of fibrin fibers per unit volume in dry state; the parameter $n$ denotes the number of the fibrin monomer in a fiber; $\epsilon_{fm} = \hat{\epsilon}_{fm}\pqty{\lambda_{fm}}$ represents the internal energy in a single fibrin monomer due to monomer stretching; $\eta_{eq}^{fib} = \hat{\eta}_{eq}^{fib}\pqty{\Lambda^{fib}}$ and $\eta_{neq}^{fib} = \hat{\eta}_{neq}^{fib}\pqty{\Lambda^{e,fib}}$ denote the change in the configurational entropy for a single fibrin fiber in the equilibrium and non-equilibrium states, respectively. $\Lambda^{fib}$ and $\Lambda^{e,fib}$ represent the overall stretches of fibrin fibers in the equilibrium and non-equilibrium deformations, respectively. The entropy-change-related energy contributions $\psi_{eq0,\eta}^{fib}$ and $\psi_{neq0,\eta}^{fib}$ can be expressed as
\begin{equation}
	\label{equ:free_energy_entropy_eq_neq}
	\psi_{eq0,\eta}^{fib} = - T^{S} \mathcal{H}_{eq}^{fib} = - T^{S} \frac{ n_{0S}^{S}}{\rho_{0S}^{S}} \sum_{1}^{N_{t0}^{fib}} \eta_{eq}^{fib} \qq{and} \psi_{neq0,\eta}^{fib} = - T^{S} \mathcal{H}_{neq}^{fib} = - T^{S} \frac{n_{0S}^{S}}{\rho_{0S}^{S}} \sum_{1}^{N_{t0}^{fib}} \eta_{neq}^{fib} \,,
\end{equation}
respectively.
\subsubsection{Internal energy of fibrin monomers}
\label{sec:internal_energy}
The protein unfolding behavior in the fibrin fibers of blood clots, arising from the stretch of fibrin monomers in fibrin fibers, has been confirmed recently \cite{brown2009multiscale}. By assuming that fibrin fibers in the fibrin network distribute uniformly, each fibrin fiber in the network has the same number of fibrin monomers $n$, and the stretch of each fibrin monomer in a fibrin fiber is identical, the total damage-free internal energy density $\psi_{fm0}^{S}$ of all fibrin monomers in \cref{equ:free_energy_entropy_comp}$_1$ can be expressed as
\begin{equation}
	\label{equ:internal_energy_total}
	\psi_{fm0}^{S} = \frac{n_{0S}^{S}}{\rho_{0S}^{S}}  N_{t0}^{fib} n \epsilon_{fm}\,.
\end{equation}
The fibrin monomer stretch $\lambda_{fm}$ is one in the reference configuration and increases with the rise of deformation. A simple expression is utilized for the internal energy, which is given as
\begin{equation}
	\label{equ:internal_energy_a_monomer}
	\epsilon_{fm} = \frac{E_{fm}}{2} \pqty{\ln \lambda_{fm}}^2 \,, \qq{with} \lambda_{fm} \coloneqq \frac{L}{L_{0}} \,,
\end{equation}
where $E_{fm} > 0$ is a material parameter, representing the stiffness of a fibrin monomer, $L_{0}$ and $L$ denote the length of a fibrin monomer in initial and stretched states, respectively (see \Cref{fig:4}).
\subsubsection{Equilibrium part of the constitutive model for the fibrin network}
\label{sec:constitituve_model_equilibrium}
The average end-to-end length of a fibrin fiber in the unstretched state is $r_0=\sqrt{n} L_{0}$ according to the classical freely joint chain model \cite{treloar1975physics}, where it is assumed that two neighboring fibrin monomers in a fibrin fiber can freely rotate relative to each other. When stretched, the end-to-end length of the fiber increases to $r$, and the stretch of each fibrin monomer evolves. The length of the stretched fibrin monomer can be expressed as $L = L_{0} \lambda_{fm}$ according to \cref{equ:internal_energy_a_monomer}$_2$. The stretch of a fiber is defined as
\begin{equation}
	\label{equ:stretch_fibrin_modified}
	\Lambda^{fib} \coloneqq \frac{r}{{r_0} \lambda_{fm}} = \frac{r}{\sqrt{n} L_{0} \lambda_{fm}} = \frac{r}{\sqrt{n} L} \,,
\end{equation}
where the relations $r_0=\sqrt{n} L_{0}$ and $L = L_{0} \lambda_{fm}$ are used. Compared with the stretch in \Cref{equ:stretch_fibrin_modified}, the stretch in the classical freely jointed chain theory is expressed as
\begin{equation}
	\label{equ:stretch_fibrin}
	\Lambda_{0}^{fib} \coloneqq \frac{r}{{r_0}} = \frac{r}{\sqrt{n} L_{0}} \,,
\end{equation}
where the monomer is assumed to be rigid.
\begin{figure}[H]
	\centering
	\begin{tikzpicture}[boot/.style={draw,align=center,label=below:#1}]
		\node[circle,fill=gray1,text width=1.5cm,boot=Reference configuration] (A) {$r_{0} \coloneqq \sqrt{n}L_{0}$};
		\node[above right=1.5cm and 4cm of A,circle,fill=gray1,text width=1.5cm,boot=Intermediate configuration 1] (B1) {$\hat{r}_{0} \coloneqq \sqrt{n}L_{0}\lambda_{fm}$};	
		\node[below right=1cm and 4cm of A,circle,fill=gray1,text width=1.5cm,boot=Intermediate configuration 2] (B2) {$\hat{r}$};
		\node[right=9cm of A,circle,fill=gray1,text width=1.5cm,boot=Actual configuration] (C) {$r$};
		\draw[-latex] (A) to node [midway,above,sloped] {$\Lambda_{unf}^{fib} \coloneqq \hat{r}_{0}/r_{0}$} (B1);
		\draw[-latex] (B1) to node [midway,above,sloped] {$\Lambda_{ent}^{fib} \coloneqq r/\hat{r}_{0}$} (C);
		\draw[-latex] (A) to node [midway,above,sloped] {$\Lambda_{0}^{fib}=\Lambda_{unf}^{fib}\Lambda_{ent}^{fib}=\bar{\Lambda}_{ent}^{fib}\bar{\Lambda}_{unf}^{fib}$} (C);
		\draw[-latex] (A) to node [midway,above,sloped] {$\bar{\Lambda}_{ent}^{fib} \coloneqq \hat{r}/r_{0}$} (B2);
		\draw[-latex] (B2) to node [midway,above,sloped] {$\bar{\Lambda}_{unf}^{fib} \coloneqq r/\hat{r}$} (C);
	\end{tikzpicture}
	\caption{Schematic of the relationship between the counter length of a fibrin fiber at different configurations. The counter lengths $r_{0}$ and $r$ are defined in the reference and actual configurations, respectively, which are connected by the stretch $\Lambda_{0}^{fib}$. Two intermediate configurations of the counter length are introduced to show the different evolution routes.}
	\label{fig:counter_length}
\end{figure}
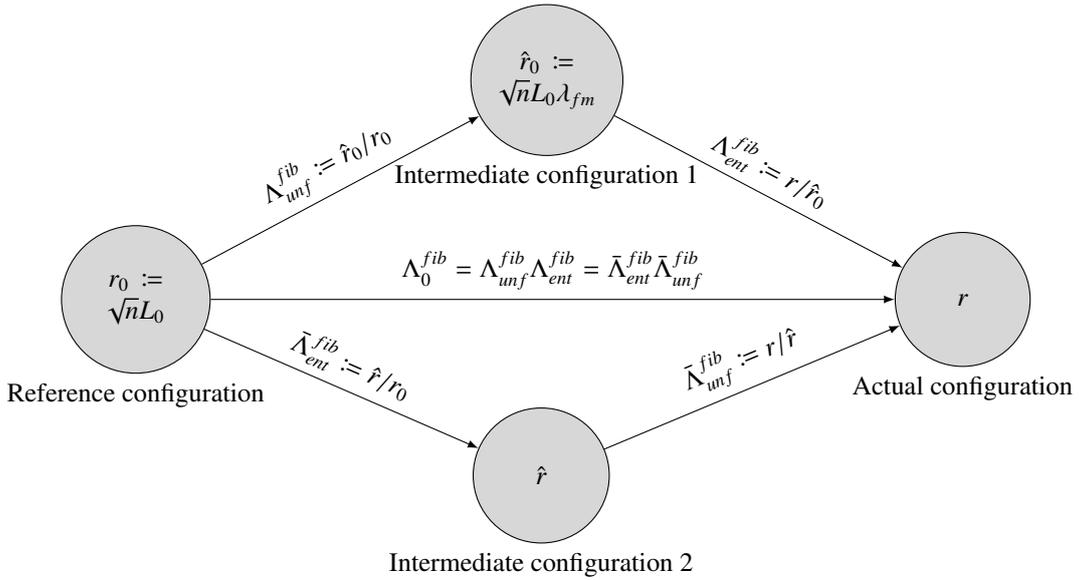
\par
To better explain the relationship between the stretches defined in \Cref{equ:stretch_fibrin_modified,equ:stretch_fibrin}, four configurations of the counter length of the fiber are defined, as shown in \Cref{fig:counter_length}. The initial length $r_{0}$ can evolve to the actual length $r$ via stretch $\Lambda_{0}^{fib}$ following two different routes, \ie, $r_{0} \rightarrow \hat{r_{0}} \rightarrow r$ and $r_{0} \rightarrow \hat{r} \rightarrow r$, with $\hat{r_{0}}$ and $\hat{r}$ defined in the virtual intermediate configuration 1 and 2, respectively. Correspondingly, the stretch $\Lambda_{0}^{fib}$ is multiplicatively decomposed into an unfolding stretch $\Lambda_{unf}^{fib} \coloneqq \hat{r_{0}}/r_{0} = \lambda_{fm}$ or $\bar{\Lambda}_{unf}^{fib} \coloneqq r/\hat{r} = \lambda_{fm}$, and a stretch $\Lambda_{ent}^{fib} \coloneqq r/\hat{r}_{0}$ or $\bar{\Lambda}_{ent}^{fib} \coloneqq \hat{r}/r_{0}$ caused by the purely entropic contribution, \ie, $\Lambda_{0}^{fib}=\Lambda_{unf}^{fib} \Lambda_{ent}^{fib}=\bar{\Lambda}_{ent}^{fib} \bar{\Lambda}_{unf}^{fib}$. Recalling the definition of $\Lambda^{fib}$, see \Cref{equ:stretch_fibrin_modified}, one obtains $\Lambda^{fib}=\Lambda_{ent}^{fib}$, indicating that the stretch $\Lambda^{fib}$ results solely from the change of configurational entropy. With these relationships at hand, the stretch $\Lambda^{fib}$ can be correlated with $\Lambda_{0}^{fib}$ via $\Lambda^{fib}=\Lambda_{ent}^{fib}=\Lambda_{0}^{fib}/\Lambda_{unf}^{fib}= r/\hat{r_{0}}$, where $\hat{r_{0}}$ is regarded as an evolving initial fiber length due to the evolution of $\lambda_{fm}$. When the initial and actual lengths $r_{0}$ and $r$ are fixed, \ie, $\Lambda_{0}^{fib}$ is fixed, the increase of the unfolding stretch $\lambda_{fm}$ results in the decrease of the stretch $\Lambda^{fib}$, thereby reducing the entropic contribution to the free energy.
\begin{remark}
	\label{remark:stretch_fiber_monomer}
	It is noted that two concepts of stretch are defined, \ie, the stretch of a fibrin fiber and the stretch of a fibrin monomer, which have different physical meanings. The stretch of a fibrin fiber causes the change of the end-to-end length of the fiber, which is defined in \cref{equ:stretch_fibrin_modified}, while the stretch of a fibrin monomer leads to protein unfolding and thereby resulting in the increase of the monomer length, which is defined in \cref{equ:internal_energy_a_monomer}$_2$. The stretch of a fibrin fiber triggers the extension of its involved fibrin monomers. The monomer stretch impacts the calculation of fibrin fiber stretch. 
\end{remark}
\par
As a highly hydrated tissue, the fibrin fibers in a blood clot can be stretched due to fluid-transport-induced swelling during clotting and deformation \cite{jimenez2023multiscale}. Therefore, the total equilibrium stretch of a fibrin fiber is decomposed into three contributions: (1) an isotropic and homogeneous swelling stretch $\lambda_{sw}$, (2) a mechanical-deformation-induced stretch $\lambda_{md}$, (3) a fibrin monomer stretch $\lambda_{fm}$ associated with protein unfolding, \ie,
\begin{equation}
	\label{equ:decomposition_stretch}
	\Lambda^{fib} = \lambda_{sw} \lambda_{md} \lambda_{fm}^{-1} \,.
\end{equation}
\par
Proceeding from statistical mechanics, the change in the configurational entropy for a single fibrin fiber in the equilibrium state $\eta_{eq}^{fib}$ in \cref{equ:free_energy_entropy_eq_neq}$_1$ is defined as
\begin{equation}
	\label{equ:entropy_equilibrium}
	\eta_{eq}^{fib} \coloneqq k_{b} \ln p_{eq} \,, \qq{with} p_{eq} = \hat{p}_{eq} \pqty{\Lambda^{fib}} \,,
\end{equation}
where $k_{b}$ denotes the Boltzmann constant, and $p_{eq}$ represents a probability density function. Modifying the classical non-Gaussian probability density function, which is approximated by \citet{kuhn1942beziehungen}, through incorporating the monomer stretch effect and swelling stretch effect yields an extended probability density function for freely jointed molecular chain, \viz \cite{kuhn1942beziehungen,dal2020extended},
\begin{equation}
	\label{equ:PDF_equilibrium}
	p_{eq} \coloneqq p_{0} \exp[-n \pqty{\frac{\Lambda^{fib}}{\sqrt{n}}\beta^{fib} + \ln \frac{\beta^{fib}}{\sinh \beta^{fib}}}] \,, \qq{with} p_{0} \coloneqq \exp[-n \pqty{\frac{\Lambda_{0}^{fib}}{\sqrt{n}}\beta_{0}^{fib} + \ln \frac{\beta_{0}^{fib}}{\sinh \beta_{0}^{fib}}}] \,,
\end{equation}
where $\beta^{fib} \coloneqq \mathcal{L}^{-1} \pqty{\Lambda^{fib}/\sqrt{n}}$ denotes the inverse function of the Langevin function with $\mathcal{L}(x) \coloneqq \mathrm{coth}x -1/x$, and  $\beta_{0}^{fib} \coloneqq \mathcal{L}^{-1} \pqty{\Lambda_{0}^{fib}/\sqrt{n}}$ represents the value of $\beta^{fib}$ in the reference state with $\Lambda_{0}^{fib} \coloneqq \lambda_{sw} \lambda_{fm}^{-1}$ the stretch of a fibrin fiber in the undeformed state. Inserting \cref{equ:PDF_equilibrium} into \cref{equ:entropy_equilibrium}$_1$, one obtains
\begin{equation}
	\label{equ:entropy_equilibrium_specific}
	\eta_{eq}^{fib} = - k_{b} \bqty{n \pqty{\frac{\Lambda^{fib}}{\sqrt{n}}\beta^{fib} + \ln \frac{\beta^{fib}}{\sinh \beta^{fib}}} - n \pqty{\frac{\Lambda_{0}^{fib}}{\sqrt{n}}\beta_{0}^{fib} + \ln \frac{\beta_{0}^{fib}}{\sinh \beta_{0}^{fib}}}} \,.
\end{equation}
\par
To link the microscopic stretch of fibrin fibers and the macroscopic deformation of the fibrin network, a proper network model is needed \cite{treloar1979non,marckmann2002theory}. In this work, the eight-chain model developed by \citet{arruda1993three} is adopted and extended. The eight-chain model has been used in modeling mechanical deformation of blood clots \cite{purohit2011protein,yesudasan2020multiscale,spiewak2022biomechanical}, showing that this model can reasonably capture the mechanical behavior of fibrin networks. The eight-chain model assumes that a fibrin network is composed of many cubic unit cells. For each cell, eight fibrin fibers are aligned along the eight half diagonals of the cube, connecting the center of the cubic cell and the eight vertices separately. When stretched, all fibrin fibers are assumed to undergo an identical stretch. In the deformed state, the fibrin network undergoes a deformation with principal stretches $\lambda_1$, $\lambda_2$ and $\lambda_3$. The relation between the purely mechanical deformation-induced fiber stretch $\lambda_{md}$ and the macroscopic deformation of the fibrin network can be expressed by
\begin{equation}
	\label{equ:stretch_chain}
	\lambda_{md} = \sqrt{\frac{\lambda_1^2 + \lambda_2^2 +\lambda_3^2}{3}} = \sqrt{\frac{I_{S1}}{3}} \,,
\end{equation}
where the first principal invariant $I_{S1}$ is correlated with the principal stretches by $I_{S1}=\lambda_1^2 + \lambda_2^2 +\lambda_3^2$.
\par
By assuming an isotropic and homogeneous swelling, the swelling stretch $\lambda_{sw}$ is given as \cite{tang2017fatigue,zhao2012theory}
\begin{equation}
	\label{equ:swelling_stretch}
	\lambda_{sw} \coloneqq \pqty{n^{S}}^{-\frac{1}{3}}\,.
\end{equation}
Moreover, swelling leads to decreasing the number density of fibrin fibers, which is incorporated by scaling the number density $N_{t0}^{fib}$ of fibrin fibers in dry state with the volume fraction $n^{S}$ of solid volume concentration, \viz, $N_{t}^{fib} = n^{S} N_{t0}^{fib}$. Incorporating \cref{equ:swelling_stretch,equ:stretch_chain} into \cref{equ:decomposition_stretch} yields the total equilibrium stretch of a fibrin fiber, \ie,
\begin{equation}
	\label{equ:decomposition_stretch_specific}
	\Lambda^{fib} = \pqty{n^{S}}^{-\frac{1}{3}} \sqrt{\frac{I_{S1}}{3}} \lambda_{fm}^{-1} \,,
\end{equation}
where $\lambda_{fm}$ can be calculated by solving the implicit evolution equation \cref{equ:evolution_equ_monomer_stretch_reduced}.
\par
The equilibrium part of the total damage-free energy density of a uniform non-Gaussian fibrin network can be split into three contributions, \ie,
\begin{equation}
	\label{equ:equilibrium-split}
	\psi_{eq0}^{fib} = \psi_{eq0,\eta}^{fib} + \psi_{eq0,J}^{fib} + \psi_{eq0,0}^{fib} \,,
\end{equation}
where the energy $\psi_{eq0,\eta}^{fib}$ is induced by the change in configurational entropy, the energy $\psi_{eq0,J}^{fib}$ is introduced to describe the compressibility of the solid skeleton, and the energy $\psi_{eq0,0}^{fib}$ enforces the zero-value stress in the undeformed state. The specific form of $\psi_{eq0}^{fib}$ is given by
\begin{equation}
	\label{equ:free_energy_component_physical_chain_eq}
	\begin{aligned}
		\psi_{eq0}^{fib} = &\underbrace{\frac{H_{TCA}}{\rho_{0S}^{fib}} n_{0S}^{S} \mu^{fib} \bqty{n \pqty{\frac{\Lambda^{fib}}{\sqrt{n}}\beta^{fib}+\ln \frac{\beta^{fib}}{\sinh \beta^{fib}}} - n \pqty{\frac{\Lambda_{0}^{fib}}{\sqrt{n}}\beta_{0}^{fib} + \ln \frac{\beta_{0}^{fib}}{\sinh \beta_{0}^{fib}}}}}_{\psi_{eq0,\eta}^{fib}} \\ 
		& - \underbrace{\frac{H_{TCA}}{\rho_{0S}^{fib}} \frac{n_{0S}^{S} \mu^{fib} \pqty{n_{0S}^{S}}^{-\frac{2}{3}} \pqty{\lambda_{fm}}^{-2} \sqrt{n} \beta_0^{fib}}{3\Lambda_{0}^{fib}}\ln J_{S}}_{\psi_{eq0,0}^{fib}} + \underbrace{\frac{H_{TCA}}{\rho_{0S}^{fib}} \lambda_{cp}^{fib} \bqty{\frac{1}{4} \bqty{\pqty{\ln J_{S}}^2+\pqty{J_{S}-1}^2} + \xi^{fib}}}_{\psi_{eq0,J}^{fib}} \,.
	\end{aligned}
\end{equation}
Herein, the energy $\psi_{eq0,\eta}^{fib}$ is obtained by substituting $\eta_{eq}^{fib} $ in \cref{equ:free_energy_entropy_eq_neq}$_1$ with \cref{equ:entropy_equilibrium_specific}. The constant $\mu^{fib} \coloneqq N_{t0}^{fib} k_b T^{S}$ denotes the shear modulus. The energy $\psi_{eq0,J}^{fib}$ is a volumetric extension part. The parameter $\lambda_{cp}^{fib}$ represents a constant related to the Lam$\acute{e}$ parameter $\lambda_{l}^{fib}$. For a blood clot, the porous material is in general macroscopically compressible due to pore fluid influx or efflux, while each single constituent is materially incompressible. The fluid-transport-induced volumetric variation in the biphasic aggregate is characterized by $J_S$ and is related to the volume fraction via $n_{0S}^{S}=J_{S} n^{S}$. At a critical state, where all fluid flows out and all pores are closed, the solid volume fraction becomes $n^{S}=1$, thereby obtaining the volume change measurement $J_S = n_{0S}^S$. This state is named as the compaction point. The volume change at the compaction point is defined as $J_{cp} \coloneqq J_S = n_{0S}^S$ which characterizes the volume change between the initial configuration and the compaction point. Since the bulk stiffness of the dense solid is always greater than that of fluid, the volumetric extension term in the free energy density for the solid skeleton must restrict the maximum volumetric contraction to the compaction point and incorporate the volumetric stiffening in the course of material densification \cite{Bluhm2002}. To this end, the quantities $\xi$ and $\lambda_{cp}^{fib}$ in \cref{equ:free_energy_component_physical_chain_eq} are introduced as functions of $J_{cp}$, \ie,
\begin{equation}
	\label{equ:XI_lambda_cp}
	\xi^{fib}=J_{cp}\ln J_{S} + \frac{1-J_{cp}}{J_{cp}-2} \bqty{\ln \frac{J_{cp}-J_{S}}{J_{S} \pqty{J_{cp}-1}- J_{cp}}- \ln (1-J_{cp})} \qq{and} \lambda_{cp}^{fib}=\lambda_{l}^{fib} \bqty{1+J_{cp} \pqty{1+\frac{J_{cp}^2}{1-J_{cp}}}}^{-1} \,,
\end{equation}
respectively.
\par
\begin{figure}[H]
	\centering
	\subfloat[]{
		\begin{minipage}[b]{0.35\linewidth}
			\includegraphics[width=1\linewidth]{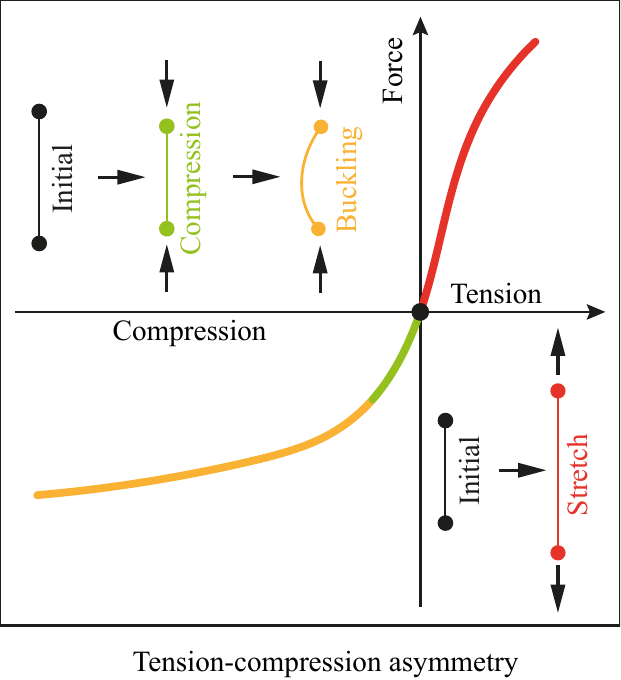} 
		\end{minipage}
		\label{}
	}
	\qquad
	\subfloat[]{
		\begin{minipage}[b]{0.47\linewidth}
			\includegraphics[width=1\linewidth]{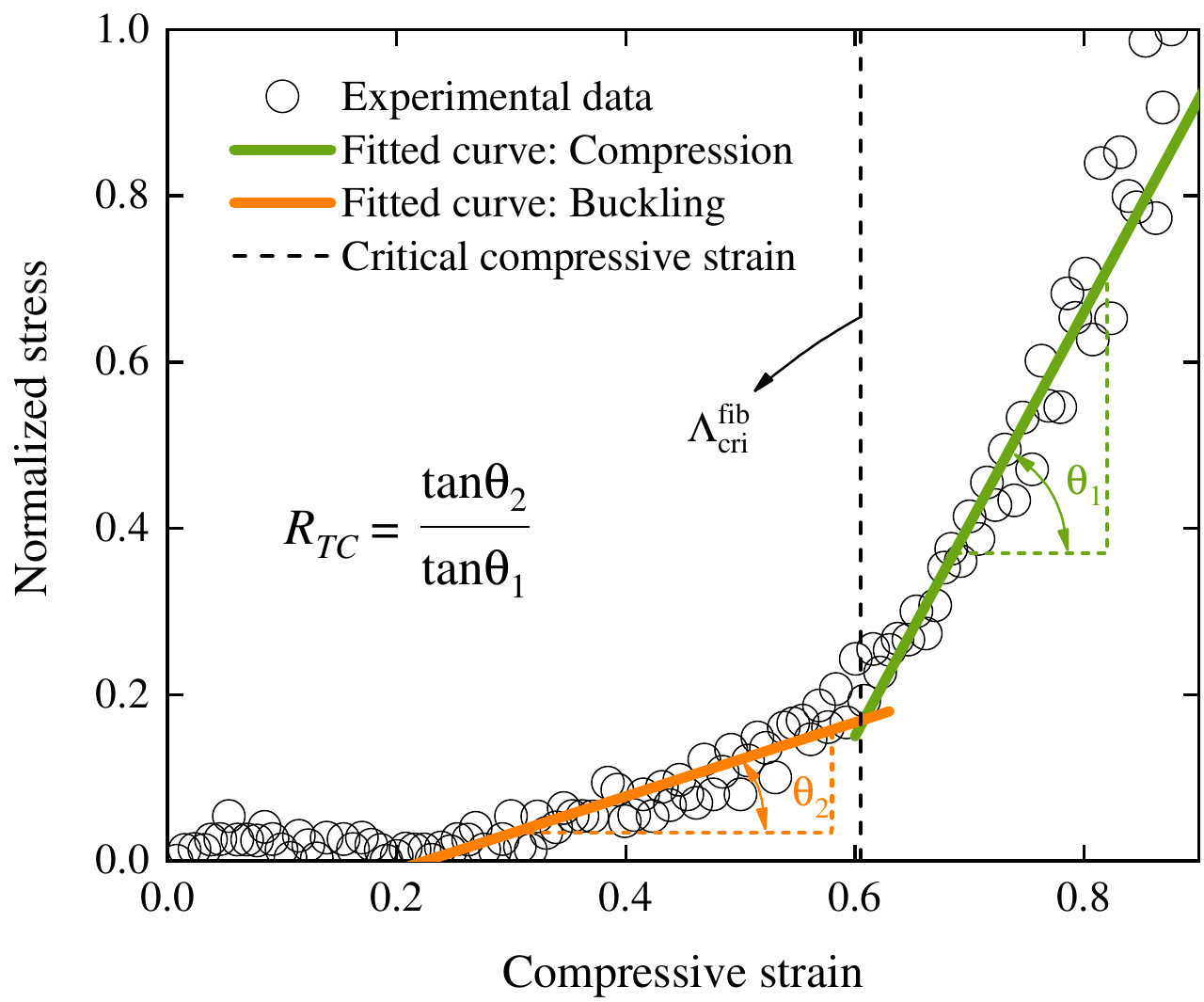}
		\end{minipage}
		\label{}
	}
	\caption{Tension-compression asymmetry of fibrin fibers. (a) A schematic representation of the reaction force of a fibrin fiber subject to axial strain. Three deformation regimes are displayed according to the applied axial strain. (b) Subject to compressive strain, the relationship between moduli of fibrin networks in distinct regimes can be identified by using the stress-strain curves in experiments \cite{Zakharov2024}. The values of the compression-buckling switch $\Lambda_{cri}^{fib}$ and ratio $R_{TC}$ are identified as $\Lambda_{cri}^{fib}=0.605$ and ratio $R_{TC}=0.174$, respectively.}
	\label{fig:6}
\end{figure}

Fibrin fibers demonstrate tension-compression asymmetry subject to axial loading \cite{Zakharov2024,Rosakis2015}. This phenomenon attributes to two effects: the compression-weakening behavior and the different degradation rate. The former is correlated with the deformation mode of fibrin fiber. As illustrated in \Cref{fig:6}a, the reaction force of a fibrin fiber, subject to axial strain, exhibits three deformation stages: stiff extension, stiff compression and weak buckling. To be able to describe these deformation modes, a step function $H_{TCA}$ is introduced into the free energy function \Cref{equ:free_energy_component_physical_chain_eq}, \ie, $H_{TCA} = 1$ if $\Lambda_{0}^{fib} \geqslant \Lambda_{cri}^{fib}$; $H_{TC}^{fib} = R_{TC}$ if $\Lambda_{0}^{fib} < \Lambda_{cri}^{fib}$, where the critical compressive strain $\Lambda_{cri}^{fib}$ denotes the compression-buckling switch, and $R_{TC}$ represents the ratio of soft buckling stiffness to the stiff compression stiffness. The values of $\Lambda_{cri}^{fib}$ and $R_{TC}$ are identified using the experimental data by \citet{Zakharov2024} (see \Cref{fig:6}b). The material degradation-induced tension-compression asymmetry is discussed in \Cref{sec:damage_derive}.
\par
Combining \cref{equ:free_energy_component_physical_chain_eq} and \cref{equ:free_energy_solid_parts}$_1$ yields the equilibrium free energy density for the fibrin network, \ie,
\begin{equation}
	\label{equ:free_energy_fibrin_eq}
	\psi_{eq}^{fib} = n^{fib} f^{fib} \psi_{eq0}^{fib} \,.
\end{equation}
The parameter $\alpha^{fib}$ in the damage function $f^{fib}$ for the fibrin network, see \Cref{sec:damage_derive}, is assumed to be one for simplicity. Subject to positive stretch, fibrin fibers break when the stretch exceeds a critical value. Unlike the stretch deformation, compression cannot lead to irreversible damage of fibrin fibers \cite{Kim2014}. This tension-compression asymmetric damage behavior of fibrin fibers is taken into account by coupling a step function $H_{TC}^{fib}$ to the damage function $f^{fib}$, \ie, $H_{TC}^{fib} = 1$ if $\Lambda_{0}^{fib} \geqslant 1$; $H_{TC}^{fib} = \hbar_{TC}^{fib}$ if $\Lambda_{0}^{fib} < 1$. The parameter $\hbar_{TC}^{fib}$ $\pqty{\in [0,1]}$ modulates the compression-induced damage. Therefore, the damage function is expressed as $f^{fib} = 1 - H_{TC}^{fib}\tilde{d}$. The equilibrium effective solid stress $\vb{T}_{E,eq}^{fib}$ for the fibrin network can be derived by incorporating \cref{equ:free_energy_fibrin_eq} into \cref{equ:effective_solid_stress}$_1$, which is given as
\begin{equation}
	\label{equ:effective_solid_stress_eq}
	\begin{aligned}
		\vb{T}_{E,eq}^{fib} = \frac{H_{TCA} n^{fib} f^{fib}}{J_{S}} \Bigg\{ &\frac{n_{0S}^{S} \mu^{fib} \pqty{n_{0S}^{S}}^{-\frac{2}{3}} \pqty{\lambda_{fm}}^{-2} \sqrt{n} \beta^{fib}}{3 \Lambda^{fib} } \vb{B}_{S} - \frac{n_{0S}^{S} \mu^{fib} \pqty{n_{0S}^{S}}^{-\frac{2}{3}} \pqty{\lambda_{fm}}^{-2} \sqrt{n} \beta_0^{fib}}{3\Lambda_{0}^{fib}} \vb{I} \\ 
		& + \Bqty{ \lambda_{cp}^{fib} \bqty{\frac{1}{2} \bqty{\ln J_{S}+J_{S}\pqty{J_{S}-1}} + \zeta^{fib}} }\vb{I} \Bigg\}  \,,
	\end{aligned}
\end{equation}
with
\begin{equation}
	\label{equ:zeta}
	\zeta^{fib} = J_{S} \pdv{\xi^{fib}}{J_{S}} = J_{cp} \bqty{1-\cfrac{J_{S}}{J_{S}^2+\cfrac{J_{cp}^2}{1-J_{cp}} \pqty{J_{S}-1}}} \,.
\end{equation}
\subsubsection{Non-equilibrium part of the constitutive model for the fibrin network}
\label{sec:constitituve_model_nonequilibrium}
The constitutive relation for the non-equilibrium part of the fibrin network can be obtained by carrying out a similar procedure as that for the equilibrium counterpart. The total non-equilibrium stretch of a fibrin fiber is defined as, \ie,
\begin{equation}
	\label{equ:decomposition_stretch_specific_neq}
	\Lambda^{e,fib} = \lambda_{sw} \lambda_{md}^{e} \lambda_{fm}^{-1} \qq{with} \lambda_{md}^{e} \coloneqq \sqrt{\frac{I_{S1}^{e,fib}}{3}} \,.
\end{equation}
The change in the configurational entropy for a single fibrin fiber in the non-equilibrium state $\eta_{neq}^{fib}$ in \cref{equ:free_energy_entropy_eq_neq}$_2$ is defined as
\begin{equation}
	\label{equ:entropy_non_equilibrium}
	\eta_{neq}^{fib} \coloneqq k_{b} \ln p_{neq} \,, \qq{with} p_{neq} \coloneqq p_{0} \exp[-n \pqty{\frac{\Lambda^{e,fib}}{\sqrt{n}}\beta^{e,fib} + \ln \frac{\beta^{e,fib}}{\sinh \beta^{e,fib}}}] \,,
\end{equation}
where the inverse Langevin function is defined as $\beta^{e,fib} \coloneqq \mathcal{L}^{-1} \pqty{\Lambda^{e,fib}/\sqrt{n}}$. Inserting $\eta_{neq}^{fib}$ into \cref{equ:free_energy_entropy_eq_neq}$_2$, and extending the formulation by considering the compressibility and the zero-value stress in the unformed state, the non-equilibrium, damage-free energy density for the uniform non-Gaussian fibrin network is given by
\begin{equation}
	\label{equ:free_energy_component_physical_chain_neq}
	\begin{aligned}
		\psi_{neq0}^{fib} = &\underbrace{\frac{H_{TCA}}{\rho_{0S}^{S}} n_{0S}^{S} \mu^{e,fib} \bqty{n \pqty{\frac{\Lambda^{e,fib}}{\sqrt{n}}\beta^{e,fib}+\ln \frac{\beta^{e,fib}}{\sinh \beta^{e,fib}}} - n \pqty{\frac{\Lambda_{0}^{fib}}{\sqrt{n}}\beta_{0}^{fib} + \ln \frac{\beta_{0}^{fib}}{\sinh \beta_{0}^{fib}}}}}_{\psi_{neq0,\eta}^{fib}} \\ 
		& - \underbrace{\frac{H_{TCA}}{\rho_{0S}^{S}} \frac{n_{0S}^{S} \mu^{e,fib} \pqty{n_{0S}^{S}}^{-\frac{2}{3}} \pqty{\lambda_{fm}}^{-2} \sqrt{n} \beta_0^{fib}}{3\Lambda_{0}^{fib}}\ln J_{S}^{e,fib}}_{\psi_{neq0,0}^{fib}} + \underbrace{\frac{H_{TCA}}{\rho_{0S}^{S}} \lambda_{cp}^{e,fib} \bqty{\frac{1}{4} \bqty{\pqty{\ln J_{S}^{e,fib}}^2+\pqty{J_{S}^{e,fib}-1}^2} + \xi^{e,fib}}}_{\psi_{neq0,J}^{fib}} \,,
	\end{aligned}
\end{equation}
with
\begin{equation}
	\label{equ:XI_lambda_cp_neq}
	\xi^{e,fib} = J_{cp}^{e} \ln J_{S}^{e,fib} + \frac{1-J_{cp}^{e}}{J_{cp}^{e}-2} \bqty{\ln \frac{J_{cp}^{e}-J_{S}^{e,fib}}{J_{S}^{e,fib} \pqty{J_{cp}^{e}-1}-J_{cp}^{e}}- \ln (1-J_{cp}^{e})} \qq{and} \lambda_{cp}^{e,fib} = \lambda_{l}^{e,fib} \Bqty{1+J_{cp}^{e} \bqty{1+\frac{\pqty{J_{cp}^{e}}^2}{1-J_{cp}^{e}}}}^{-1} \,,
\end{equation}
where the interpretations of the energy contributions $\psi_{neq0,\eta}^{fib}$, $\psi_{neq0,0}^{fib}$ and $\psi_{neq0,J}^{fib}$ $\lambda_{l}^{e,fib}$ are similar to $\psi_{neq0,\eta}^{fib}$, $\psi_{neq0,0}^{fib}$ and $\psi_{neq0,J}^{fib}$ $\lambda_{l}^{e,fib}$ in \Cref{equ:equilibrium-split}, respectively. The parameters $\lambda_{l}^{e,fib}$ and $\mu^{e,fib}$ are material constants. $J_{cp}^{e}$ is defined as $J_{cp}^{e} \coloneqq n_{Sv}^{S} = n_{0S}^{S} \pqty{J_{S}^{v,fib}}^{-1}$, with $J_{S}^{v,fib} \coloneqq \det \vb{F}_{S}^{v,fib}$ and $n_{Sv}^{S}$ the inelastic solid volume fraction with respect to the intermediate configuration. Recalling the non-equilibrium free energy density for the fibrin network in \cref{equ:free_energy_solid_parts}$_2$, \ie,
\begin{equation}
	\label{equ:free_energy_fibrin_neq}
	\psi_{neq}^{fib} = n^{fib} f^{fib} \psi_{neq0}^{fib} \,,
\end{equation}
one obtains the non-equilibrium effective solid stress $\vb{T}_{E,neq}^{fib}$ for the fibrin network by inserting $\psi_{neq}^{fib}$ into \cref{equ:effective_solid_stress}$_2$, \ie,
\begin{equation}
	\label{equ:effective_solid_stress_neq}
	\begin{aligned}
		\vb{T}_{E,neq}^{fib} = \frac{H_{TCA} n^{fib} f^{fib}}{J_{S}} \Bigg\{ &\frac{n_{0S}^{S} \mu^{e,fib} \pqty{n_{0S}^{S}}^{-\frac{2}{3}} \pqty{\lambda_{fm}}^{-2} \sqrt{n} \beta^{e,fib}}{3 \Lambda^{e,fib} } \vb{B}_{S}^{e} -\bqty{ \frac{n_{0S}^{S} \mu^{e,fib} \pqty{n_{0S}^{S}}^{-\frac{2}{3}} \pqty{\lambda_{fm}}^{-2} \sqrt{n} \beta_0^{fib}}{3\Lambda_{0}^{fib}} }\vb{I} \\ 
		& + \Bqty{ \lambda_{cp}^{e,fib} \bqty{\frac{1}{2} \bqty{\ln J_{S}^{e,fib}+J_{S}^{e,fib}\pqty{J_{S}^{e,fib}-1}} + \zeta^{e,fib}} }\vb{I} \Bigg\}  \,,
	\end{aligned}
\end{equation}
with
\begin{equation}
	\label{equ:zeta_neq}
	\zeta^{e,fib} = J_{S}^{e,fib} \pdv{\xi^{e,fib}}{J_{S}^{e,fib}} = J_{cp}^{e} \bqty{1-\cfrac{J_{S}^{e,fib}}{\pqty{J_{S}^{e,fib}}^2+\cfrac{\pqty{J_{cp}^{e}}^2}{1-J_{cp}^{e}}\pqty{J_{S}^{e,fib}-1}}} \,.
\end{equation}
\subsubsection{Evolution equation for fibrin monomer stretch}
\label{sec:evolution_equ_monomer_stretch_specified}
The framework of the evolution equation for the fibrin monomer stretch is given in \Cref{sec:evolution_equ_monomer_stretch}. It is assumed that the fibrin-monomer-stretch-induced protein unfolding only impacts the fibrin network and has no effect on blood cells. As a consequence, the evolution equation \cref{equ:evolution_equ_monomer_stretch_reduced} can be reduced to
\begin{equation}
	\label{equ:evolution_equ_monomer_stretch_reduced_reduced}
	\frac{H}{H_{fm}} \pqty{- \rho_{0S}^{S} \pdv{\psi_{eq}^{fib}}{\lambda_{fm}} - \rho_{0S}^{S} \pdv{\psi_{neq}^{fib}}{\lambda_{fm}} - \rho_{0S}^{S} \pdv{\psi_{fm}^{S}}{\lambda_{fm}}} = \pqty{\lambda_{fm}}'_{S}\,.
\end{equation}
The specific form of the involved terms can be derived by using \cref{equ:free_energy_fibrin_eq,equ:free_energy_fibrin_neq,equ:internal_energy_total,equ:free_energy_solid_parts}, which are given as
\begin{equation}
	\label{equ:evolution_equ_monomer_stretch_component}
	\begin{aligned}
		\rho_{0S}^{S} \pdv{\psi_{eq}^{fib}}{\lambda_{fm}} &= n^{fib} f^{fib} n_{0S}^{S} \mu^{fib} n \frac{1}{\lambda_{fm}} \Bqty{-\frac{\Lambda^{fib}}{\sqrt{n}} \beta^{fib} + \frac{\pqty{n_{0S}^{S}}^{-\frac{1}{3}}}{\sqrt{n} \lambda_{fm}} \beta_0^{fib} + \Bqty{\frac{\pqty{n_{0S}^{S}}^{-\frac{1}{3}}}{\sqrt{n} \lambda_{fm}} \beta_0^{fib} + \bqty{\frac{\pqty{n_{0S}^{S}}^{-\frac{1}{3}}}{\sqrt{n} \lambda_{fm}}}^2 \pdv{\beta_0^{fib}}{\bar{r}_{0}}} \frac{\ln J_{S}}{3} } \,, \\
		\rho_{0S}^{S} \pdv{\psi_{neq}^{fib}}{\lambda_{fm}} &= n^{fib} f^{fib} n_{0S}^{S} \mu^{e,fib} n \frac{1}{\lambda_{fm}} \Bqty{-\frac{\Lambda^{e,fib}}{\sqrt{n}} \beta^{e,fib} + \frac{\pqty{n_{0S}^{S}}^{-\frac{1}{3}}}{\sqrt{n} \lambda_{fm}} \beta_0^{fib} + \Bqty{\frac{\pqty{n_{0S}^{S}}^{-\frac{1}{3}}}{\sqrt{n} \lambda_{fm}} \beta_0^{fib} + \bqty{\frac{\pqty{n_{0S}^{S}}^{-\frac{1}{3}}}{\sqrt{n} \lambda_{fm}}}^2 \pdv{\beta_0^{fib}}{\bar{r}_{0}}} \frac{\ln J_{S}^{e,fib}}{3} } \,, \\
		\rho_{0S}^{S} \pdv{\psi_{fm}^{S}}{\lambda_{fm}} &= n^{fib} f^{fib} n_{0S}^{S} N_{t0}^{fib} n E_{fm} \frac{1}{\lambda_{fm}} \ln \lambda_{fm},
	\end{aligned} 
\end{equation}
with $\bar{r}_{0} \coloneqq \pqty{n_{0S}^{S}}^{-{1}/{3}} / \pqty{\sqrt{n} \lambda_{fm}}$. The fibrin monomer stretch $\lambda_{fm}$ can be obtained by integrating the evolution equation \cref{equ:evolution_equ_monomer_stretch_reduced_reduced}.
\par
\begin{figure}[H]
	\centering
	\subfloat[]{
		\begin{minipage}[b]{0.47\linewidth}
			\includegraphics[width=1\linewidth]{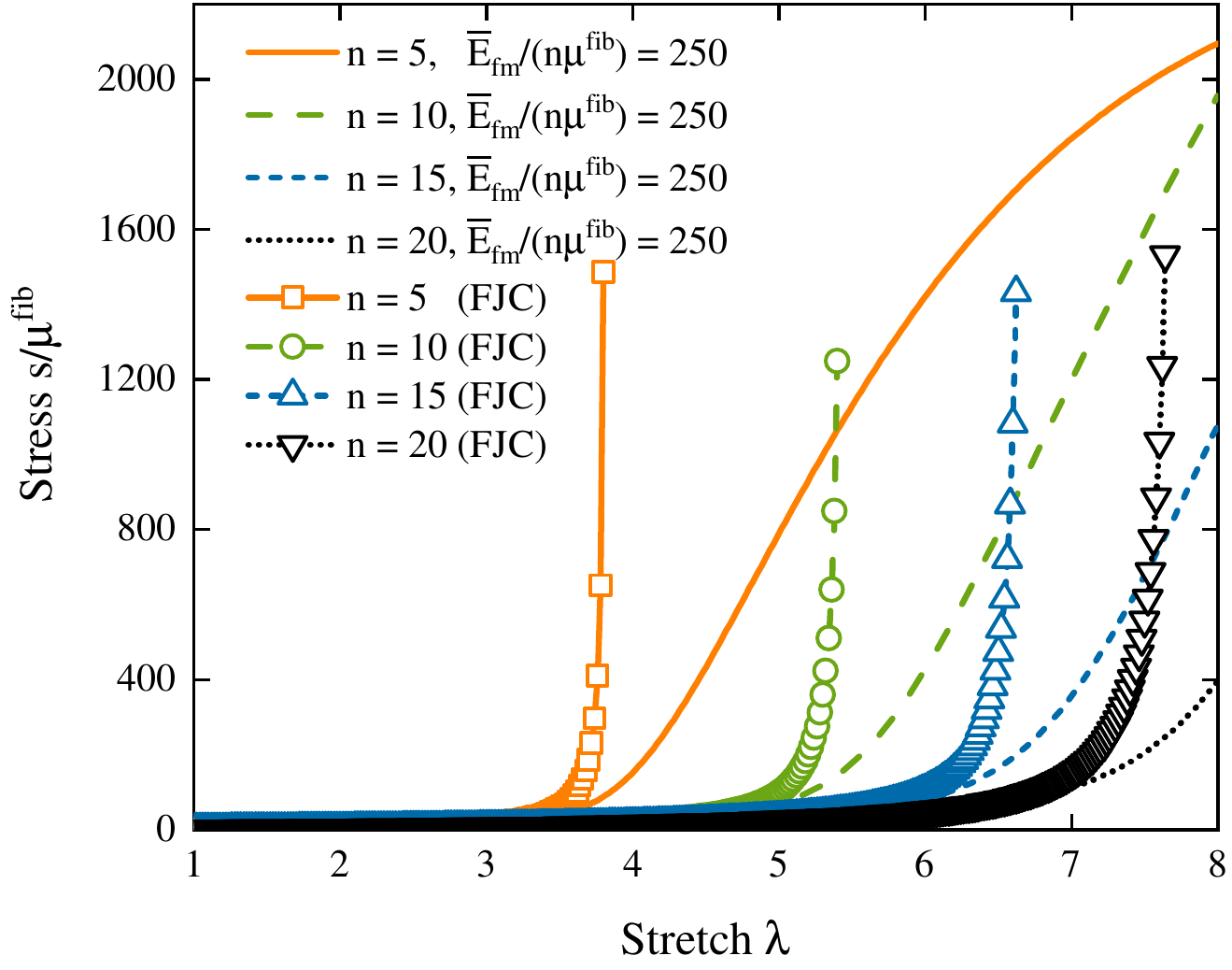} 
		\end{minipage}
		\label{}
	}
	\quad
	\subfloat[]{
		\begin{minipage}[b]{0.46\linewidth}
			\includegraphics[width=1\linewidth]{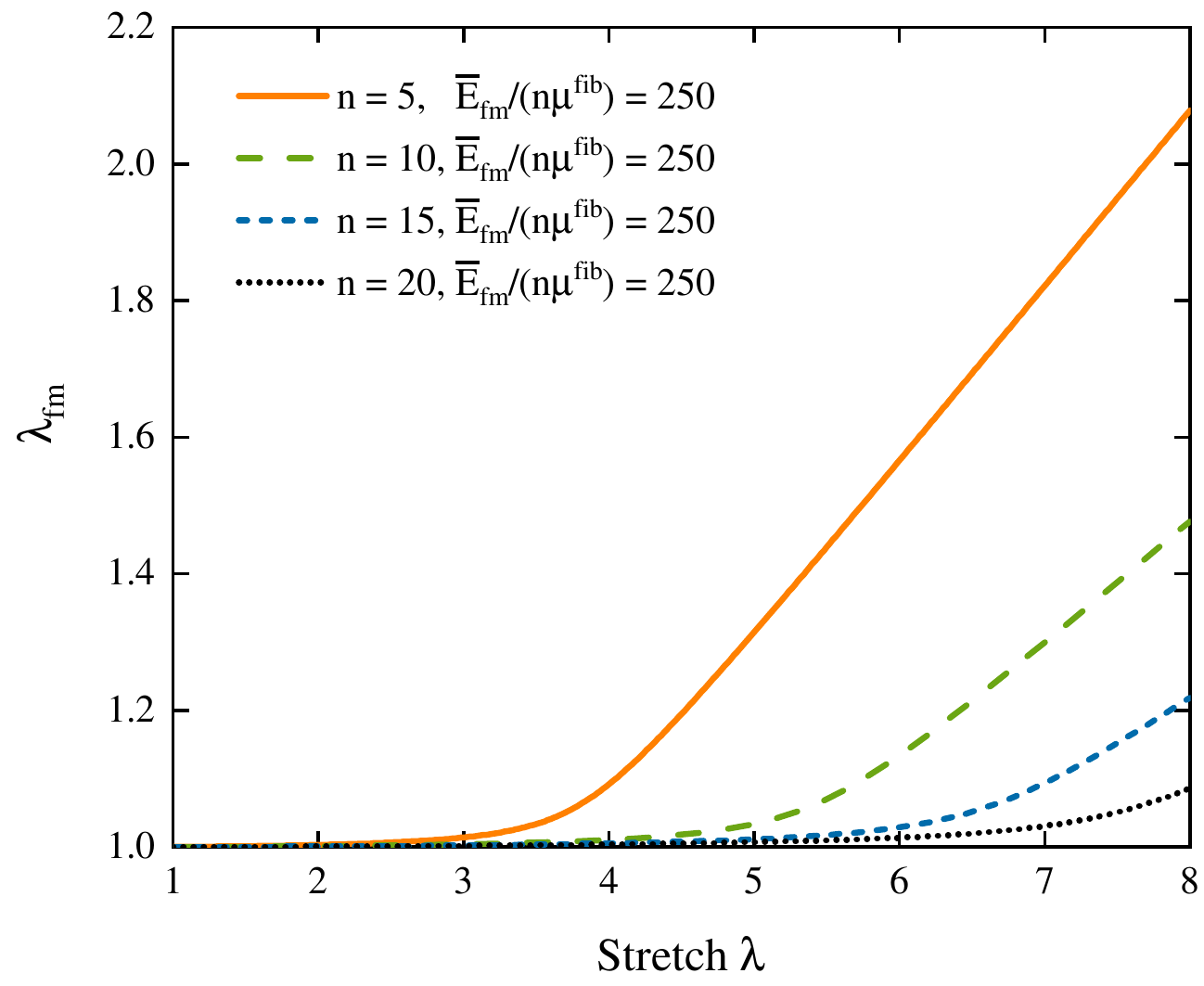}
		\end{minipage}
		\label{}
	}
	\caption{The effect of the number of fibrin monomer on (a) stress-stretch response and (b) monomer stretch. The corresponding lines with symbols in (a) denote the stresses calculated using the freely jointed chain (FJC) model.}
	\label{fig:7}
\end{figure}

Using this evolution law, the influence of the parameters ($n, \bar{E}_{fm} \coloneqq N_{t0}^{fib} n E_{fm}$) on the unfolding behavior is studied, and the stress-stretch response of the proposed model under uniaxial tension is compared with the classical freely jointed chain model. The influence of monomer number $n$ on the protein unfolding is shown in \Cref{fig:7}. When the fiber length approaches the fiber length limit $\sqrt{n}$, the fibrin monomers are stretched by $\lambda_{fm}$(\Cref{fig:7}b), which results from the protein molecule unfolding due to the change of molecular structure, see \Cref{fig:4}. The molecular unfolding leads to the softening of the fibrin fiber and avoids the nonphysical singularity in the classical freely jointed chain model where the freely jointed segments are assumed to be non-stretchable (\Cref{fig:7}a).
\par
\begin{figure}[H]
	\centering
	\subfloat[]{
		\begin{minipage}[b]{0.47\linewidth}
			\includegraphics[width=1\linewidth]{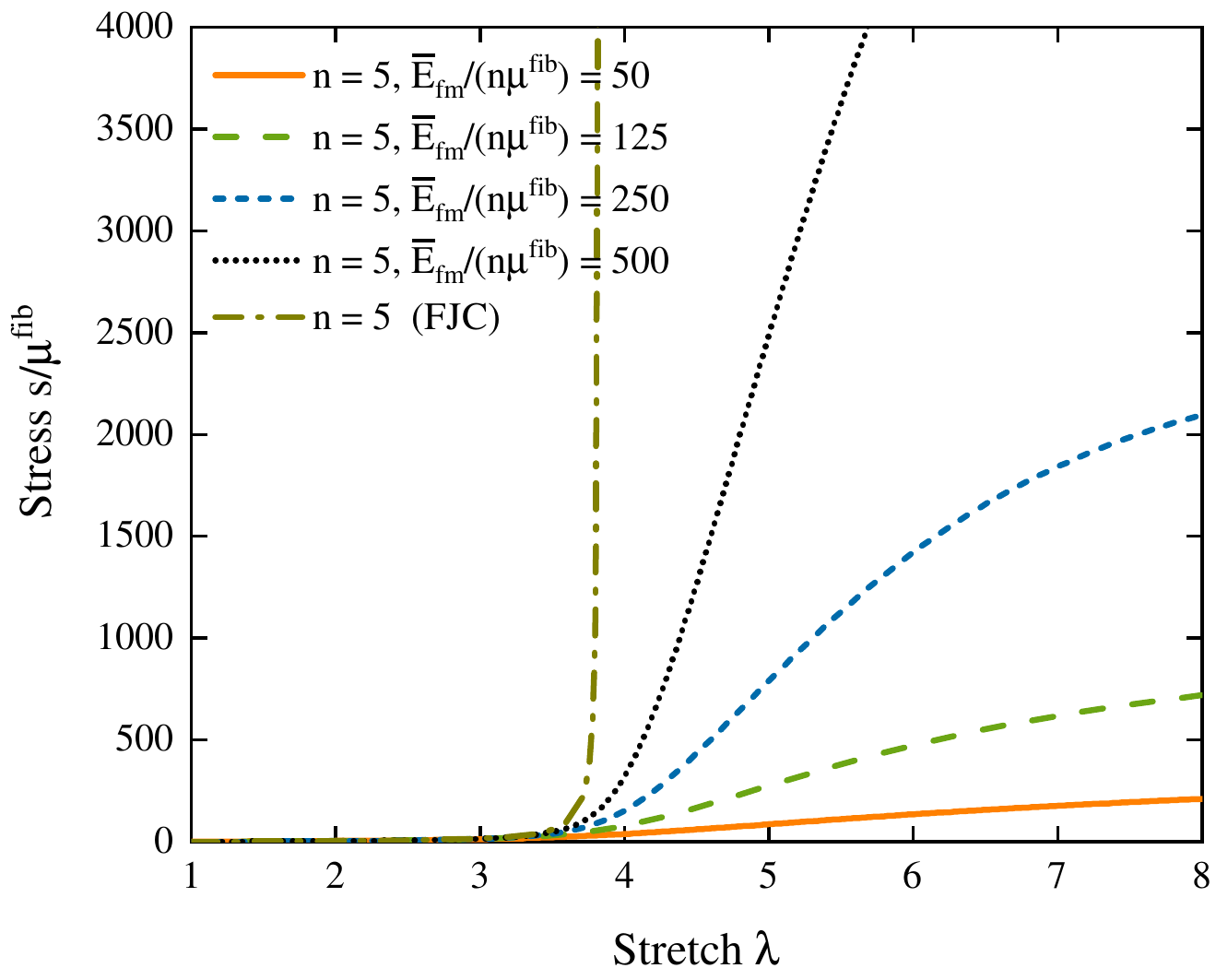} 
		\end{minipage}
		\label{}
	}
	\quad
	\subfloat[]{
		\begin{minipage}[b]{0.46\linewidth}
			\includegraphics[width=1\linewidth]{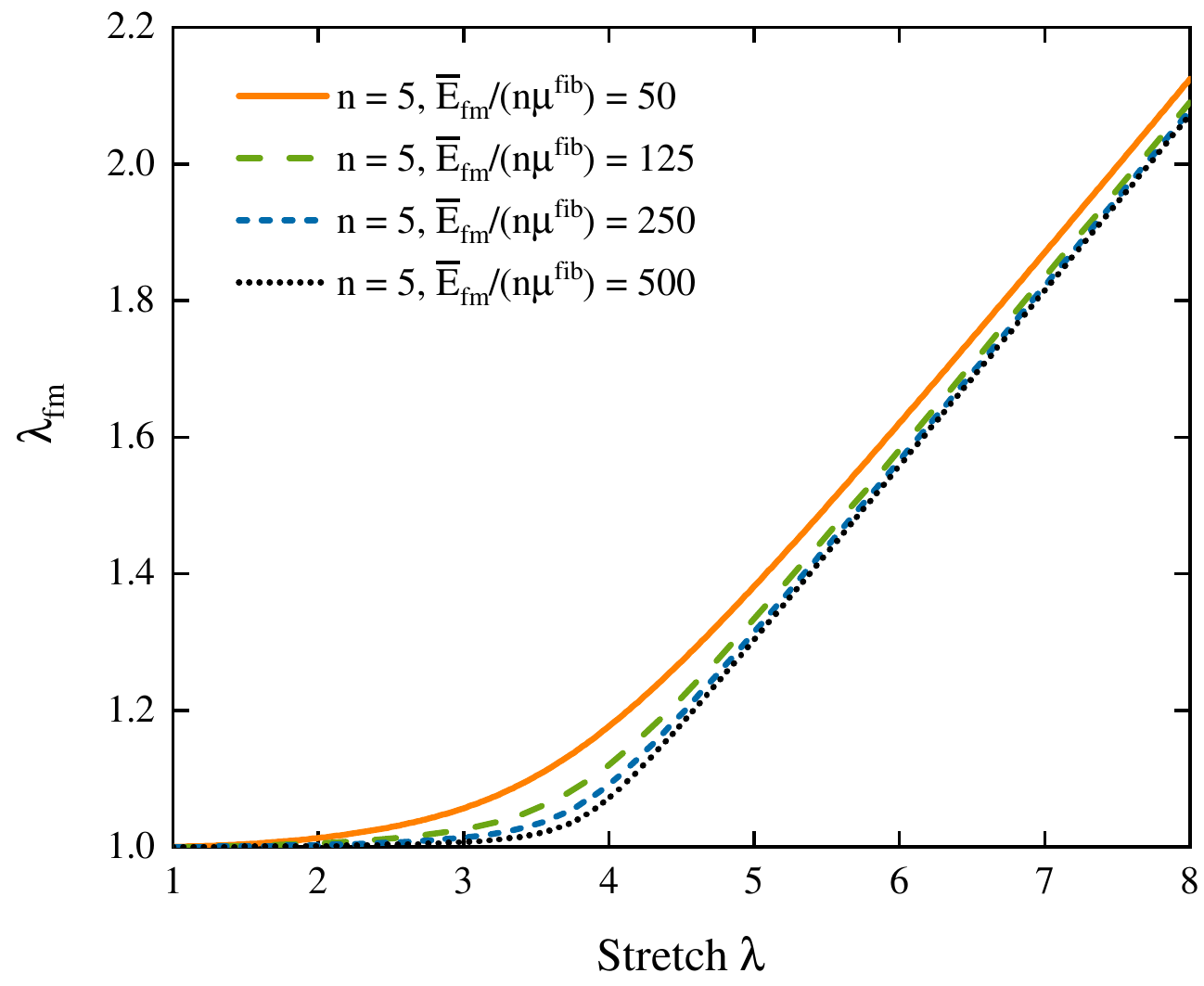}
		\end{minipage}
		\label{}
	}
	\caption{The effect of the fibrin monomer modulus on (a) stress-stretch response and (b) monomer stretch. The dash-dot line in (a) denotes the stresses calculated using the freely jointed chain model.}
	\label{fig:8}
\end{figure}

\Cref{fig:8} shows the impact of the fibrin monomer modulus on the fibrin unfolding behavior. Subject to the same total stretch, the stress increases with the stiffening of the fibrin monomer (\Cref{fig:8}a). A larger fibrin modulus results in a smaller fibrin monomer stretch $\lambda_{fm}$ because the fibrin monomer is more difficult to stretch (\Cref{fig:8}b). When the monomer modulus increases to infinity, the fibrin monomer is treated as a rigid segment and cannot be stretched. In this critical situation, the freely jointed chain model is recovered, where the stress is sufficiently large when stretch approaches $\sqrt{n}$ (see the dash-dot line in \Cref{fig:8}a).
\subsection{Constitutive relation for blood cells}
\label{sec:constitituve_model_blood_cell}
For a whole blood clot, the blood cells mainly include red blood cells and platelets. The contribution of the transient deformation of blood cells to the viscoelastic deformation of the whole blood clot is negligible compared with that of the fibrin networks \cite{puig2007viscoelasticity,liu2010mechanical}. Therefore, the contribution of the non-equilibrium free energy density of blood cells is ignored in this work, and the total free energy density of blood cells stems from the equilibrium part, \ie,
\begin{equation}
	\label{equ:free_energy_cell_eq}
	\psi_{eq}^{m} = n^{m} f^{m} \psi_{eq0}^{m} \,.
\end{equation}
The damage function $f^{m}$ describes the mechanical degradation of blood cells due to overloading, induced by shear stress, stretch and compression \cite{chernysh2020structure,leverett1972red}. Their damage evolution law under stretch and compression is assumed to be identical by adopting a unit step function $H_{TC}^{m}=1$. For simplicity, the parameter $\alpha^{m} = 1$ is used for $f^{m}$, thereby obtaining the damage function $f^{m} = 1 - \tilde{d}$.
\par
The Helmholtz free energy density for the blood cells is given as
\begin{equation}
	\label{equ:free_energy_cell_eq_damage_free}
	\psi_{eq0}^{m} = \frac{1}{\rho_{0S}^{S}} \Bqty{\frac{1}{2} \mu^{m} \pqty{I_{S1}-3} - \mu^{m} \ln J_{S} + \lambda_{cp}^{m} \bqty{\frac{1}{2}\pqty{\ln J_{S}}^2+\xi^{m}}} \,,
\end{equation}
where $\mu^{m}$ is the shear modulus of the blood cells. The compaction effect, namely the volumetric stiffening towards the compaction point, is incorporated by including the term $\xi^{m}$ and extending the Lam$\acute{e}$ constant $\lambda_{cp}^{m}$ as a function of $J_{cp}^{m}$, which are expressed as
\begin{equation}
	\label{equ:XI_lambda_cp_cell}
	\xi^{m} = \xi^{fib} \qq{and} \lambda_{cp}^{m}=\lambda_{l}^{m} \bqty{1+J_{cp} \pqty{1+\frac{J_{cp}^2}{1-J_{cp}}}}^{-1} \,.
\end{equation}
Combining \cref{equ:free_energy_cell_eq_damage_free,equ:free_energy_cell_eq,equ:effective_solid_stress_1}$_1$, the effective solid stress for blood cells is calculated as
\begin{align}
	\label{equ:Cauchy_effective_PG_comp}
	\vb{T}_{E,eq}^{m} = \frac{n^{m} f^{m}}{J_{S}} \bqty{\mu^{m} \vb{B}_{S}  -\mu^{m} \vb{I} + \lambda_{cp}^{m} \pqty{\ln J_{S}+\zeta^{m}} \vb{I}} \qq{with} \zeta^{m} = \zeta^{fib} \,.
\end{align}
\par
The involved parameters in the proposed model are summarized as,
\begin{itemize}
	\item constituent parameters:
	$\rho^{SR}, \rho^{FR}, n^{fib}, n_{0S}^{S}$;
	\item fibrin network parameters:
	$\mu^{fib}, \lambda_{l}^{fib}, \mu^{e,fib}, \lambda_{l}^{e,fib}, n, \phi^{fib}, \bar{E}_{fm} \coloneqq N_{t0}^{fib} n E_{fm}, H_{fm}, J_{cp}$;
	\item blood cell parameters:
	$\mu^{m}, \lambda_{l}^{m}$;
	\item non-local damage parameters:
	$c_{d0}/\beta_{d}, h, d_{cri}, b^{S}, Y_{0}^{fib}, Y_{0}^{m}, \varpi, \hbar,\hbar_{TC}^{fib}$.
\end{itemize}
Herein, $\rho^{SR}$ and $\rho^{FR}$ denote the realistic density of solid and fluid, respectively; $n^{fib}$ represents the volume fraction of fibrin network in the solid phase; $n_{0S}^{S}$ is the initial solid volume concentration in the solid reference configuration; $\mu^{fib}$ and $\mu^{e,fib}$ are the shear modulus of the fibrin network in the equilibrium and non-equilibrium states, respectively; $\lambda_{l}^{fib}$ and $\lambda_{l}^{e,fib}$ are the Lam$\acute{e}$ parameter of the fibrin network in the equilibrium and non-equilibrium states, respectively; $n$ denotes the number of fibrin monomer in a single fiber; $\phi^{fib}$ is the viscosity of the fibrin network; $\bar{E}_{fm}$ represents the modulus of the fibrin monomer for the fibrin network; $H_{fm}$ is the viscosity of the fibrin monomer stretch; $J_{cp}$ is the volume change of blood clots at the compaction point; $\mu^{m}$ and $\lambda_{l}^{m}$ are the shear modulus and Lam$\acute{e}$ parameter of blood cells,respectively; $c_{d0}/\beta_{d}$ is a normalized regularization parameter for the non-local quantity $\overline{Y}_{eqv}^{S}$; $h$ denotes a coupling modulus; $d_{cri}$ denotes a critical damage value; $b^{S}$ is the damage development parameter; $Y_{0}^{fib}$ and $Y_{0}^{m}$ are the damage threshold; $\varpi$ is a weighting parameter; $\hbar$ represents a model parameter in the equivalent damage driving force function; $\hbar_{TC}^{fib}$ controls the damage-induced tension-compression asymmetry of fibrin fibers. 
\section{Numerical simulations}
\label{sec:computational_example}
The proposed gradient-enhanced poro-visco-hyperelastic damage model is implemented in the commercial finite element program Abaqus/Standard (2021) by writing our own user material subroutine UMAT. The solution of the non-local variable $\overline{Y}_{eqv}^{S}$ is obtained by writing the user subroutine HETVAL. In this section, the mesh objectivity and the effect of the regularization parameter on damage bandwidth are first presented. The model is validated by comparing the numerical results with the fracture tests carried out on single-edge cracked blood clot specimens with different fibrin contents. Subsequently, the time-dependent clot fracture is investigated, and the mechanisms of the time-dependent fracture are analyzed. Finally, the mixed-mode fracture is demonstrated.
\subsection{Square plate with a central circular hole}
\label{sec:square-hole}
\begin{figure}[H]
	\centering
	\includegraphics[width=0.946\linewidth]{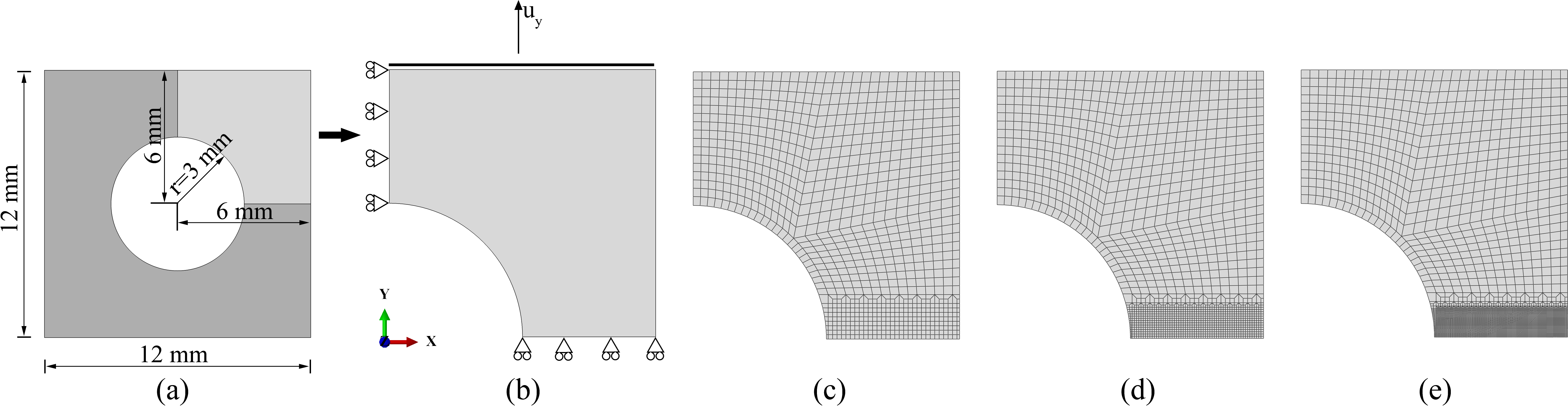}
	\caption{The geometry, boundary conditions and meshes of the square plate with a central circular hole: (a) geometry and dimensions of the plate in the initial configuration with a thickness of $6.5$ mm; (b) boundary conditions of one quarter of the plate; (c) coarse mesh with the smallest element size of $0.2$ mm; (d) moderate mesh with the smallest element size of $0.1$ mm and (e) fine mesh with the smallest element size of $0.05$ mm. The computational model is meshed with CPE4RPT elements. The damage parameters $b^{S} = 2.5$ and $c_{d0}/\beta_{d} = 0.02$ mm$^2$ are utilized to show the mesh insensitivity. Other model parameters of a whole blood clot with a middle fibrin content are adopted in the simulations (see \Cref{table:parameter_whole_blood_clot})}
	\label{fig:9}
\end{figure}

A square plate with a central circular hole is used to investigate the mesh objectivity of numerical simulations. The dimensions of the plate are shown in \Cref{fig:9}a. Due to symmetry, only one quarter of the plate is used in simulations (\Cref{fig:9}b). Symmetry boundary conditions are applied to the cross-sectional surfaces. The top surface is moved in the y-axis directions at a rate of $1.2$ mm/s. The external surface microforce $k$, associated with the non-local term $\overline{Y}_{eqv}^{S}$, is assumed to be zero. Three mesh densities are considered in simulations, as shown in \Cref{fig:9}c to e.
\par
\begin{figure}[H]
	\centering
	\includegraphics[width=0.73\linewidth]{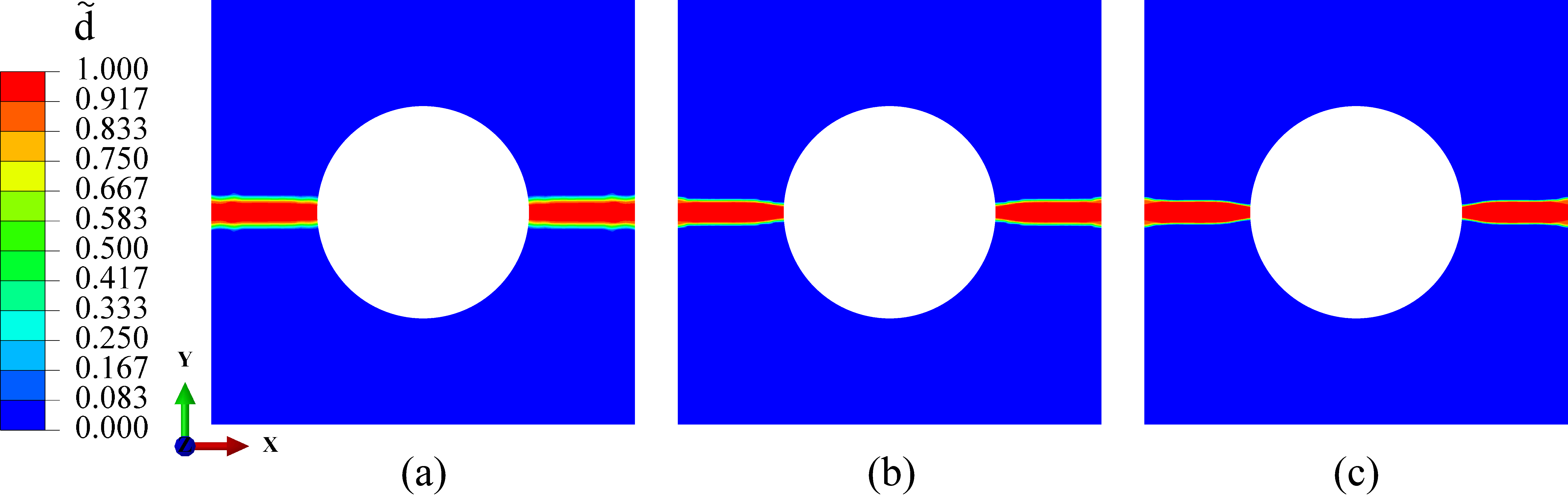}
	\caption{Fracture patterns of the plate with different mesh densities: (a) coarse; (b) medium and (c) fine mesh density.}
	\label{fig:10}
\end{figure}
\begin{figure}[H]
	\centering
	\subfloat[]{
		\begin{minipage}[b]{0.44\linewidth}
			\includegraphics[width=1\linewidth]{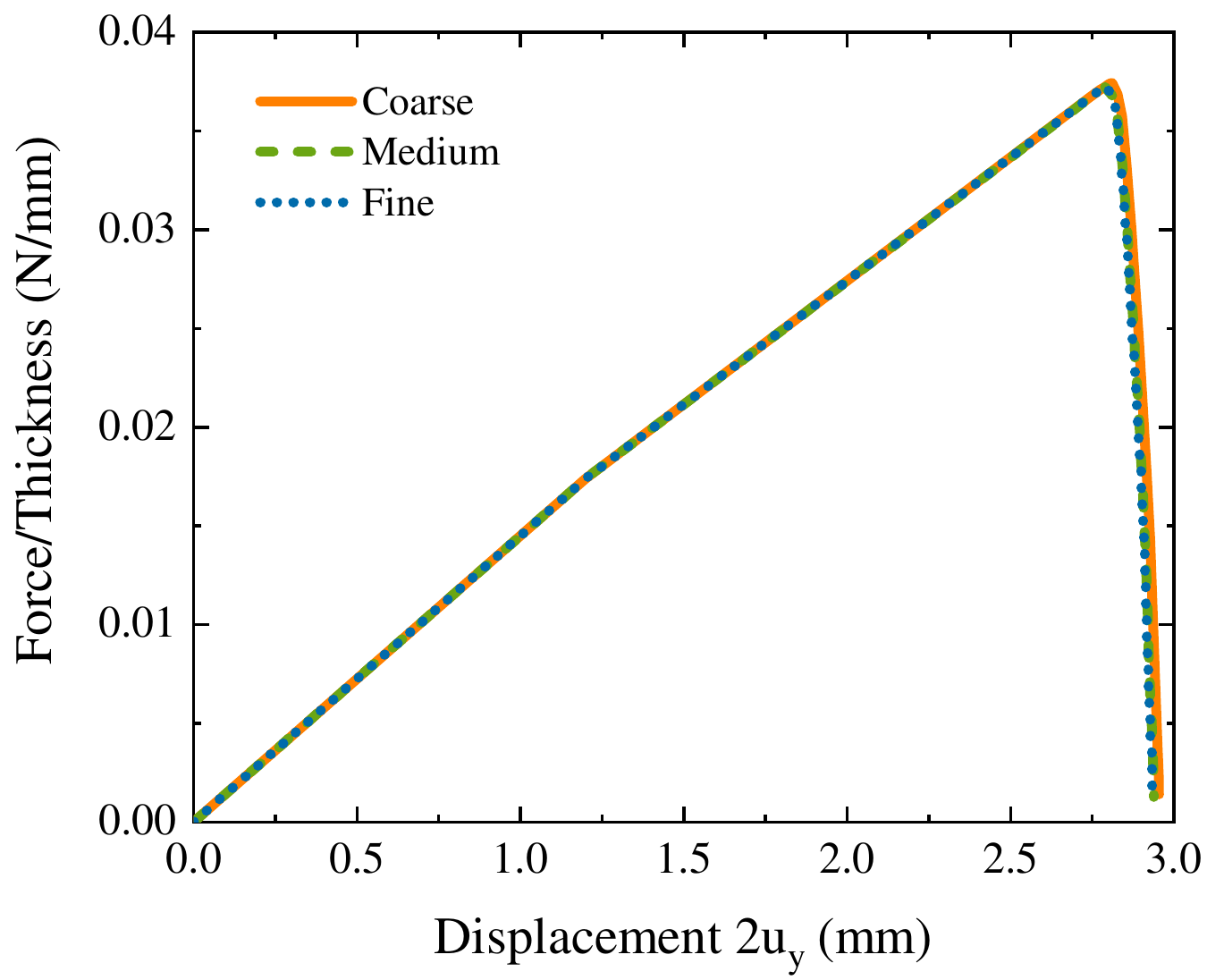} 
		\end{minipage}
		\label{}
	}
	\quad
	\subfloat[]{
		\begin{minipage}[b]{0.44\linewidth}
			\includegraphics[width=1\linewidth]{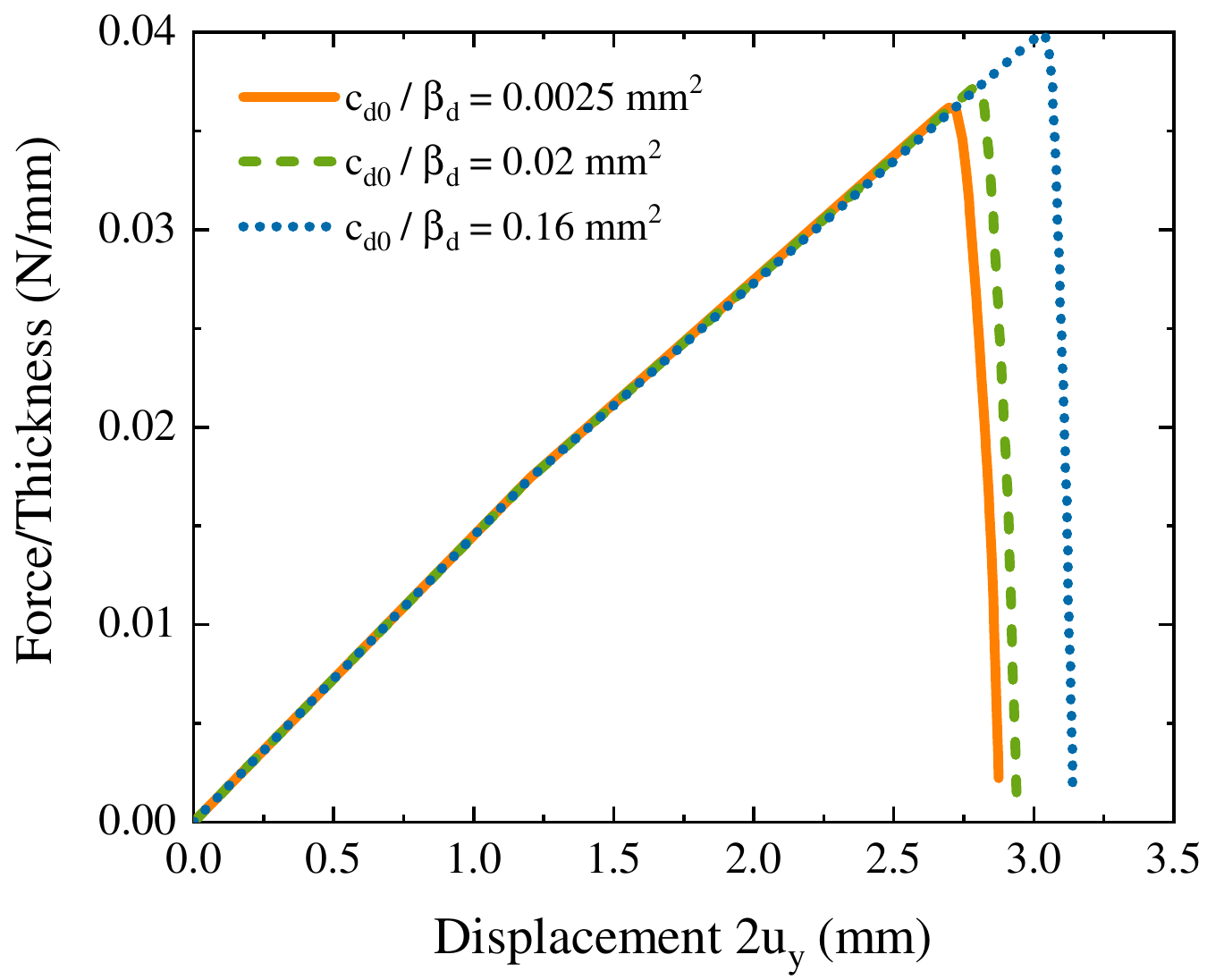}
		\end{minipage}
		\label{}
	}
	\caption{The force-displacement curves obtained from the plate model with (a) different mesh densities and (b) different values of $c_{d0}/\beta_{d}$.}
	\label{fig:11}
\end{figure}

The damage contours in the square plate with three different refinements are demonstrated in \Cref{fig:10}. The damage in the three cases locates at nearly the same region with a nearly constant damage zone bandwidth, which is independent of the mesh density. The mesh objectivity of the numerical simulations can also be evaluated by the resulting force-displacement relationships, as plotted in \Cref{fig:11}a, which shows that the difference of the reaction force in the simulations with different mesh densities is insignificant.
\begin{figure}[H]
	\centering
	\includegraphics[width=0.73\linewidth]{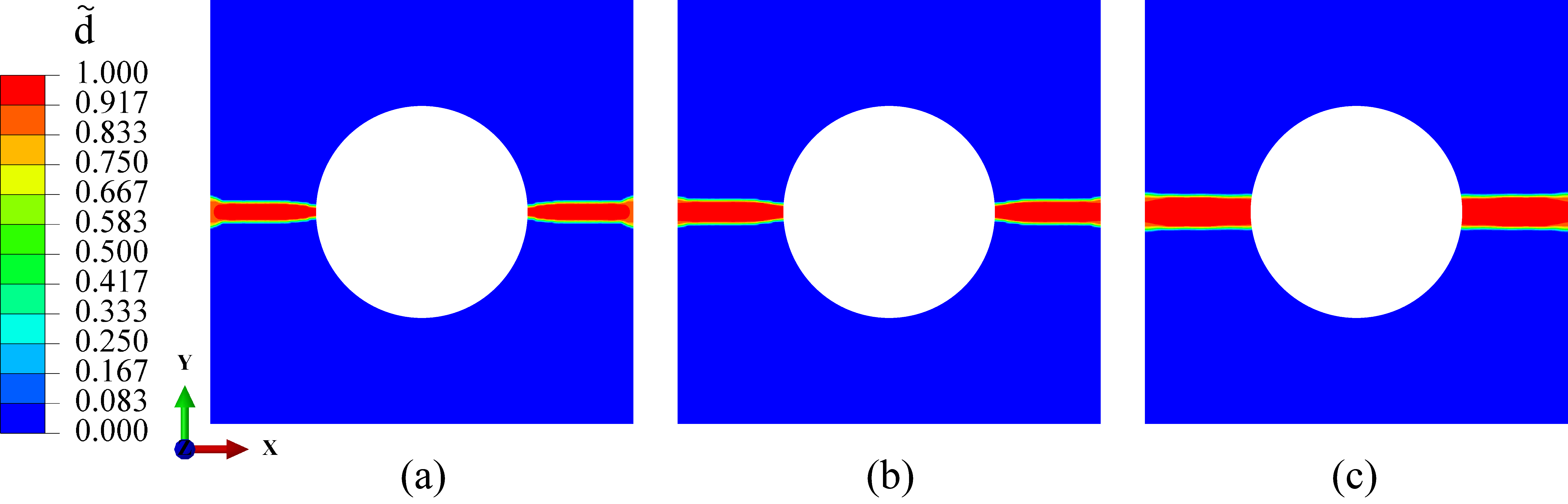}
	\caption{Fracture patterns of the plate with different values of $c_{d0}/\beta_{d}$: (a) $c_{d0}/\beta_{d} = 0.0025$ mm$^2$; (b) $c_{d0}/\beta_{d} = 0.02$ mm$^2$ and (c) $c_{d0}/\beta_{d} = 0.16$ mm$^2$. The moderate mesh density is used in these simulations.}
	\label{fig:12}
\end{figure}
\par
The influence of the regularization parameter on the damage zone bandwidth is investigated using the square plate model with the moderate mesh density. As shown in \Cref{fig:12}, the damage zone bandwidth broadens with the increase of $c_d/\beta_{d}$. The simulation results demonstrate the proportional relationship between the damage bandwidth and the regularization parameter. The corresponding force-displacement curves are presented in \Cref{fig:11}b. The maximum force and the corresponding critical displacement, where the force reaches the maximum value, increase with increasing $\sqrt{c_d/\beta_{d}}$.
\subsection{Single-edge cracked whole blood clots with different fibrin contents}
\label{sec:validation}
\begin{figure}[H]
	\centering
	\includegraphics[width=0.45\linewidth]{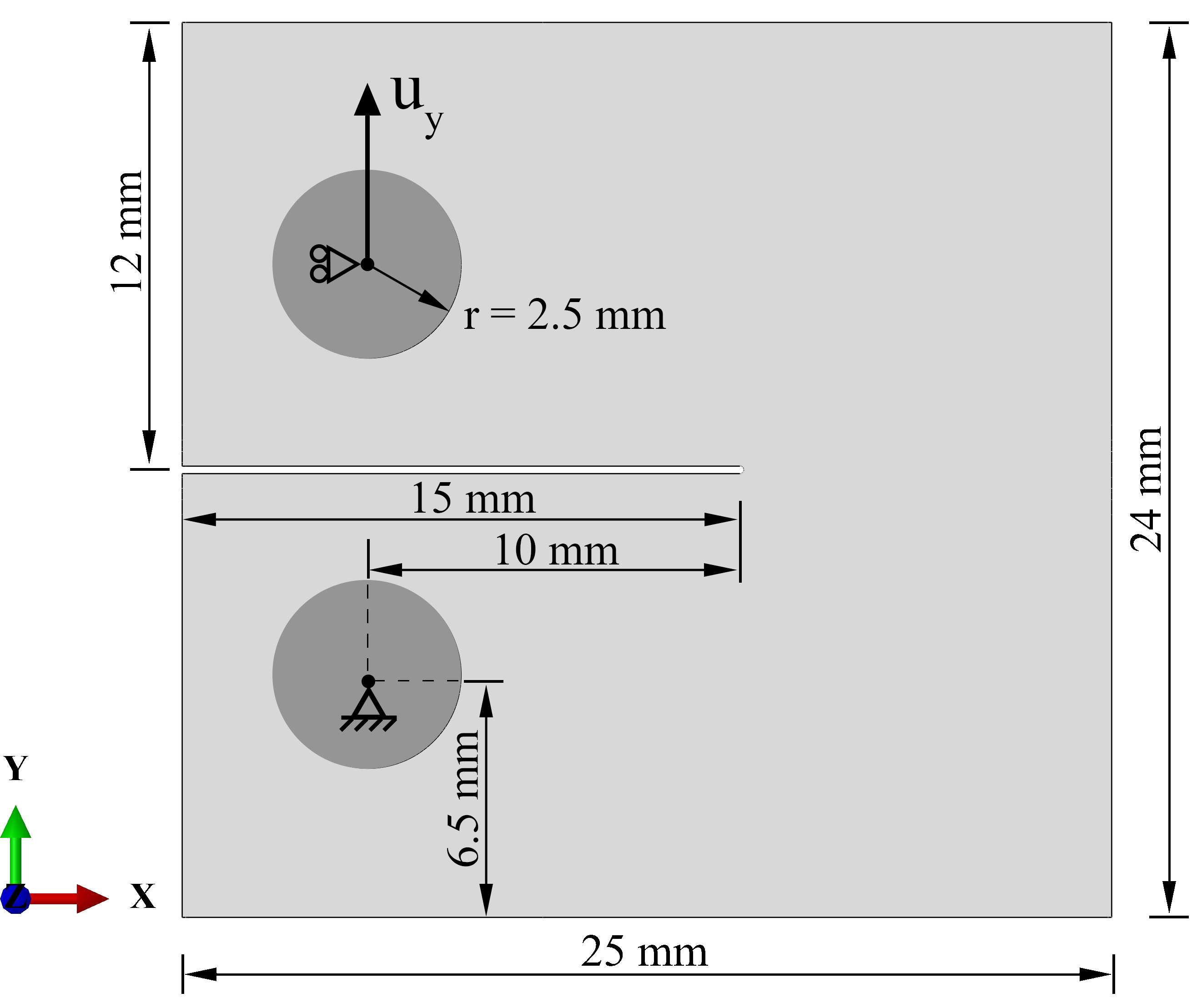}
	\caption{Geometry, dimensions and boundary conditions of the whole blood clot specimens. The top and bottom dark circles represent the load grip and fixture grip, respectively. The internal surfaces of the holes in the specimens contact with the external surface of the grips, and a frictionless contact between them is considered. The grips are modeled as rigid bodies. The fixture grip is fixed in all directions, and the load grip moves in the y-axis direction to stretch the specimens. The computational model is meshed with CPE4RPT elements. The smallest element size is 50 $\mu$m.}
	\label{fig:13}
\end{figure}

In this section, the deformation and fracture of whole blood clots with different fibrin contents are simulated and compared with experimental data to validate the proposed model. In simulations, single-edge cracked clot specimens are used. The geometry, dimensions and boundary conditions of the computational model are shown in \Cref{fig:13}. The thickness of the specimens is 6.5 mm. Plane-strain condition is assumed in the simulations \cite{fereidoonnezhad2021blood}. The top load grip moves in the y-axis direction at a loading rate of 10 mm/min.
\par
In the whole blood clots, the solid skeleton is composed of a fibrin network and blood cells \cite{fereidoonnezhad2021blood}. The composition parameters ($\rho^{SR}, \rho^{FR}, n^{fib}, n_{0S}^{S}$) are obtained from experimental measurement and histological analysis of blood clots \cite{wufsus2013hydraulic,fereidoonnezhad2021blood}. The fibrin network parameters
($\mu^{fib}, \lambda_{l}^{fib}, \mu^{e,fib}, \lambda_{l}^{e,fib}, n, \phi^{fib}, \bar{E}_{fm}, H_{fm}, J_{cp}$) and the blood cell parameters ($\mu^{m}, \lambda_{l}^{m}$) are obtained based on the work by \cite{fereidoonnezhad2021blood,he2022viscoporoelasticity,rausch2021hyper,yesudasan2020multiscale}. The damage parameters ($c_{d0}, \beta_{d}, h, d_{cri}, b^{S}, Y_{0}^{fib}, Y_{0}^{m}, \varpi, \hbar, \hbar_{TC}^{fib}$) are calculated and fitted according to the work by \cite{liu2021fracture,fereidoonnezhad2021blood}. The parameters are summarized in \Cref{table:parameter_whole_blood_clot}.
\begin{table}[H]
	\centering
	\captionsetup{width=1\textwidth}
	\caption{Material parameters of the whole blood clots in the simulations.}
	\label{table:parameter_whole_blood_clot}
	\setlength\tabcolsep{9.5pt}
	\footnotesize
	\renewcommand{\arraystretch}{1}
	\begin{threeparttable}
		\begin{tabular}{ c c c c c c c c } 
			\toprule
			$\rho^{SR}$ (g/cm$^3$) & $\rho^{FR}$ (g/ml) & $n^{fib}$ (-) & $n_{0S}^{S}$ (-) & $\mu^{fib}$ (MPa) & $\lambda_{l}^{fib}$ (MPa) & $\mu^{e,fib}$ (MPa) & $\lambda_{l}^{e,fib}$ (MPa)\\
			\midrule 
			1.4 & 1.0 & 0.643, 0.35, 0.15 & 0.526 & 0.0123 & 0.0185 & 0.0101 & 0.0151 \\
			\bottomrule
			\toprule
			$n$ (-) & $\phi^{fib}$ (MPa$\cdot$s) & $\bar{E}_{fm}/(n \mu^{fib})$ (-) & $H_{fm}$ (-) & $J_{cp}$ (-) & $\mu^{m}$ (MPa) & $\lambda_{l}^{m}$ (MPa) &  $c_{d0}/\beta_{d}$ (mm$^2$)  \\
			\midrule 
			216 & 0.1058 & 3 & 0.01 & 0.526 & 0.0003077 & 0.0004615 & 0.0289\\
			\bottomrule
			\toprule
			$h$ (-) & $d_{cri}$ (-) & $b^{S}$ (-) & $Y_{0}^{fib}$ (MPa) & $Y_{0}^{m}$ (MPa) & $\varpi$ (-) & $\hbar$ (-) & $\hbar_{TC}^{fib}$ (-) \\ 
			\midrule
			1 $\times$ 10$^{-10}$ & 0.8 & 0.07 & 0.0020 & 0.00025 & 0.2 & 1.1 & 0\\
			\bottomrule
		\end{tabular}
	\end{threeparttable}
\end{table}
\par
\begin{figure}[H]
	\centering
	\includegraphics[width=0.5\linewidth]{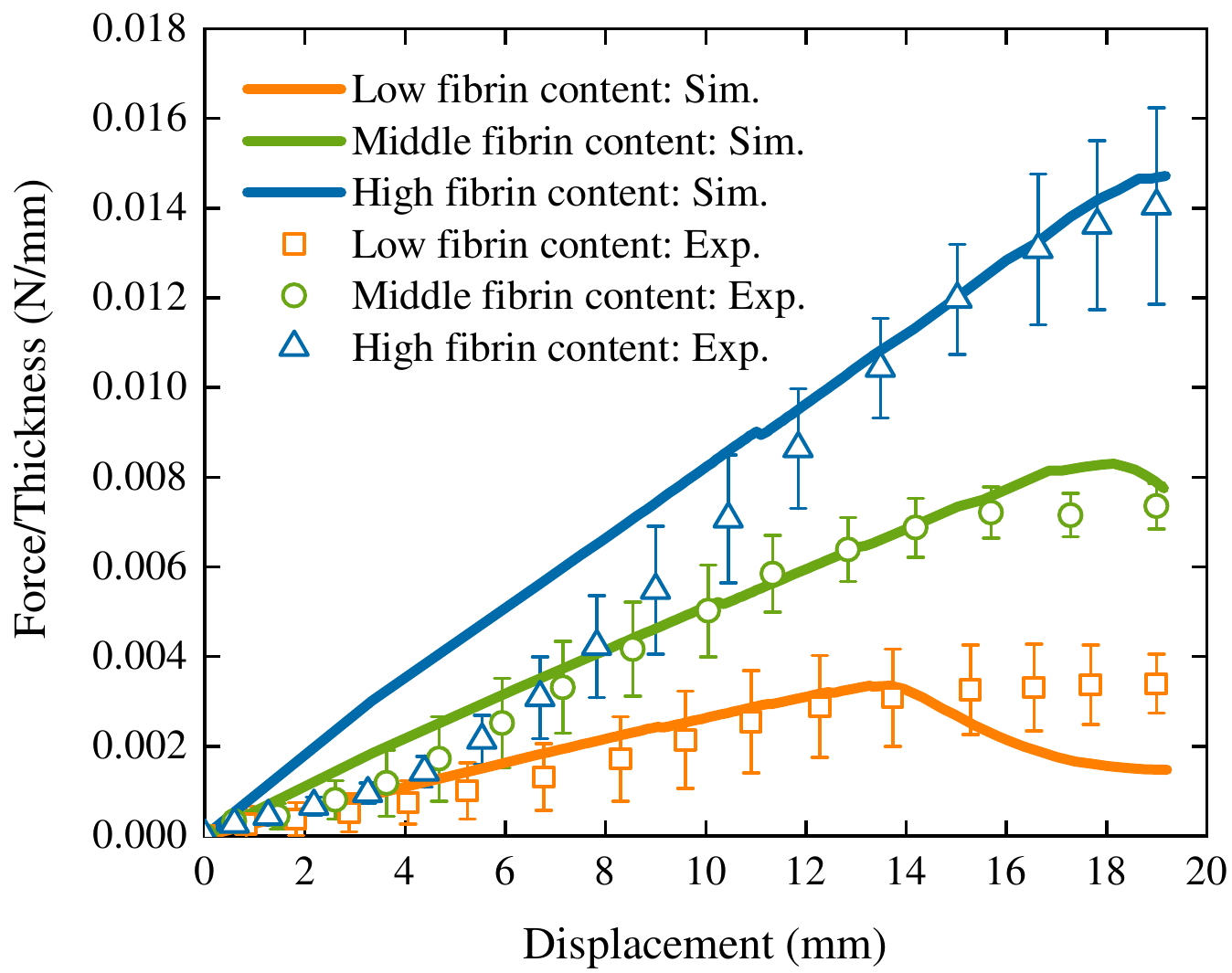}
	\caption{Comparison of the force-displacement relationships between simulation results and experimental data from fracture tests using single-edge crack specimens of whole blood clots with different compositions. The fibrin content denotes the fibrin volume fraction $n^{fib}$.}
	\label{fig:14}
\end{figure}
\begin{figure}[H]
	\centering
	\includegraphics[width=1\linewidth]{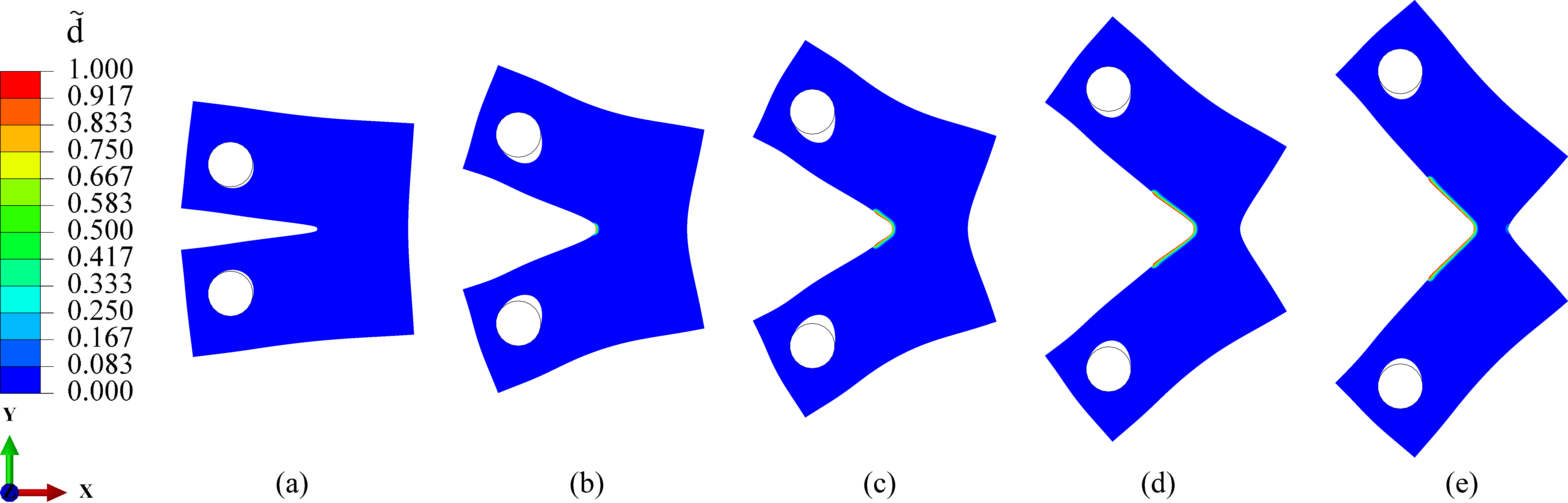}
	\caption{Crack patterns in the whole blood clot specimen with low fibrin content under different stretch ratios: (a) $u_{y} = 3.33$ mm; (b) $u_{y} = 10.0$ mm; (c) $u_{y} = 15.0$ mm; (d) $u_{y} = 20.0$ mm and (e) $u_{y} = 24.0$ mm.}
	\label{fig:15}
\end{figure}
\begin{figure}[H]
	\centering
	\includegraphics[width=1\linewidth]{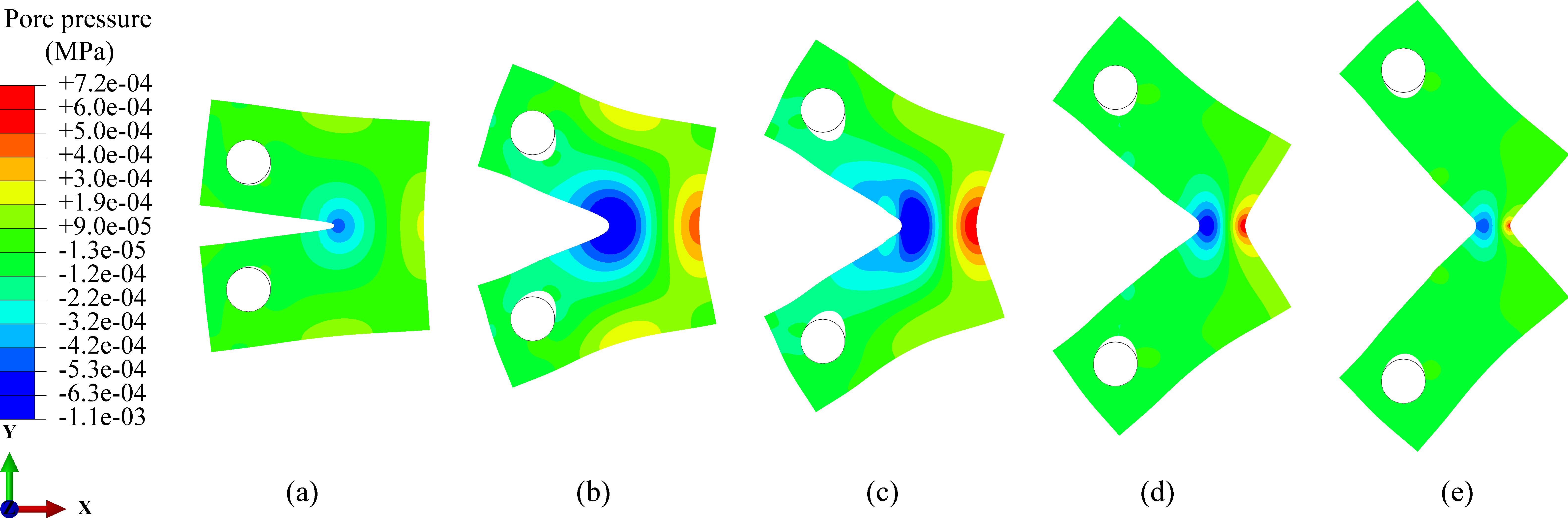}
	\caption{Distribution of pore pressure in the whole blood clot specimen with a low fibrin content under different stretch ratios: (a) $u_{y} = 3.33$ mm; (b) $u_{y} = 10.0$ mm; (c) $u_{y} = 15.0$ mm; (d) $u_{y} = 20.0$ mm and (e) $u_{y} = 24.0$ mm.}
	\label{fig:16}
\end{figure}
\begin{figure}[H]
	\centering
	\includegraphics[width=0.7\linewidth]{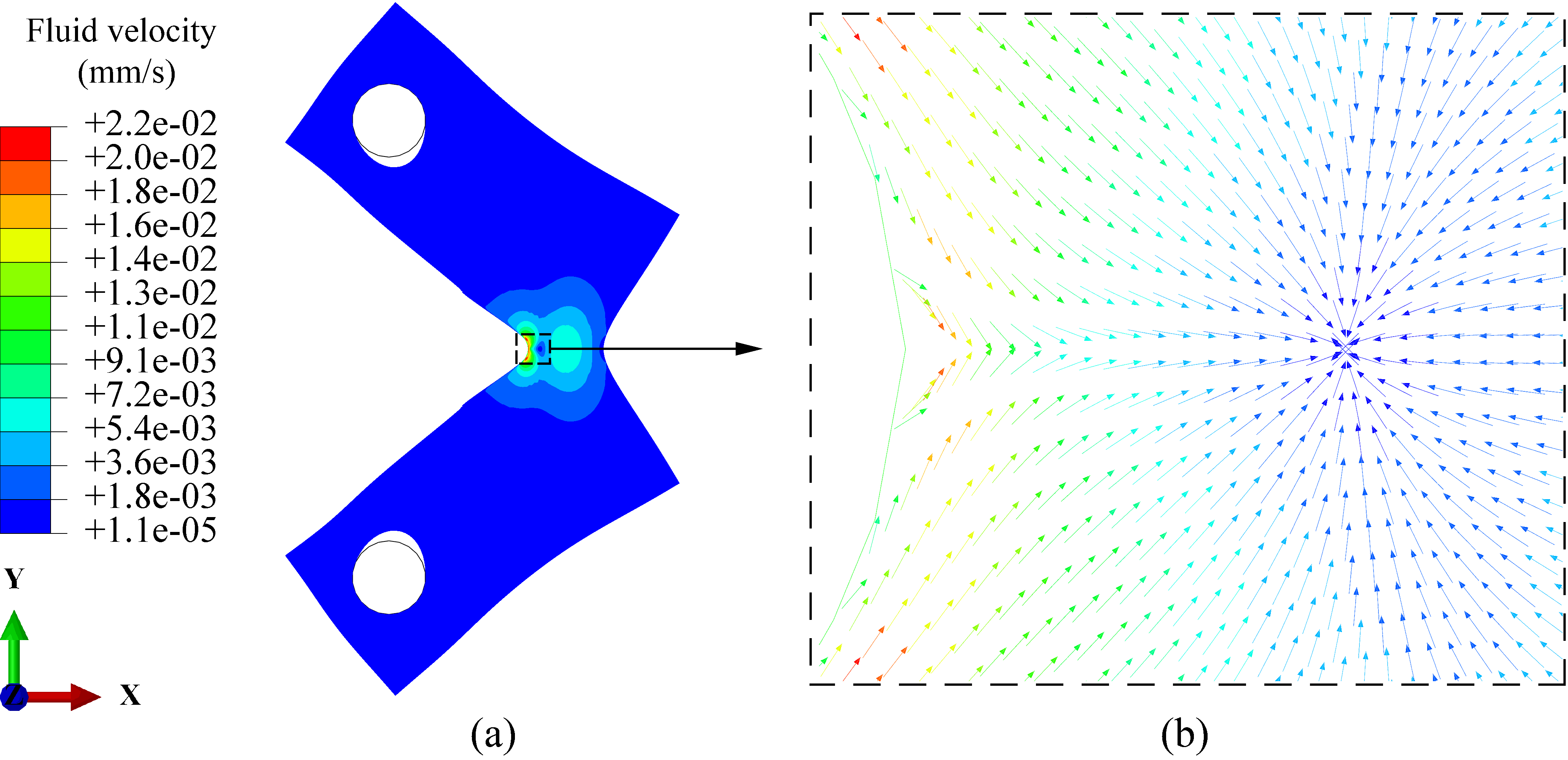}
	\caption{The velocity of fluid flow in the blood clot under the displacement of $u_{y} = 20.0$ mm: (a) field distribution and (b) a detailed view of the direction and magnitude of fluid flow velocity at the crack tip region.}
	\label{fig:17}
\end{figure}
\begin{figure}[H]
	\centering
	\includegraphics[width=1\linewidth]{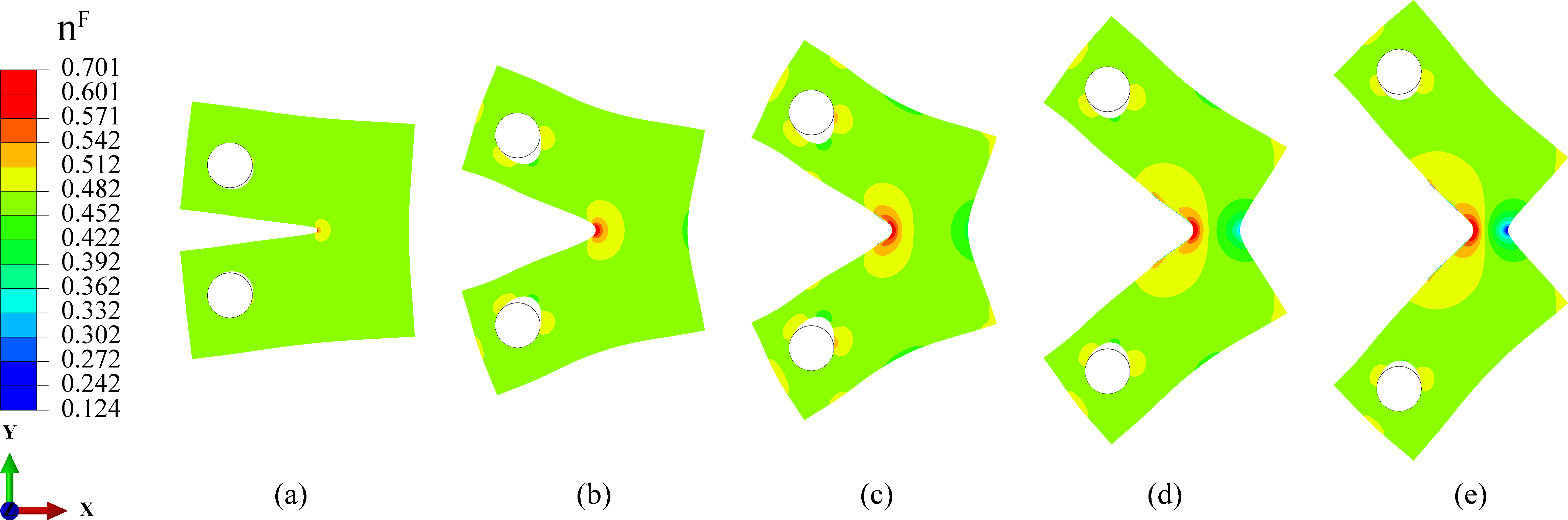}
	\caption{Distributions of fluid volume fraction in the whole blood clot specimen with a low fibrin content under different stretch ratios: (a) $u_{y} = 3.33$ mm; (b) $u_{y} = 10.0$ mm; (c) $u_{y} = 15.0$ mm; (d) $u_{y} = 20.0$ mm and (e) $u_{y} = 24.0$ mm.}
	\label{fig:18}
\end{figure}

\Cref{fig:14} illustrates the force-displacement relationship of the whole blood clot specimens with different compositions obtained from simulations and experimental data \cite{fereidoonnezhad2021blood}. The fibrin fiber content has a significant impact on the deformation and damage of blood clots. Since the fibrin network is significantly stiffer than blood cells \cite{fereidoonnezhad2021blood}, the fibrin network dominates the deformation and damage resistance of a whole blood clot. As shown in \Cref{fig:14}, a clot specimen with a higher fibrin content demonstrates a larger reaction force under the same stretch. Furthermore, the fibrin network enhances the load-bearing capacity and fracture resistance as the strength and critical stretch in a clot with higher fibrin content are obviously greater.
To demonstrate the fracture procedure, the crack propagation of the clot with low fibrin content under five stretch levels is presented in \Cref{fig:15}. In \Cref{fig:15}$a$, the notch is opened to a small angle, but no damage has initiated. When the specimen is stretched further, damage initiates and accumulates at the crack tip until it reaches a critical level, after which fracture initiates, and the crack propagates (see \Cref{fig:15}b to e). The fracture patterns of the specimen are identical to those of experimental observations \cite{fereidoonnezhad2021blood}. The simulation results show that both the force-displacement curves and fracture patterns agree well with the experimental data and observations, indicating the capacity of the proposed model to capture and predict the fracture behavior of whole blood clots with different fibrin/blood cell ratios.
Due to the biphasic property, the stress in the fibrin network and blood cells, and the pore pressure of interstitial fluid concurrently resist external loading. Subject to high stretches, the solid skeleton undergoes locally non-linear and non-uniform deformation, leading to the inhomogeneous distribution of pore pressure. \Cref{fig:16} shows the pore pressure distributions in the same clot specimen and under the same stretch levels as those in \Cref{fig:15}. Subject to a relatively smaller stretch (\Cref{fig:16}b), the magnitude of the pore pressure at the crack tip region is lower, therefore bearing more external loading. With the further stretch, stress in the fibrin network at the crack tip increases and resists more load, which leads to increased pore pressure. Furthermore, the pore pressure gradient drives the flow of fluid through the fibrin network (\Cref{fig:17}). With the increase of loading time, more fluid migrates to the crack tip region, resulting in the decline of the pore pressure gradient. Therefore, the pore pressure at the crack tip region first increases with the increasing stretch and decreases after the displacement is higher than 15 mm (see \Cref{fig:16}).
In the initial state, fluid in a blood clot distributes uniformly. Upon loading, the locally non-uniform deformation causes the generation of a pore pressure gradient, which drives fluid transport towards the crack tip region. The distribution evolution of the fluid volume fraction in the clot specimen is illustrated in \Cref{fig:18}. At an early loading stage, the fluid content at the crack tip becomes higher than that of the remaining region (\Cref{fig:18}a). The non-uniform distribution of fluid evolves into a more prominent stage with further stretch as fluid continues to flow to the crack tip and deposits at this region (\Cref{fig:18}b to e).  
\subsection{Single-edge cracked whole blood clots under different loading rates}
\label{sec:fracture_multi_loading_rate}
\par
Recent experimental studies have shown the time-dependent deformation and fracture characteristics of blood clots \cite{he2022viscoporoelasticity,ghezelbash2022blood,Liu2024}, which concluded that the time-dependent deformation was associated with the viscoelastisity of the solid skeleton, fluid transport through the solid skeleton and the solid/fluid interactions. In this section, the time-dependent deformation and fracture of a whole blood clot specimen with middle fibrin content are simulated, and the underlying mechanisms of the time-dependent behaviors are analyzed in detail.
\par
The geometry, boundary conditions and material parameters of the whole blood clot specimen with a middle fibrin content, as described in \Cref{sec:validation}, are used in the simulations. Since the two time-dependent procedures, \ie, viscoelasticity and fluid transport, have their own timescales, they are sensitive to loading rates. By stretching the blood clot at different rates, the above two procedures can be separated, and their contributions to the time-dependent fracture can be observed and analyzed individually. In simulations, six different stretch rates are considered, \ie, $1/3$ mm/s, $1/6$ mm/s, $1/20$ mm/s, $1/40$ mm/s, $1/80$ mm/s, and $1/200$ mm/s.
\par
\begin{figure}[H]
	\centering
	\subfloat[]{
		\begin{minipage}[b]{0.43\linewidth}
			\includegraphics[width=1\linewidth]{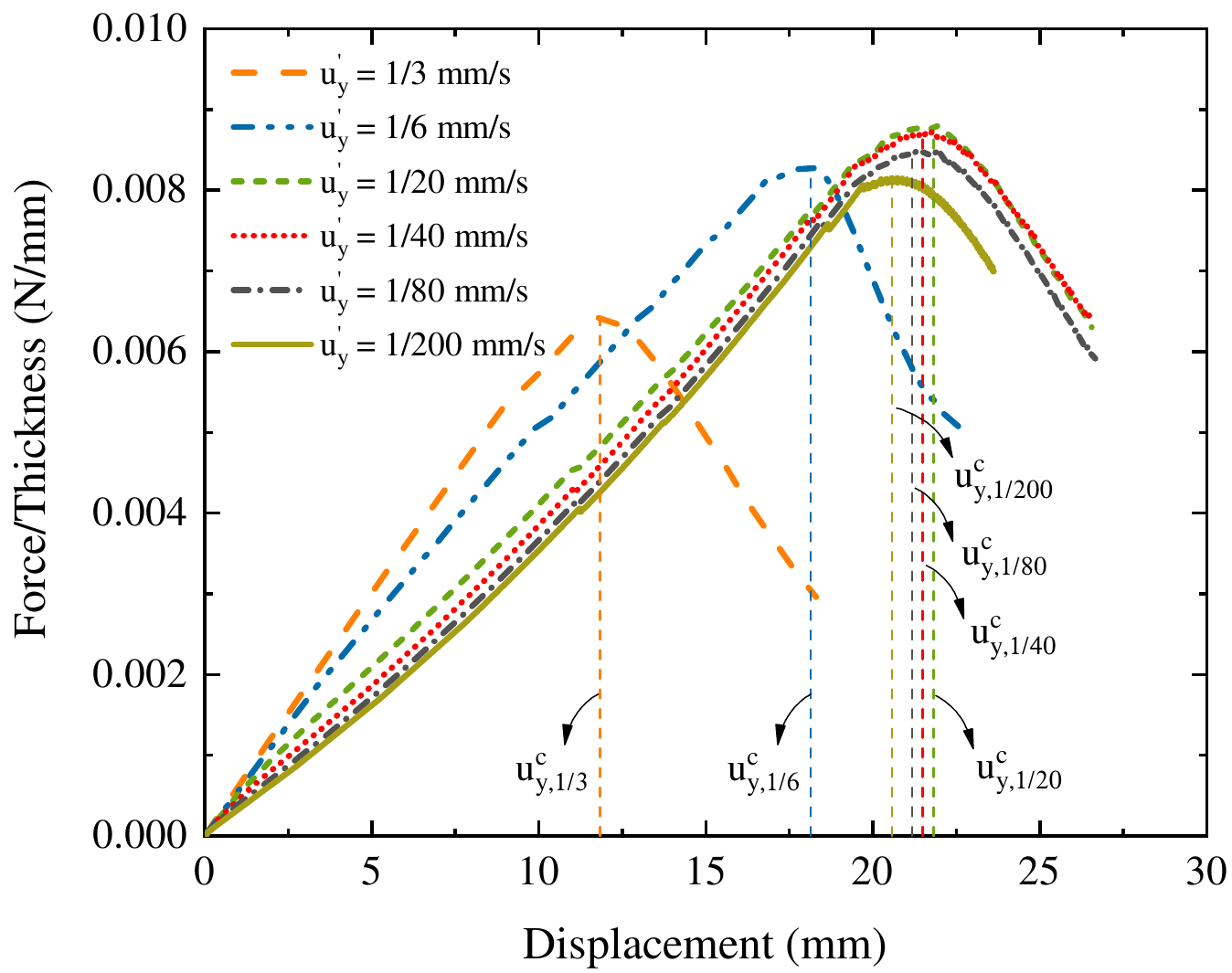} 
		\end{minipage}
		\label{}
	}
	\quad
	\subfloat[]{
		\begin{minipage}[b]{0.47\linewidth}
			\includegraphics[width=1\linewidth]{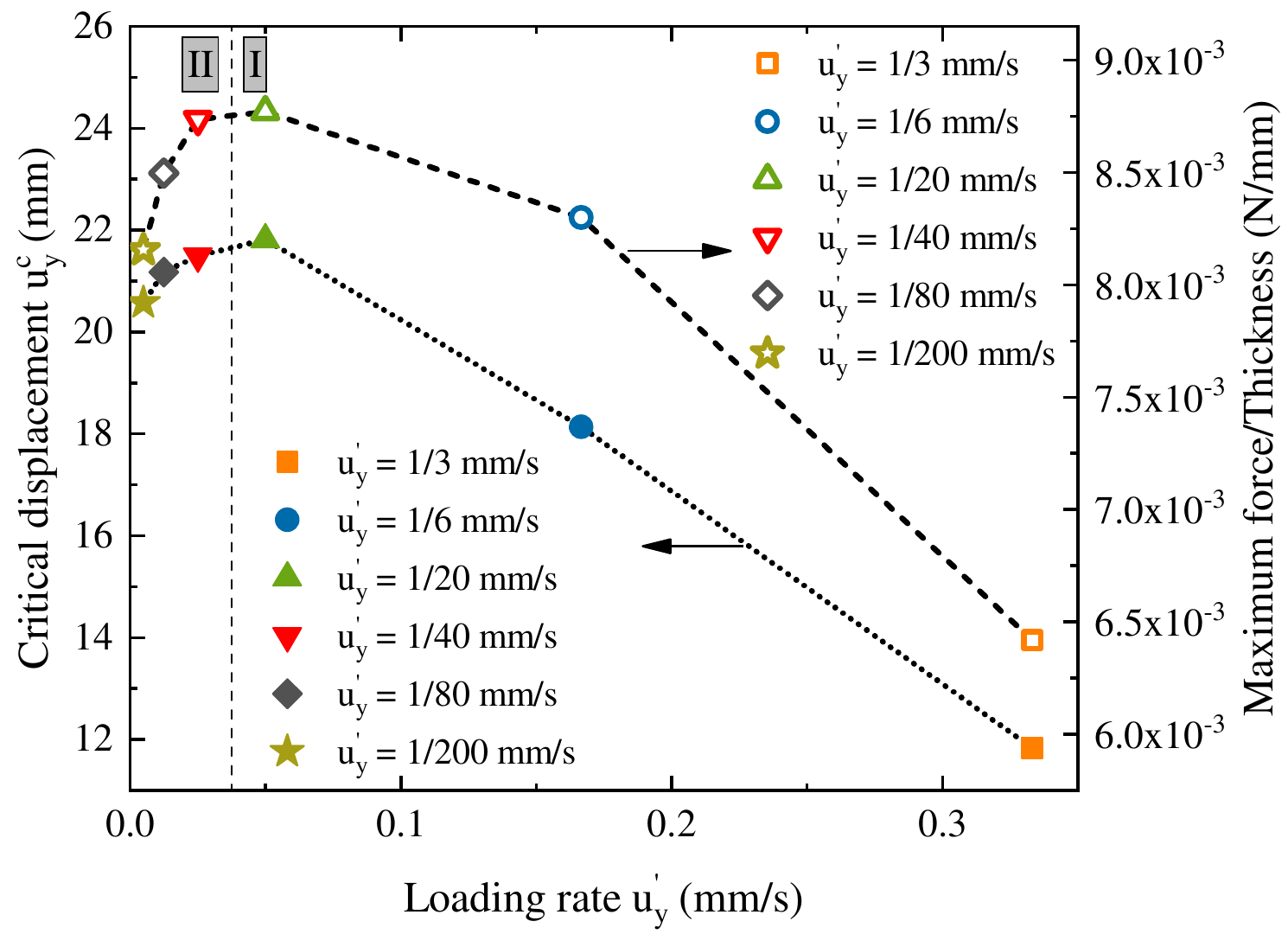}
		\end{minipage}
		\label{}
	}
	\caption{The predicted (a) force-displacement curves obtained from the blood clot with middle fibrin content subject to different loading rates and (b) corresponding critical displacement and maximum force at the selected loading rates.}
	\label{fig:19}
\end{figure}

 
\Cref{fig:19}a plots the resulting force-displacement curves, which show significant time dependency. First, the deformation of the blood clots is time-dependent. When a blood clot is stretched at a high loading rate, the viscoelastic stress relaxation and fluid migration are limited. The viscoelastic stress and pore pressure can effectively contribute to the resistance of external loading applied to the clot. Therefore, the stiffness of the force-displacement curve increases when the tissue deforms at a higher rate, see \Cref{fig:19}a.
Furthermore, the damage and fracture of the specimens demonstrate time dependency. Interestingly, the variation trend of the maximum force and the critical displacement $u_{y}^{c}$ exhibit an obvious switch at the loading rate of $1/40$ mm/s. The critical displacement $u_{y}^{c}$, which is marked in \Cref{fig:19}a, is defined as the displacement value at which the force reaches its maximum magnitude. The maximum forces and $u_{y}^{c}$ at the selected loading rates are plotted in \Cref{fig:19}b. For convenience, two stages are defined according to this switch point: the loading rates $1/3$ mm/s, $1/6$ mm/s and $1/20$ mm/s belong to stage I; the loading rates $1/40$ mm/s, $1/80$ mm/s, and $1/200$ mm/s are within stage II, as marked in \Cref{fig:19}b. In stage I, the maximum force and critical displacement increase with the reduction of the loading rate, while in stage II, their variations with loading rates demonstrate an opposite trend. The dissipation of the non-equilibrium energy in stage I, which is associated with fibrin viscoelasticity (see \Cref{fig:2}), is relatively low at higher loading rates. This leads to more energy storage in the clot specimens and thereby triggers the onset of damage and fracture at a lower stretch level.
\par
\begin{figure}[H]
	\centering
	\includegraphics[width=1\linewidth]{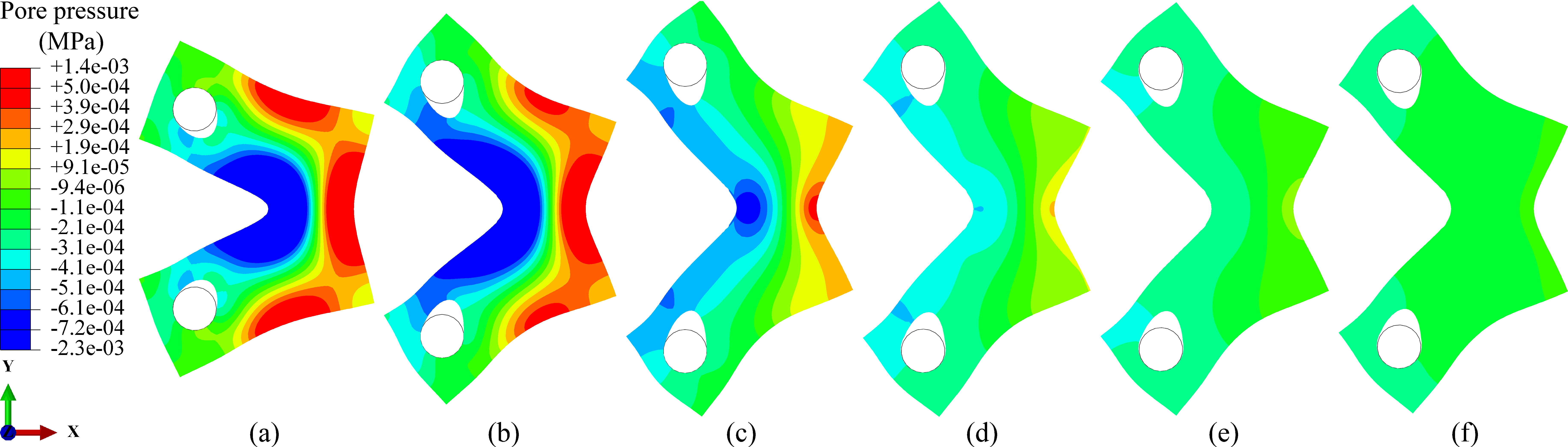}
	\caption{Distributions of pore pressure in blood clots subject to different stretch rates: (a) $1/3$ mm/s; (b) $1/6$ mm/s; (c) $1/20$ mm/s; (d) $1/40$ mm/s; (e) $1/80$ mm/s and (f) $1/200$ mm/s.}
	\label{fig:20}
\end{figure}

Due to the low permeability of blood clots, the timescale of fluid flow within the tissue is larger than that of the viscoelasticity of a fibrin network \citep{ghezelbash2022blood,Liu2024}. Directly measuring the accurate contribution of each mechanism is challenging. Instead, the critical length scales of fluid transport in clots subject to different loading rates are estimated. The significant fluid migration is confined in a region surrounding the crack tip. The length scale of this region is calculated by $D/V_{c}$ \citep{Long2016}, where $D = 2 \times 10^{-7}$ m$^{2}$/s is the effective diffusivity \cite{ghezelbash2022blood,Liu2024}, and $V_{c}$ is the average crack propagation rate. The approximated length scales are $0.63$ mm, $1.21$ mm, $4.72$ mm, $9.59$ mm, $17.94$ mm and $47.2$ mm for the clot specimens under the selected loading rate of $1/3$ mm/s through $1/200$ mm/s, respectively. In stage I, the length scale of fluid transport is limited. Therefore, the time-dependent fracture is dominated by the viscoelasticity of the fibrin network. In stage II, the non-equilibrium energy significantly dissipates due to the low loading rate, and its contribution to the time-dependent fracture weakens. In this stage, fluid sufficiently flows to the region ahead of the crack tip, resulting in a large critical length scale of the fluid transport. \Cref{fig:20} shows the distribution of pore fluid pressure at different loading rates. In stage I, the pore pressure at the crack tip region remains low, while it significantly increases and approaches zero in stage II with the drop in loading rates.
\par
\begin{figure}[!htb]
	\centering
	\includegraphics[width=1\linewidth]{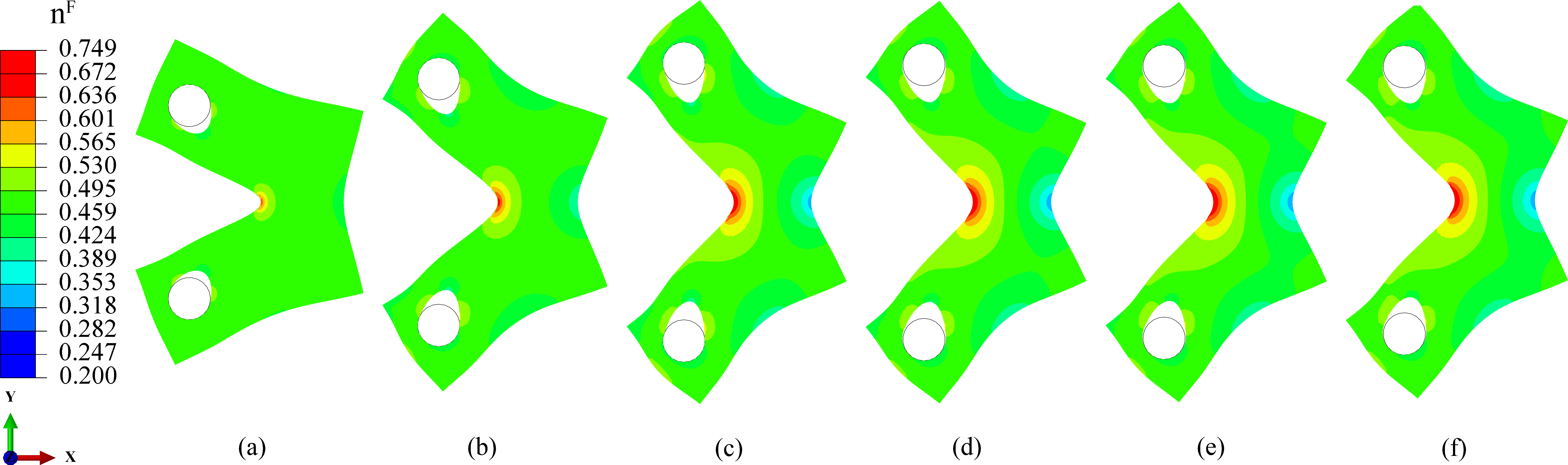}
	\caption{Distributions of fluid volume fraction in blood clots subject to different stretch rates: (a) $1/3$ mm/s; (b) $1/6$ mm/s; (c) $1/20$ mm/s; (d) $1/40$ mm/s; (e) $1/80$ mm/s and (f) $1/200$ mm/s.}
	\label{fig:21}
\end{figure}

The pore pressure gradient drives the flow of fluid through the fibrin network and induces fluid redistribution, as shown in \Cref{fig:21}. With the fluid continuing to migrate to the crack tip region, fluid accumulates at this region, causing the increase of the fluid volume fraction, \ie, tissue swelling. The swelling leads to further stretch of fibrin fibers, facilitating the crack propagation. At a lower stretch rate, more water flows to the crack tip, which results in a higher swelling stretch of the fibrin fiber, promoting the fracture of the tissue. Therefore, the critical displacement and maximum force decrease with stretch rate declining in stage II (see \Cref{fig:19}), and fluid transport is the dominant mechanism of the time-dependent fracture. 
\par
\begin{figure}[H]
	\centering
	\includegraphics[width=1\linewidth]{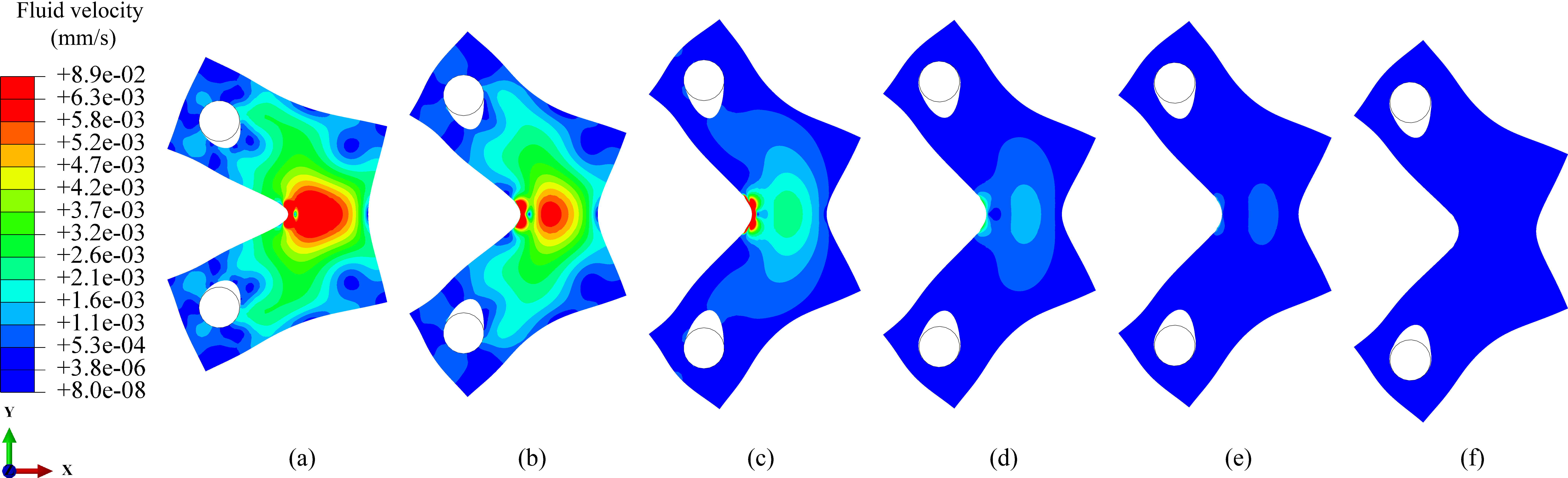}
	\caption{Distributions of flow velocity of the fluid in blood clots subject to different stretch rates: (a) $1/3$ mm/s; (b) $1/6$ mm/s; (c) $1/20$ mm/s; (d) $1/40$ mm/s; (e) $1/80$ mm/s and (f) $1/200$ mm/s.}
	\label{fig:22}
\end{figure}

The redistribution of fluid and evolution of pore pressure gradient can be reflected by the fluid flow velocity (\Cref{fig:22}). Since the velocity is associated with the pore fluid pressure, the fluid flow velocity (\Cref{fig:22}a to c) is large in the high stretch rate regime (stage I). Nevertheless, the significant redistribution of fluid in the low stretch rate stage (\Cref{fig:21}e and f) induces the uniformity of pore pressure in the whole field (\Cref{fig:20}e and f) and thereby leads to the decrease of the fluid flow velocity.
\par
The time-dependent fracture of blood clots was recently experimentally studied by \citet{Liu2024}, where the clot specimens are stretched at three loading rates. The critical stretch shows a pattern where it initially rises and then falls as the loading rate increases. The experimentally observed critical stretch-loading rate relation agrees with the simulated two-stage variation pattern of the critical displacement with respect to loading rates. Using the proposed model, the underlying mechanisms of the time-dependent fracture of blood clots can be well understood.
\subsection{Double-edge cracked whole blood clots under mixed-mode loading}
\label{sec:fracture_shear}
The loading condition of blood clots in a realistic physiological environment is complex. As blood clots adhere to the blood vessel wall, deformation of the vessel wall can cause stretching of the clots. Furthermore, blood flows over the surface of a blood clot. The interfacial interaction between the blood clot and blood, \ie, viscous traction, leads to a shear strain of the clot. In this section, the main loading applied to a blood clot is assumed to be tensile force and shear force (\Cref{fig:1}). The mixed-mode fracture of a double-edge cracked clot, subject to the two dominant loading conditions, is predicted.
\par
\begin{figure}[!htb]
	\centering
	\includegraphics[width=0.5\linewidth]{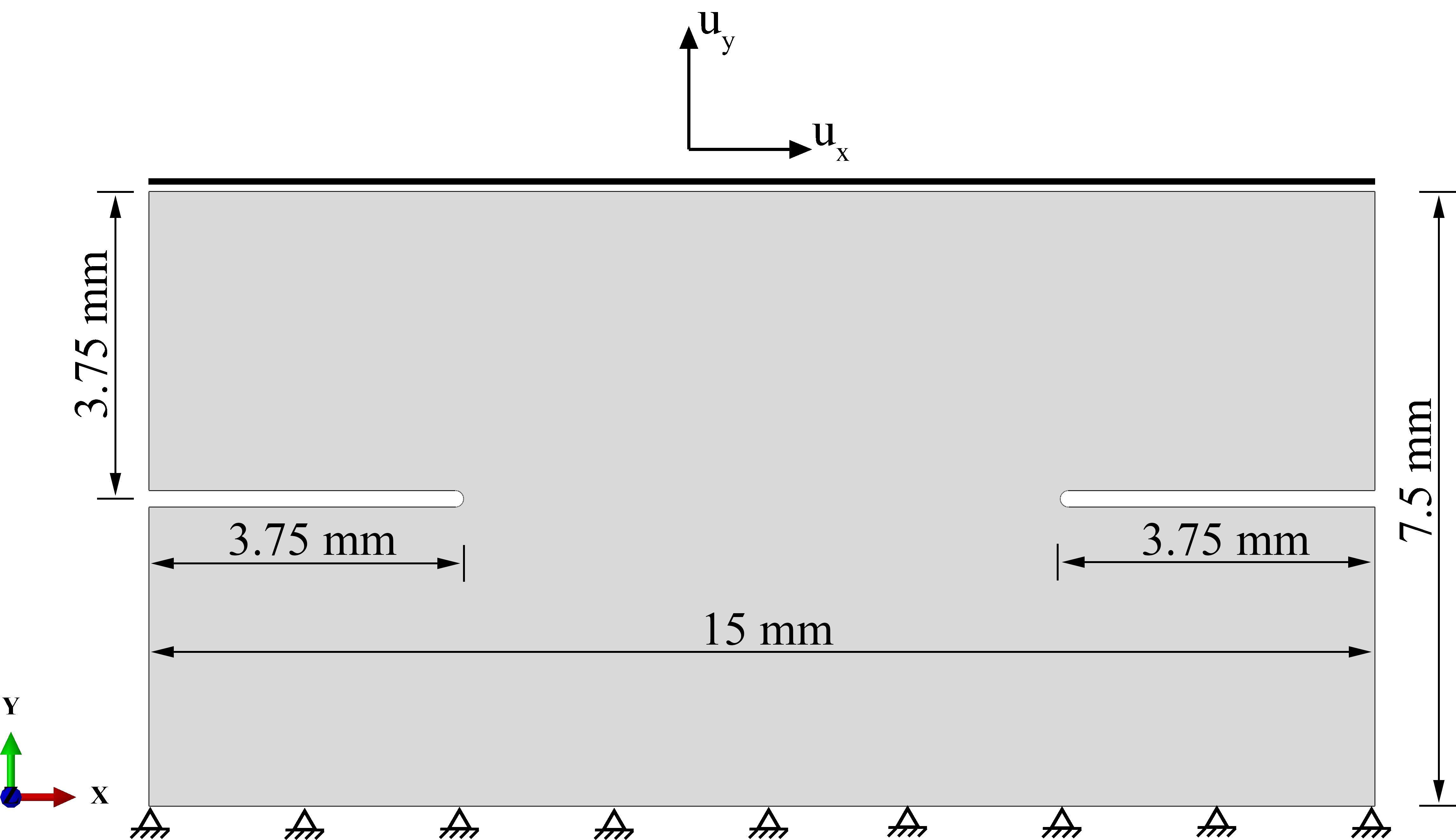}
	\caption{Geometry, dimensions and boundary conditions of a double-edge cracked blood clot specimen in mixed-mode fracture simulations. The thickness of the specimens is 6.5 mm.}
	\label{fig:23}
\end{figure}

The geometry, dimensions and boundary conditions of the computational model are shown in \Cref{fig:23}. A blood clot specimen with two initial cracks is utilized in the simulations. The bottom of the model is fixed, and the top surface is subject to displacement in the y- and x-axis directions, \ie, $u_{y}$ and $u_{x}$, respectively, at a fixed shear rate of 1/6 mm/s. Three values of the ratio of $u_{y}$ to $u_{x}$, \ie, $u_{y}/u_{x}$= 0.5, 0.75 and 1, are considered to predict the clot fracture patterns in different physiological loading situations. The model is meshed by CPE4RPT elements, with the smallest element size of 50 $\mu$m. The model parameters of the clot specimen with a low fibrin content, as depicted in \Cref{sec:validation}, are used.
\par
\begin{figure}[H]
	\centering
	\subfloat[]{
		\begin{minipage}[b]{0.44\linewidth}
			\includegraphics[width=1\linewidth]{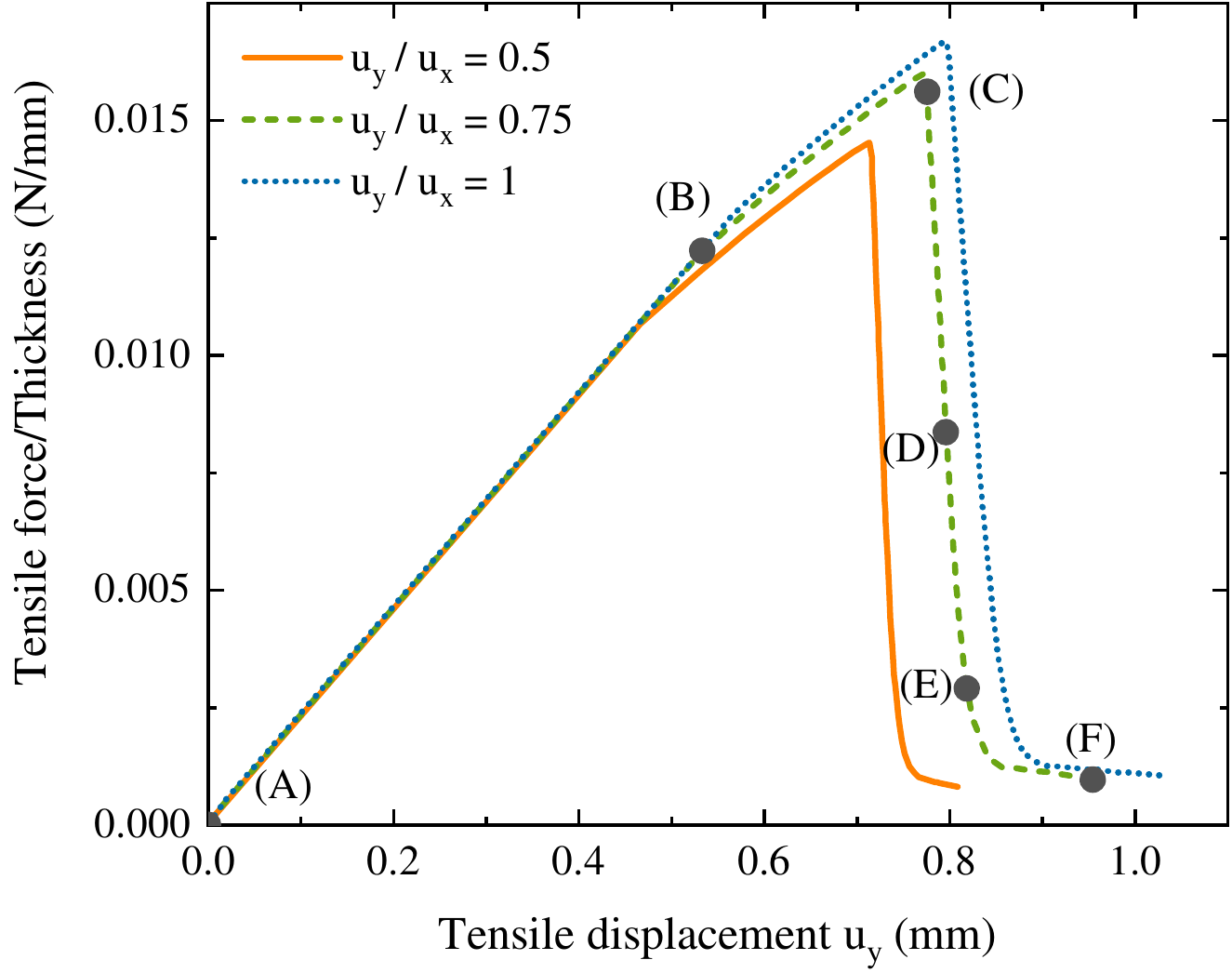} 
		\end{minipage}
		\label{}
	}
	\quad
	\subfloat[]{
		\begin{minipage}[b]{0.445\linewidth}
			\includegraphics[width=1\linewidth]{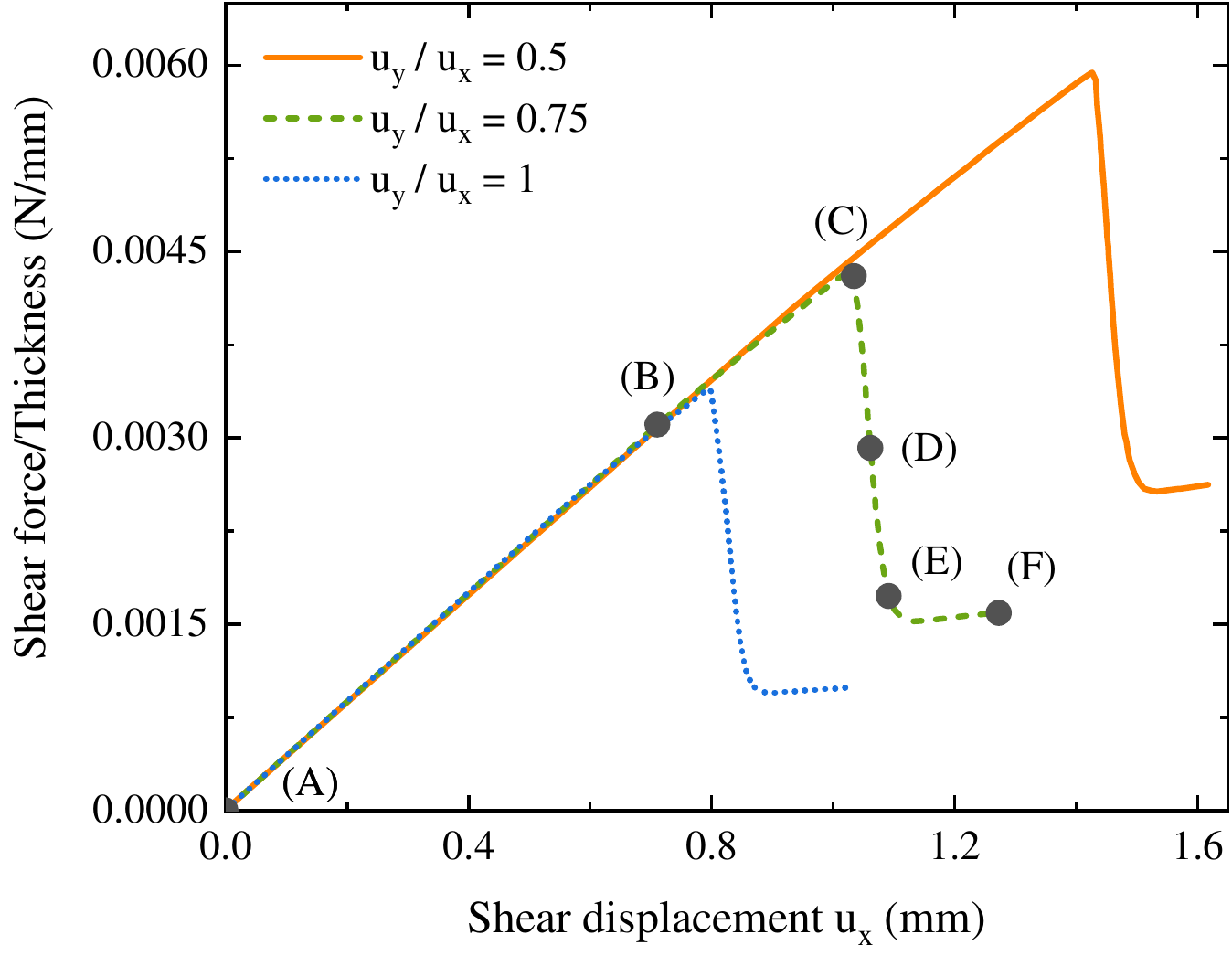}
		\end{minipage}
		\label{}
	}
	\caption{Predicted force-displacement curves obtained from a blood clot subject to different tension/shear ratios: (a) tensile force-tensile displacement $u_{y}$ and (b) shear force-shear displacement $u_{x}$.}
	\label{fig:24}
\end{figure}
\begin{figure}[H]
	\centering
	\includegraphics[width=1.0\linewidth]{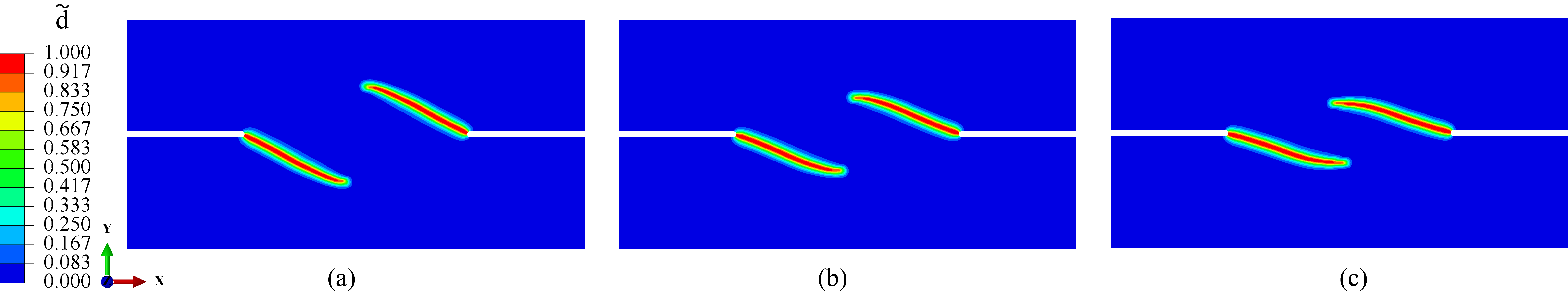}
	\caption{Comparison of fracture patterns in blood clot specimens subject to the loading of (a) $u_{y}/u_{x} = 0.5$, (b) $u_{y}/u_{x} = 0.75$ and (c) $u_{y}/u_{x} = 1.0$. For a better comparison and visualization, the undeformed contours of the clot damage are displayed.}
	\label{fig:25}
\end{figure}

The force-displacement responses in the specimens under different tension/shear loading conditions are plotted in \Cref{fig:24}. The tensile strength of the clot is roughly proportional to the ratio $u_{y}/u_{x}$, as shown in \Cref{fig:24}a, while the shear strength demonstrates an opposite trend, see \Cref{fig:24}b. The corresponding crack paths in the clot specimens subject to different $u_{y}/u_{x}$ ratios are displayed in \Cref{fig:25}. The crack propagation path is sensitive to the ratio $u_{y}/u_{x}$. Subject to pure stretch in the y-axis direction, \ie, $u_{y}/u_{x} = +\infty$, crack extends in the x-axis direction (\Cref{fig:9}). Upon the coexistence of tension and shear, \ie, $u_{x}\neq0$ and $u_{y}\neq0$, the propagation direction deviates from the initial orientation of the crack, \ie, the x-axis direction. In the case of $u_{y}/u_{x} = 0.5$, where the tensile displacement is the least among the three situations, the crack deviation is the most significant, see \Cref{fig:25}a. With the increase of the ratio $u_{y}/u_{x}$, the deviation angle reduces, and the crack path deflects to the x-axis direction (\Cref{fig:25}b and c). In a blood clot, multiple cracks may exist simultaneously. Their interaction and merging result in clot fragmentation. The simulation results indicate that the variation of the physiological loading changes the propagation path of the cracks, and thereby impacting the clot fragmentation mode, which is crucial for the evolution of thromboembolism-induced diseases.
\par
\begin{figure}[H]
	\centering
	\includegraphics[width=1.0\linewidth]{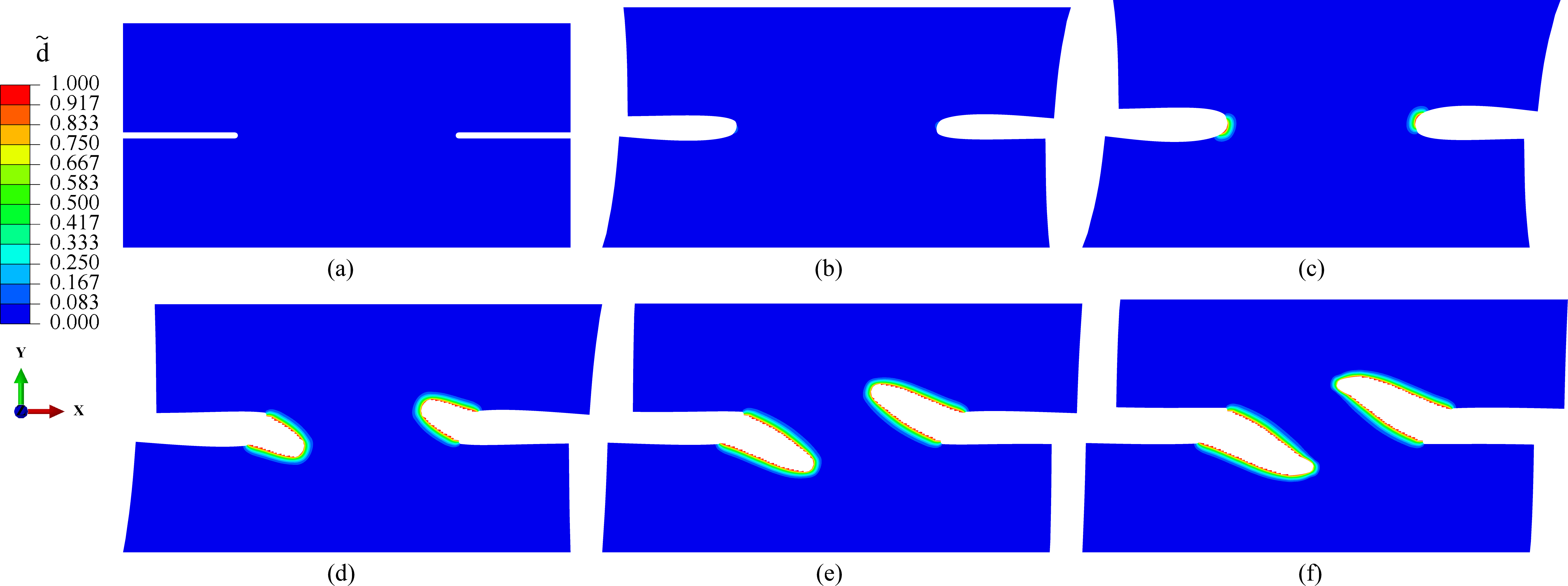}
	\caption{Contour plots of damage evolution and crack propagation in the blood clot specimen subject to the loading of $u_{y}/u_{x} = 0.75$ at the points (A) through (F) in \Cref{fig:24}.}
	\label{fig:26}
\end{figure}

To show the crack evolution procedure, the counter plots for the clot damage under $u_{y}/u_{x} = 0.75$ at the points (A) through (F) on the force-displacement curves in \Cref{fig:24} are demonstrated in \Cref{fig:26}. \Cref{fig:26}a shows the initial configuration with no damage. As the specimen is stretched, the energy at the crack tip increases. When it exceeds the energy threshold, damage initiates, and further stretch causes damage accumulation at the crack tip (\Cref{fig:26}b). As the damage accumulates to a critical level due to increasing loading, fracture initiates. \Cref{fig:26}c shows the onset of crack extension. The subsequent fracture process is presented in \Cref{fig:26}d to f.
\begin{figure}[H]
	\centering
	\includegraphics[width=1.0\linewidth]{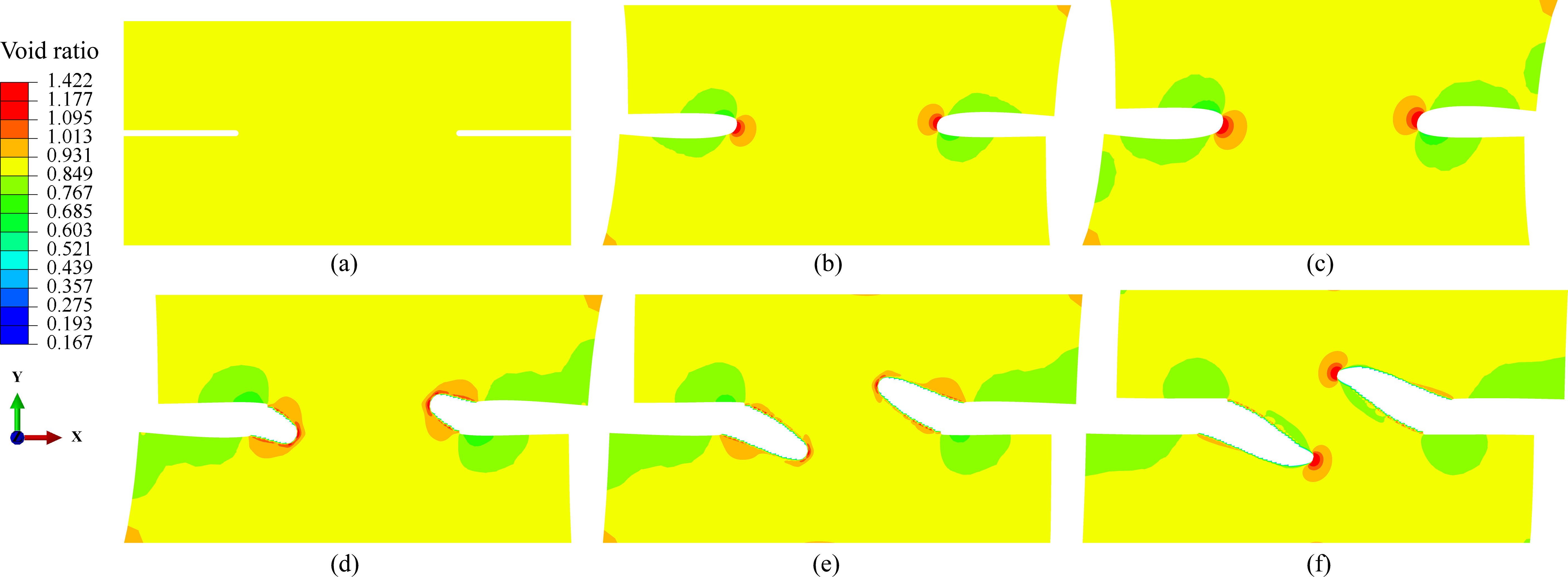}
	\caption{Contour plots of the void ratio distribution in the blood clot specimen subject to the loading of $u_{y}/u_{x} = 0.75$ at the points (A) through (F) in \Cref{fig:24}.}
	\label{fig:27}
\end{figure}
\par
The counter plots of the void ratio distribution in the clot specimen under $u_{y}/u_{x} = 0.75$ are demonstrated in \Cref{fig:27}. The selected contours correspond to the points (A) through (F) on the force-displacement curves in \Cref{fig:24}. In the initial state, the void ratio distributes uniformly in the whole specimen (\Cref{fig:27}a). Upon stretching, fluid flows towards the crack tip region, causing the increase of the void ratio at the crack tip (\Cref{fig:27}b and c). With further loading, the crack propagates, and the region of high void ratio moves with the crack tip (\Cref{fig:27}d and f). Since fluid influx leads to swelling of the tissue, the swelling stretch of the fibrin network contributes to the fracture and aggravates the deflection of the crack path.
\par
\begin{figure}[H]
	\centering
	\includegraphics[width=0.75\linewidth]{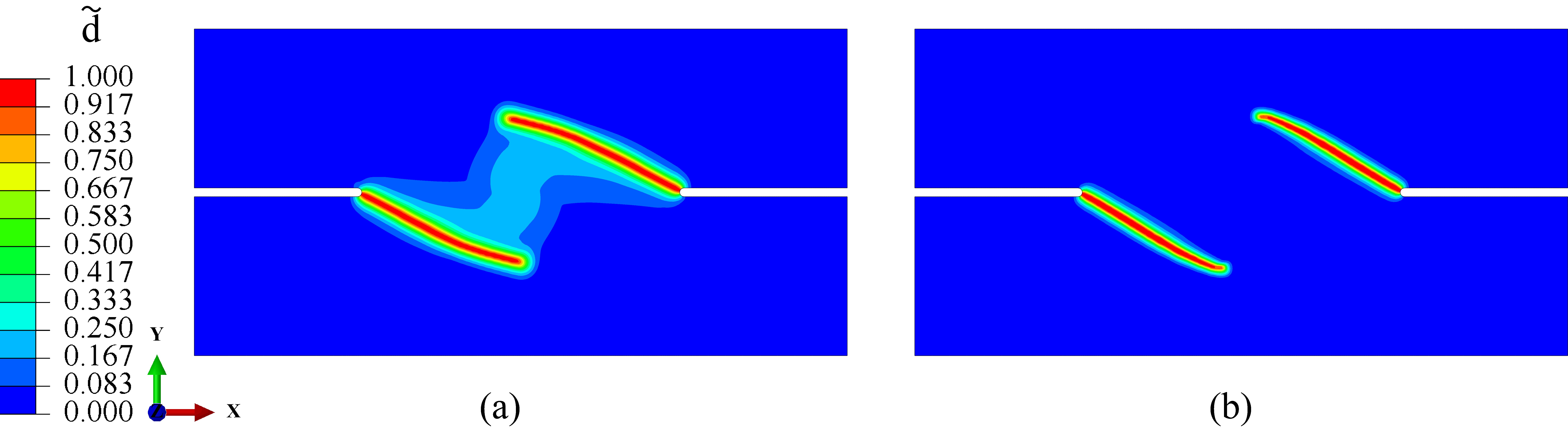}
	\caption{Comparison of the predicted damage evolution in the clot specimen subjected to $u_{y}/u_{x} = 0.5$ using different gradient enhancement algorithms: (a) conventional gradient-enhanced model with a constant regularization parameter and (b) improved gradient-enhanced model with an evolving regularization parameter. For a better comparison and visualization, the undeformed contours are displayed.}
	\label{fig:28}
\end{figure}

The conventional gradient-enhanced damage models are known to demonstrate non-physically broadened damage zones due to the assumption of a constant regularization parameter. A new evolving regularization parameter function is proposed in this work to confine the damage zone. \Cref{fig:28} shows the comparison between the predicted cracks using different gradient enhancement algorithms. \Cref{fig:28}a illustrates the simulated crack propagation using the conventional gradient-enhanced method, which manifests spurious damage growth. The two cracks extend and interact due to the unrealistically broadened damage zone. \Cref{fig:28}b presents the crack pattern using the improved gradient-enhanced model. By employing the evolving regularization parameter function, the damage evolution is confined to a finite bandwidth.
\section{Conclusion}
\label{sec:conclusion}
Fracture of blood clots is associated with many thromboembolic diseases. To study their complex fracture mechanisms, a thermodynamically consistent multi-physics, gradient-enhanced damage model was developed for blood clots within the framework of porous media theory at finite strains. The main features of the proposed model involve: (\rom{1}) solid/fluid interaction and the fluid transport through the solid skeleton; (\rom{2}) an energy-based damage model considering tension-compression asymmetry and the energy contributions from different solid constituents; (\rom{3}) evolving regularization function in the gradient-type enhancement of the free energy density function; (\rom{4}) the concept of unfolding of fibrin molecules at the microscopic levels and (\rom{5}) non-linear viscoelasticity of the fibrin network by multiplicatively decomposing the solid deformation gradient tensor into elastic and inelastic parts. Furthermore, a physics-based network model was extended by incorporating fluid-transport-induced compressibility to describe the constitutive behavior of the solid skeleton in blood clots. The proposed model was implemented into a finite element program by writing user subroutines. It was validated by comparing the computational results to data from fracture tests. The effect of the main model parameters was investigated. Using the proposed model, the fracture behavior of blood clots was studied systematically. The main conclusions and contributions of this work are summarized as follows according to the results and discussions:
\begin{itemize}
	\item The computational results are in good agreement with experimental data from fracture tests using whole blood clot specimens with different fibrin contents. The experimentally measured force-displacement responses, fracture patterns and deformations can be well simulated using the proposed model.
	\item The experimentally observed time-dependent large deformation and fracture of blood clots are simulated, and the underlying mechanisms are captured by the proposed model. The dominant mechanism changes according to the timescale of loading. When the loading time approaches the characteristic time of the intrinsic viscoelasticity of the fibrin network and fluid transport, these two physical processes significantly contribute to clot fracture.
	\item In mixed-mode fracture simulations, the crack propagation path is affected by the loading modes. Subject to physiological loading conditions, initial cracks can be deflected from their initial direction. These cracks interplay and can cause the clot fragmentation. The high void ratio region moves with the crack tip, facilitating the crack propagation by increasing the swelling stretch of the fibrin fibers. The proposed method can capture the complex fracture process under mixed-mode loading and can provide insights into the underlying fracture mechanisms.  
	\item The damage processing zone is confined to a reasonably small bandwidth by incorporating the proposed evolving regularization parameter. The mesh dependency of damage in simulations is significantly alleviated by introducing the gradient-type enhancement of free energy function.
\end{itemize}
\par
The proposed theoretical framework can describe the multiple physical processes and their interactions in blood clots, which is able to accurately predict the complex time-dependent fracture of blood clots under physiological loading. The findings in the present work provide deep insight into the mechanisms of clot fracture and facilitate the understanding of physiological hemostasis and pathological thrombosis. Therefore, the developed model can advance the development of new strategies to prevent, diagnose and treat thrombotic diseases. Furthermore, the developed computational method can also contribute to developing novel wound-dressing biomaterials.
\appendix
\section{Material time derivatives in Clausius--Plank inequality}
\label{app:time_derivative}
According the definition of $\psi^{S}$ and $\psi^{F}$ in \Cref{equ:free_energy_solid_fluid}, and recalling the decomposition of $\psi^{S}$ in \Cref{equ:free_energy_solid_total}, the material time derivatives of $\psi^{S}$ and $\psi^{F}$ are calculated as
\begin{equation}
	\label{equ:time_derivative_energy}
	\begin{aligned}
		\pqty{\psi^{S}}'_{S} &= \pqty{\psi_{eq}^{S}}'_{S} + \pqty{\psi_{neq}^{S}}'_{S} + \pqty{\psi_{fm}^{S}}'_{S} + \pqty{\psi_{nl}^{S}}'_{S} \\
		&= \sum_{i}^{m,fib} n^{i} \pqty{\psi_{eq}^{i}}'_{S} + \sum_{i}^{m,fib} n^{i} \pqty{\psi_{neq}^{i}}'_{S} + \pqty{\psi_{fm}^{S}}'_{S} + \pqty{\psi_{nl}^{S}}'_{S}  \qq{and} \\ 
		\pqty{\psi^{F}}'_{F} &= 0 \,,
	\end{aligned}
\end{equation}
respectively. Recalling the expressions in \cref{equ:free_energy_solid_parts}, the material time derivatives in \cref{equ:time_derivative_energy}$_1$ are written as
\begin{align}
	\label{equ:time_derivative_energy_eq_neq}
	\pqty{\psi_{eq}^{i}}'_{S} &= 2\pqty{\vb{F}_{S} \pdv{\psi_{eq}^{i}}{\vb{C}_{S}} \vb{F}_{S}^\intercal} \vdot \vb{D}_{S} + \pdv{\psi_{eq}^{i}}{d} \pqty{d}'_{S} + \pdv{\psi_{eq}^{i}}{\lambda_{fm}} \pqty{\lambda_{fm}}'_{S} \,, \notag \\
	\pqty{\psi_{neq}^{i}}'_{S} &= 2\pdv{\psi_{neq}^{i}}{\widehat{\vb{C}}_{S}^{e,i}} \vdot \pqty{\widehat{\vb{\Gamma}}_{S}^{e,i}}'_{S} + \pdv{\psi_{neq}^{i}}{d} \pqty{d}'_{S} + \pdv{\psi_{neq}^{i}}{\lambda_{fm}} \pqty{\lambda_{fm}}'_{S} \notag \\
	&= 2\pdv{\psi_{neq}^{i}}{\widehat{\vb{C}}_{S}^{e,i}} \vdot \pqty{\widehat{\vb{\Gamma}}_{S}^{i}}_{S}^{v \vartriangle} - 2\pqty{\widehat{\vb{C}}_{S}^{e,i} \pdv{\psi_{neq}^{i}}{\widehat{\vb{C}}_{S}^{e,i}}} \vdot \widehat{\vb{L}}_{S}^{v,i} + \pdv{\psi_{neq}^{i}}{d} \pqty{d}'_{S} + \pdv{\psi_{neq}^{i}}{\lambda_{fm}} \pqty{\lambda_{fm}}'_{S} \,, \\
	\pqty{\psi_{fm}^{S}}'_{S} &= \pdv{\psi_{fm}^{S}}{\lambda_{fm}} \pqty{\lambda_{fm}}'_{S} + \pdv{\psi_{fm}^{S}}{d} \pqty{d}'_{S} \qq{and} \notag \\
	\pqty{\psi_{nl}^{S}}'_{S} &= \pdv{\psi_{nl}^{S}}{\overline{Y}_{eqv}^{S}} \pqty{\overline{Y}_{eqv}^{S}}'_{S} + \pdv{\psi_{nl}^{S}}{\grad_{\vb{x}}{\overline{Y}_{eqv}^{S}}} \vdot \grad_{\vb{x}} \pqty{\overline{Y}_{eqv}^{S}}'_{S} + 2\pqty{\vb{F}_{S} \pdv{\psi_{nl}^{S}}{Y_{eqv}^{S}} \pdv{Y_{eqv}^{S}}{\vb{C}_{S}} \vb{F}_{S}^\intercal} \vdot \vb{D}_{S} \,. \notag
\end{align}
The rate of strain tensor $\widehat{\vb{\Gamma}}_{S}^{e,i}$ in \cref{equ:time_derivative_energy_eq_neq}$_2$ must be objective according to the material objectivity principle in the large deformation context. Here, the Oldroyd rates of $\widehat{\vb{\Gamma}}_{S}^{i}$ is introduced for replacing $\pqty{\widehat{\vb{\Gamma}}_{S}^{e,i}}'_{S}$ to satisfy the material objectivity principle. \cref{equ:time_derivative_energy_eq_neq}$_2$ is derived using the definition of the Oldroyd rates of $\widehat{\vb{\Gamma}}_{S}^{i}$ and $\widehat{\vb{\Gamma}}_{S}^{v,i}$, and the inelastic strain rate tensor $\widehat{\vb{D}}_{S}^{v,i}$, \ie,
\begin{equation}
	\label{equ:Oldroyd_rate_of_Gamma_5}
	\begin{aligned}
		\pqty{\widehat{\vb{\Gamma}}_{S}^{i}}_{S}^{v \vartriangle} &\coloneqq \pqty{\widehat{\vb{\Gamma}}_{S}^{i}}'_{S} + \pqty{\widehat{\vb{L}}_{S}^{v,i}}^\intercal \widehat{\vb{\Gamma}}_{S}^{i} + \widehat{\vb{\Gamma}}_{S}^{i} \widehat{\vb{L}}_{S}^{v,i} \,, \\
		\pqty{\widehat{\vb{\Gamma}}_{S}^{v,i}}_{S}^{v \vartriangle} &\coloneqq \pqty{\widehat{\vb{\Gamma}}_{S}^{v,i}}'_{S} + \pqty{\widehat{\vb{L}}_{S}^{v,i}}^\intercal \widehat{\vb{\Gamma}}_{S}^{v,i} + \widehat{\vb{\Gamma}}_{S}^{v,i} \widehat{\vb{L}}_{S}^{v,i} \,, \\
		\widehat{\vb{D}}_{S}^{v,i} &\coloneqq \frac{\widehat{\vb{L}}_{S}^{v,i} + \pqty{\widehat{\vb{L}}_{S}^{v,i}}^\intercal}{2} \,,
	\end{aligned}
\end{equation}
respectively, where $\widehat{\vb{L}}_{S}^{v,i} \coloneqq \pqty{\vb{F}_{S}^{v,i}}'_{S} \pqty{\vb{F}_{S}^{v,i}}^{-1}$ is the inelastic velocity gradient tensor. It is noted that the stress power $2\pqty{\pdv*{\psi_{neq}^{i}}{\widehat{\vb{C}}_{S}^{e,i}}} \vdot \pqty{\widehat{\vb{\Gamma}}_{S}^{i}}_{S}^{v \vartriangle}$, namely the rate of internal mechanical work, in \cref{equ:time_derivative_energy_eq_neq}$_2$ is expressed in the intermediate configuration. Since the energetically conjugate pairs stay invariant between different configurations, the stress power can be alternatively expressed in the current configuration as \citep{markert2008biphasic,liu2022computational}
\begin{equation}
	\label{equ:time_derivative_energy_eq_neq_4}
	2\pdv{\psi_{neq}^{i}}{\widehat{\vb{C}}_{S}^{e,i}} \vdot \pqty{\widehat{\vb{\Gamma}}_{S}^{i}}_{S}^{v \vartriangle} = 2\bqty{\vb{F}_{S}^{e,i} \pdv{\psi_{neq}^{i}}{\widehat{\vb{C}}_{S}^{e,i}} \pqty{\vb{F}_{S}^{e,i}}^\intercal} \vdot \vb{D}_{S} \,.
\end{equation}
Making use of \cref{equ:time_derivative_energy_eq_neq_4}, the material time derivative of non-equilibrium free energy function in \cref{equ:time_derivative_energy_eq_neq}$_2$ can be reformulated as
\begin{equation}
	\label{equ:time_derivative_energy_neq_4}
	\pqty{\psi_{neq}^{i}}'_{S} = 2\bqty{\vb{F}_{S}^{e,i} \pdv{\psi_{neq}^{i}}{\widehat{\vb{C}}_{S}^{e,i}} \pqty{\vb{F}_{S}^{e,i}}^\intercal} \vdot \vb{D}_{S} - 2\pqty{\widehat{\vb{C}}_{S}^{e,i} \pdv{\psi_{neq}^{i}}{\widehat{\vb{C}}_{S}^{e,i}}} \vdot \widehat{\vb{L}}_{S}^{v,i} + \pdv{\psi_{neq}^{i}}{d} \pqty{d}'_{S} \,.
\end{equation}
\par
The material time derivative of the saturation equation \cref{equ:saturation_condition} in the Clausius--Planck inequality \cref{equ:entropy_inequ_with_Saturatiion_1} following the solid motion is given as
\begin{equation}
	\label{equ:mat_time_diritive_Saturation_cond}
	\pqty{n^{S} + n^{F}}'_{S} = -\pqty{n^{S} \vb{I} \vdot \vb{D}_{S} + n^{F} \vb{I} \vdot \vb{D}_{F} + \grad_{\vb{x}}{n^{F}} \vdot \vb{w}_{FR}} \,,
\end{equation}
where the volume balance equation \cref{equ:volume_balance} and $\grad_{\vb{x}} \vdot \acute{\vb{x}}_{\alpha} = \vb{D}_{\alpha} \vdot \vb{I}$ are used.
\bibliographystyle{elsarticle-num-names} 
\bibliography{References}






\end{document}